\documentclass[superscriptaddress,12pt,onecolumn,nofootinbib]{revtex4-2} 

\usepackage{mathtools}
\usepackage{braket}
\usepackage{hyperref}

\usepackage{booktabs,makecell,array}
\newcolumntype{P}[1]{>{\raggedright\arraybackslash}p{#1}}

\usepackage{amsmath}  
\usepackage{amsfonts} 
\usepackage{graphicx} 
\usepackage{esvect}
\usepackage{mathtools}

\usepackage{xcolor}
\usepackage{appendix}
\usepackage{siunitx} 
\usepackage{mathrsfs}

\usepackage{placeins}

\newcounter{researchquestion}

\usepackage[caption=false]{subfig} 

\usepackage{pbox}

\usepackage{mathtools}
\makeatletter
\newcommand{\vast}{\bBigg@{3}}
\newcommand{\Vast}{\bBigg@{3.5}}
\makeatother

\begin{document}


\title{A Numerical Approach to Quantization}


\author{Grace Murray}
\thanks{These authors contributed equally.}
\affiliation{Harvey Mudd College, Claremont, CA 91711, USA.}

\author{Siyue Zhu}
\thanks{These authors contributed equally.}
\affiliation{Scripps College, Claremont, CA 91711, USA.}

\author{Angeline Hu}
\affiliation{Claremont McKenna College, Claremont, CA 91711, USA.}

\author{Kaitlyn S. Yasumura}
\affiliation{Scripps College, Claremont, CA 91711, USA.}

\author{Ioannis G. Kevrekidis}
\affiliation{Department of Chemical and Biomolecular Engineering, \\ John Hopkins University, Baltimore, MD 21218, USA}

\author{Trung V. Phan}
\email{{tphan@natsci.claremont.edu}}
\affiliation{Department of Natural Sciences, Scripps and Pitzer Colleges, \\ Claremont Colleges Consortium, Claremont, CA 91711, USA}

\begin{abstract}
We introduce a numerical approach that simultaneously searches for matrix quantizations and compatible Poisson structures on classical geometric surfaces. Our method identifies irreducible Hermitian matrix representations and a trainable Poisson multiplier by minimizing a loss function that penalizes violations of the Casimir and commutation constraints. Through numerical investigations and subsequent theoretical analysis, we establish a no-go theorem showing that, for Weyl quantized fuzzy spheres with nondegenerate axial spectrum, the matrix dimension cannot exceed the degree of the nonconstant polynomial axial Poisson multiplier. We further extend our approach to nonpolynomial Poisson multipliers, demonstrating its versatility and potential to drive novel discoveries. Although here we focus on Weyl quantized fuzzy spheres, the framework is extensible to general geometric manifolds and quantization schemes. Our approach also provides a practical starting point for investigating Schild string theory and the quantization of spacetime histories.
\end{abstract}

\date{\today}

\maketitle 

\section{Introduction}

\begin{figure*}[hbp!]
    \centering
\includegraphics[width=\linewidth]{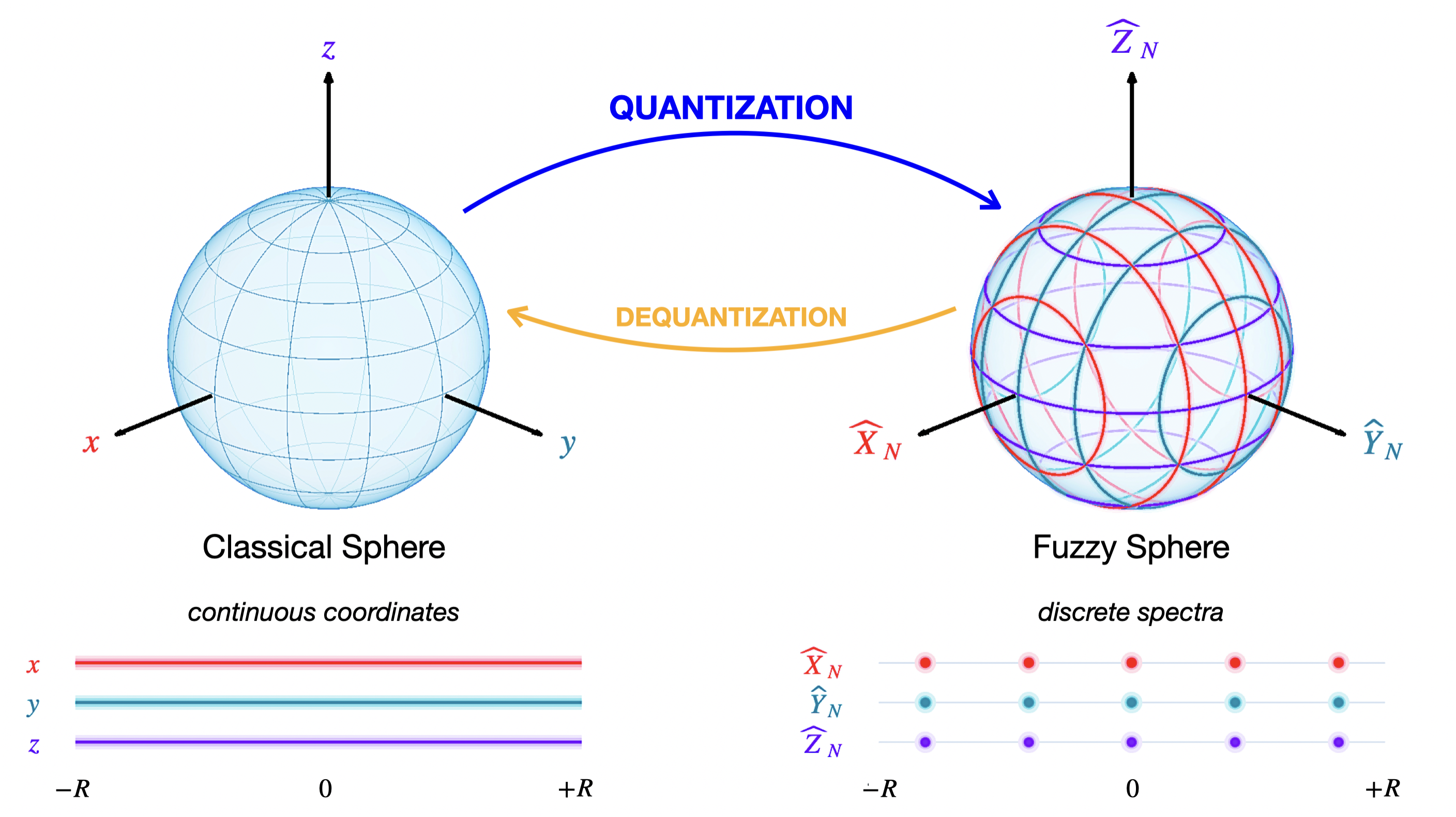}
    \caption{\textbf{Quantization of a classical surface into a fuzzy matrix geometry.} A classical sphere with continuous coordinates is quantized into a fuzzy sphere, where the coordinates become matrices with discrete spectra (here we use matrix dimension $N=5$). The reverse process, which maps a fuzzy geometry back to a classical surface, is called dequantization.}
\label{fig00}
\end{figure*}

Quantum mechanics provides our fundamental description of the physical world, but our everyday experience is effectively described by classical mechanics \cite{zurek2003decoherence}. Because quantum phenomena are far removed from our usual intuition, constructing quantum models from classical knowledges presents a profound conceptual and mathematical challenge.  Quantization offers a systematic way for such construction, but the path is generally not unique. When promoting commuting classical variables to noncommuting quantum operators representing physical observables, different operator-ordering prescriptions can lead to different quantum theories \cite{ali2005quantization}. Moreover, the Groenewold-van Hove obstruction \cite{groenewold1946principles,gotay1999groenewold} prevents any quantization from preserving all classical Poisson bracket relations under Dirac canonical quantization rules \cite{dirac1925fundamental,dirac1958quantum}, raising the question of which properties can be consistently retained. In matrix approaches to quantization \cite{arnlind2009fuzzy}, classical coordinate functions are represented by finite-dimensional Hermitian matrices satisfying algebraic constraints determined by the geometry and its Poisson structure \cite{arnlind2009noncommutative,steinacker2011non} (see Fig. \ref{fig00}). This construction is purely kinematics, i.e. quantizing the geometry directly by specifying an algebra of coordinate observables consistent with the classical Poisson structure. Describing the dynamics, however, requires additional structure, such as a Hamiltonian or an action.

Fuzzy geometries \cite{barrett2015matrix}, i.e. finite-dimensional matrix descriptions of noncommutative geometry \cite{varilly2006introduction,connes2010noncommutative}, have a broad range of applications in theoretical physics, including quantum membranes \cite{de1999quantum,hoppe1989quantum,vo2026size}, nonperturbative formulations of string theory \cite{taylor2001m,banks1999m,ishibashi1997large}, the M2–M5 brane system \cite{basu2005m2}, matrix formulations of gauge theories \cite{steinacker2004quantized}, and critical phenomena \cite{he2026fuzzy}. They also provide frameworks for emergent geometry and gravity, in which spacetime geometry arises from more fundamental matrix degrees of freedom \cite{steinacker2010emergent,steinacker2024quantum,10.1093/ptep/ptz092}. Given these applications, determining which classical geometries can have consistent finite-dimensional matrix realizations is an important question. Although highly symmetric examples  -- such as fuzzy spheres with trivial Poisson structures \cite{madore1992fuzzy} -- admit explicit constructions, finding and classifying irreducible matrix representations for more general manifolds and nonlinear constraints remains extremely difficult \cite{arnlind2020low}. Conventional approaches often rely on educated guesses and geometry-specific analytical ansatz \cite{arnlind2009fuzzy}, which become increasingly difficult to formulate and solve as the constraints grow more complex as the matrix dimension increases. These limitations motivate alternative approaches that can systematically search for admissible solutions. Numerical optimization has been used to construct classical matrix model solutions by minimizing equation of motion residuals \cite{10.1093/ptep/ptaa042}, while deep learning methods have been applied to variational wavefunctions of matrix quantum mechanics \cite{han2020deep}. Other machine-learning approaches train matrices from point-cloud data \cite{candelori2025robust,abanov2025quantum}, with the Poisson structure emerging from the learned geometry rather than being imposed as a fundamental constraint. These approaches study the dynamics of specified matrix models or infer matrix geometries from data.  To the best of our knowledge, a numerical approach to exact quantization that explicitly preserve the classical Poisson structure under a prescribed operator ordering have yet to be explored. 

Here, we investigate the potential of a numerical approach that uses an algebraic solver to approximate exact matrix quantizations. We find this approach particularly useful for guiding further theoretical developments that explain when these quantizations can exist. We begin by formulating the quantization problem as a concrete system of matrix algebraic constraints in Section \ref{sec:overview}, before specializing to Weyl quantized fuzzy spheres with axial Poisson multipliers in \ref{sec:fs}. In Section \ref{sec:num}, we introduce our numerical approach and apply it to search for solutions across different matrix dimensions. Our numerical results seemingly reveals a no-go rule, i.e. the admissible matrix dimensions appear to be bounded by the polynomial degree of the Poisson multiplier. Section \ref{sec:theory} provides a theoretical explanation of this observation under the assumption of a nondegenerate axial spectrum. We further demonstrate in Section \ref{sec:beyond} that our numerical approach can also identify new solutions even for nonpolynomial axial Poisson multipliers, whose quantization parameters agree with our subsequent theoretical analysis. We conclude with a discussion of our findings in Section \ref{sec:dis_con}. For simplicity, we have only focused on Weyl quantization of two-dimensional spherical surface $S^2$ embedding in three-dimensional Euclidean space $\mathbb{R}^3$ with axial Poisson multiplier, but our approach can be generalized. For a brief overview of classical Poisson structures and their compatibility condition with Casimir constraints, see Appendix \ref{app:Poi_Alg} and \ref{app:Poi_Cas}. For numerical Weyl quantization in more advanced geometric settings, see Appendix \ref{app:superman} for supermanifolds (fuzzy supersphere \cite{grosse1998fuzzy,hasebe2008susy}) and Appendix \ref{app:nambu} for Nambu structures (fuzzy hypersphere \cite{ho2016nambu,minic1999m}). 

\section{A Path to Quantization \label{sec:path}}

We first provide a brief overview of the algebraic constraints we adopt for quantizing a two-dimensional manifold embedded in three-dimensional Euclidean space in Section \ref{sec:overview}, then discuss their specific form for Weyl quantized fuzzy spheres with axial Poisson multipliers in Section \ref{sec:fs}.

\subsection{From Classical to Fuzzy Geometry \label{sec:overview}}

Consider a two-dimensional compact manifold embedded in three-dimensional Euclidean space $\mathbb{R}^3$, where a position in space is specified in a Cartesian coordinate system O$xyz$ via $$\vec{r}=x\vec{e}_x + y \vec{e}_y + z \vec{e}_z \ , $$
where $(\vec{e}_x,\vec{e}_y,\vec{e}_z)$ is an orthonormal basis. The manifold is described by an implicit surface: 
\begin{equation}
    F(x,y,z)=0 \ .
    \label{eq:implicit_surf}
\end{equation}
The most general Poisson structure in $\mathbb{R}^3$ compatible with the function $F(x,y,z)$ as its Casimir constraint, i.e.
$$ \{ F(x,y,z),x\} = \{ F(x,y,z),y\} = \{ F(x,y,z),z\} = 0 \ , $$
would be of the form:
\begin{equation}
    \begin{split}
        \{ x,y\} &= \lambda(x,y,z) \frac{\partial}{\partial z} F(x,y,z) \ ,
        \\
        \{ y,z\} &= \lambda(x,y,z) \frac{\partial}{\partial x} F(x,y,z) \ ,
        \\
        \{ z,x\} &= \lambda(x,y,z) \frac{\partial}{\partial y} F(x,y,z) \ ,
    \end{split}
\label{eq:gen_Poisson}
\end{equation} 
where Poisson multiplier $\lambda(x,y,z)$ is an arbitrary smooth function. We derive this in Appendix \ref{app:Poi_Cas}. The Poisson bracket is symplectic on the surface $F(x,y,z) = 0$ wherever $\lambda(x,y,z)$ is nonzero. Without loss of generality, we take the Poisson multiplier $\lambda(x,y,z)$ to be everywhere positive on the surface, i.e. $\lambda(x,y,z)>0$, so that the induced Poisson structure is everywhere nondegenerate and therefore defines a symplectic structure.

Finite-dimensional matrix quantization can be obtained by replacing the Cartesian coordinates $(x,y,z)$ with $N \times N$ Hermitian matrices $(\hat{X}_N,\hat{Y}_N,\hat{Z}_N)$, i.e.
\begin{equation}
    \hat{X}_N^\dagger = \hat{X}_N \ , \ \hat{Y}_N^\dagger = \hat{Y}_N \ , \ \hat{Z}_N^\dagger = \hat{Z}_N \ , 
    \label{eq:Herm_XYZ}
\end{equation}
where the Hermitian adjoint $^\dagger$ is equivalent to a complex conjugation $^*$ followed by matrix transposition $^T$, i.e. ${}^\dagger \equiv {}^{*T}$. This replacement introduces an matrix-ordering ambiguity when translating nonlinear scalar functions of the coordinates to matrix function. To resolve this ambiguity, an ordering prescription $\mathcal{Q}$ must be chosen. Although there is no universally preferred prescription, a natural requirement is that it preserves Hermiticity, i.e. for any real-valued function $f(x,y,z)$, its matrix counterpart $\mathcal{Q}[f](\hat{X}_N,\hat{Y}_N,\hat{Z}_N)$ should be Hermitian:
$$\mathcal{Q}[f](\hat{X}_N,\hat{Y}_N,\hat{Z}_N) = \left( \mathcal{Q}[f](\hat{X}_N,\hat{Y}_N,\hat{Z}_N) \right)^\dagger \ . $$ 
Throughout this work, for simplicity we adopt Weyl symmetric ordering \cite{weyl1927quantenmechanik,cheng2000gauge,steinacker2024quantum}, denoted by $\mathcal{Q} \rightarrow \mathcal{W}$. Furthermore, we do not additionally require the classical limit (i.e. commutative geometry as the matrix dimension goes to infinity):
\begin{equation}
    \lim_{N \rightarrow \infty} \hbar_N = 0 \ ,
    \label{eq:class_lim}
\end{equation}
nor do we demand matrix traces to converge to integrals over the classical surface with respect to the symplectic measure used in matrix regularization (i.e. Berezin-Toeplitz quantization) \cite{arnlind2012multi,bordemann1994toeplitz,schlichenmaier2010berezin} -- that is, we do not impose the asymptotic correspondence\footnote{If $f(x,y,z) = 1$ corresponds to
$\mathcal{Q}[1](\hat{X}_N,\hat{Y}_N,\hat{Z}_N)=\hat{I}_{N\times N}$, then this implies
\begin{itemize}
    \item The integration $\int_{F(x,y,z)=0} d\Sigma \ \left[ \lambda(x,y,z) \big| \vec{\nabla} F(x,y,z) \big| \right]^{-1}$ is a finite-value.
    \item The quantization parameters scale as $\hbar_N \propto N^{-1}$ at large-$N$, which agrees with Eq. \eqref{eq:class_lim}.
\end{itemize}}:
$$ \lim_{N \rightarrow \infty } \hbar_N \text{Tr} \left\{ \mathcal{Q}[f](\hat{X}_N,\hat{Y}_N,\hat{Z}_N) \right\}= \int_{F(x,y,z)=0} \frac{d\Sigma}{2\pi} \ \frac{f(x,y,z)}{\lambda(x,y,z) \big| \vec{\nabla} F(x,y,z) \big|} \ , $$
where $\text{Tr}$ denotes the matrix trace and $d \Sigma $ is the induced surface-area element on the surface. Here, we only focus on quantization defined purely by algebraic relations.

The Casimir constraint in Eq. \eqref{eq:implicit_surf}, after Weyl quantization, becomes:
\begin{equation}
    \mathcal{W}[F](\hat{X}_N,\hat{Y}_N,\hat{Z}_N) \equiv 0 \ .
    \label{eq:implicit_surf_mat_quant}
\end{equation}
Since the matrix-valued coordinates generally do not commute, they define a noncommutative geometry in which the classical coordinate algebra is replaced by a finite-dimensional noncommutative matrix algebra. It is then natural to relate the matrix commutators to the classical Poisson brackets in Eq. \eqref{eq:gen_Poisson}, motivated by their similar algebraic properties (see Appendix \ref{app:Poi_Alg}):
\begin{equation}
    \begin{split}
        [ \hat{X}_N, \hat{Y}_N] &\equiv i \hbar_N \mathcal{W}\left[ \lambda \frac{\partial}{\partial z} F\right](\hat{X}_N,\hat{Y}_N,\hat{Z}_N) \ ,
        \\
        [ \hat{Y}_N,\hat{Z}_N] &\equiv i \hbar_N \mathcal{W}\left[ \lambda \frac{\partial}{\partial x} F\right](\hat{X}_N,\hat{Y}_N,\hat{Z}_N) \ ,
        \\
        [ \hat{Z}_N,\hat{X}_N ] &\equiv i \hbar_N\mathcal{W}\left[ \lambda \frac{\partial}{\partial y} F\right](\hat{X}_N,\hat{Y}_N,\hat{Z}_N) \ .
    \end{split}
\label{eq:gen_Poisson_mat_quant}
\end{equation}
This correspondence is known as Dirac canonical quantization rule.

The quantization framework studied here shares the same geometric structures with the matrix formulation of string theory based on the Schild action \cite{hoppe1996m,yoneya1997schild,ishibashi1997large}. In this formulation, an auxiliary worldsheet density determines the Poisson brackets of the embedding coordinates, playing a geometric role analogous to the Poisson multiplier $\lambda(x,y,z)$. Matrix descriptions then replace these coordinates by matrices and Poisson brackets by commutators. Because the embedding coordinates include target-space time, the matrix formulation describes the string's worldsheet as a spacetime history, rather than only its spatial configuration at a fixed time. From this perspective, our search over $\lambda(x,y,z)$ and the coordinate matrices explores compatible Poisson structures together with their finite-dimensional algebraic representations. A concrete action would provide the additional structure needed to study the dynamics of these fuzzy geometries, some of which may even require infinite-dimensional representations  \cite{arnlind2019quantum}. Inspired by the approximation used for the D-string background in the Ishibashi-Kawai-Kitazawa-Tsuchiya (IKKT) model \cite{ishibashi1997large}, we could study such geometries numerically by truncating their infinite-dimensional matrices to finite size\footnote{Consider the canonical commutation relation $[\hat{P},\hat{Q}] \propto i\hat{I}$, which admits no finite-dimensional representation. An explicit finite-dimensional truncation modifies the commutation relation to \cite{gazeau2006finite,buchdahl1967concerning,polychronakos2001quantum}: $$[\hat{P}_N,\hat{Q}_N] \propto i \left(\hat{I}_{N\times N}-N|N-1\rangle\langle N-1| \right) \  . $$The error is confined to the highest retained state, so the canonical relation holds exactly on any fixed set of lower states once $N$ is sufficiently large. Related finite-matrix constructions underlie e.g. the fuzzy disc \cite{lizzi2003fuzzy}, its spectral analysis \cite{lizzi2005beat}, and numerical applications \cite{lizzi2012noncommutative}.}.

\subsection{Weyl Quantized Fuzzy Spheres with Axial Poisson Multiplier \label{sec:fs}}

We seek a fuzzy geometric description of a spherical surface defined implicitly by:
\begin{equation}
    F(x,y,z)= x^2+y^2+z^2 - R^2 = 0 \ ,
    \label{eq:implicit_sphere}
\end{equation}
where $R$ corresponds to the radius of the sphere. Here, we focus on fuzzy Weyl quantization with a purely axial Poisson multiplier $\lambda(z)$, which is given by a finite-degree polynomial of degree $d_\lambda \equiv \text{deg}\big( \lambda(z) \big) \in \mathbb{N}_0$ such that:
\begin{equation}
     \lambda(z) = \sum^{d_\lambda}_{\alpha = 0} \lambda_\alpha z^\alpha \ .
    \label{eq:lambda_pol}
\end{equation}
We impose the positivity condition everywhere on the surface:
\begin{equation}
    \lambda(z) > 0 \ \ \forall z \in [-R,+R] \ ,
    \label{eq:lambda_cond}
\end{equation} 
so that the induced Poisson structure is nondegenerate and has a fixed orientation, in which $[-R,+R]$ is the bound range between $z=-R$ and $z=+R$.

For products and functions involving a single variable, there is no ordering ambiguity. Thus, the Casimir function Eq. \eqref{eq:implicit_sphere} in matrix form Eq. \eqref{eq:implicit_surf_mat_quant} is given by:
\begin{equation}
    \hat{X}_N^2 + \hat{Y}_N^2 + \hat{Z}_N^2 - R^2 \hat{I}_{N \times N} = 0 \ , \label{eq:implicit_sphere_mat}
\end{equation}
where $\hat{I}_{N \times N}$ is the $N \times N$ identity matrix. We consider nontrivial representations with $N\geq 2$, for which the coordinate matrices can satisfy nonvanishing commutation relations. The corresponding commutator relation in the first line of Eq. \eqref{eq:gen_Poisson_mat_quant} takes the form:
\begin{equation}
        [ \hat{X}_N, \hat{Y}_N] = i2 \hbar_N \lambda(\hat{Z}_N) \hat{Z}_N \ .
\label{eq:gen_Poisson_axial_mat_quant_X}
\end{equation}
Define
\begin{equation}
    S_{\lambda}[\hat{A}] \equiv \frac{d}{dt} \Lambda(\hat{Z}_N + t\hat{A}) \Big|_{t=0} \ ,
    \label{eq:Weyl_def}
\end{equation}
where $\Lambda(z)$ is an antiderivative of $\lambda(z)$, i.e. $\Lambda'(z)=\lambda(z)$ and $'$ denotes differentiation with respect to the function's sole argument. For a polynomial $\lambda(z)$ as in Eq. \eqref{eq:lambda_pol}, this expression becomes\footnote{Given Eq. \eqref{eq:lambda_pol}, then $\Lambda(z)= \sum^{d_\lambda}_{\alpha = 0} \frac{\lambda_\alpha}{\alpha+1} z^{\alpha+1} $; Eq. \eqref{eq:def_S} follows from differentiating each factor in turn:
$$\frac{d}{dt}(\hat Z_N+t\hat A)^{\alpha+1}\Big|_{t=0}=\sum_{k=0}^{\alpha}(\hat Z_N+t\hat A)^k\hat A(\hat Z_N+t\hat A)^{\alpha-k}\Big|_{t=0}=\sum_{k=0}^{\alpha}\hat Z_N^k\hat A\hat Z_N^{\alpha-k} \ .$$}:
\begin{equation}
    S_{\lambda}[\hat{A}] = \sum^{d_\lambda}_{\alpha=0}  \frac{\lambda_\alpha}{\alpha+1} \sum^\alpha_{\beta=0} \hat{Z}_N^\beta \hat{A}_N \hat{Z}^{\alpha-\beta}_N \ ,
    \label{eq:def_S}
\end{equation}
which allowing us to express Weyl symmetric-ordering terms explicitly:
\begin{equation}
\begin{split}
    \mathcal{W}\left[ \lambda x \right](\hat{X}_N,\hat{Y}_N,\hat{Z}_N) = S_{\lambda}[\hat{X}_N]
    \ \ \text{and} \ \ 
    \mathcal{W}\left[ \lambda y \right](\hat{X}_N,\hat{Y}_N,\hat{Z}_N) = S_{\lambda}[\hat{Y}_N] \ .
\end{split} \label{eq:Weyl_explicit}
\end{equation} 
Then, Eq. \eqref{eq:Weyl_explicit} can be used to rewrite the second- and third-lines of Eq. \eqref{eq:gen_Poisson_mat_quant}:
\begin{equation}
    [ \hat{Y}_N,\hat{Z}_N] = i 2 \hbar_N S_{\lambda}[\hat{X}_N] \ \ \text{and} \ \
        [ \hat{Z}_N,\hat{X}_N ] = i 2 \hbar_N S_{\lambda}[\hat{Y}_N] \ .
\label{eq:gen_Poisson_axial_mat_quant_YZ}
\end{equation}
For the quantization to be possible, there must exist an irreducible triplet of Hermitian matrices $(\hat{X}_N,\hat{Y}_N,\hat{Z}_N )$ and a quantization parameters $\hbar_N$ that simultaneously satisfy Eq. \eqref{eq:implicit_sphere_mat}, Eq. \eqref{eq:gen_Poisson_axial_mat_quant_X}, and Eq. \eqref{eq:gen_Poisson_axial_mat_quant_YZ}. The axial Poisson multiplier $\lambda(z)$ may be given in advance or should be determined as part of the solution, subject to appropriate constraints.

It is important to note that we have used Fréchet derivative representation \cite{higham2008functions} to define Weyl ordering in Eq. \eqref{eq:Weyl_def}, thereby allowing its extension to nonpolynomial functions. Although such an extension is not unique in general, our prescription agrees with the standard and uniquely defined Weyl ordering formula for polynomials \cite{weyl1927quantenmechanik,cheng2000gauge}. Furthermore, as we later demonstrate in Section \ref{sec:beyond}, this simplifies the quantization with nonpolynomial $\lambda(z)$ in the specific cases of interest there.

\section{Numerical Approach \& Finding \label{sec:num}}

Traditionally, the identification of the matrices $(\hat{X}_N,\hat{Y}_N,\hat{Z}_N)$ has been obtained analytically for the trivial Poisson multiplier $\lambda(x,y,z)=1$, which corresponds to a representation of the $\mathrm{SU}(2)$ Lie algebra \cite{madore1992fuzzy,arnlind2009fuzzy}. By numerically solving the matrix equations, we can consider more general Poisson multipliers $\lambda(x,y,z)$, exploring quantizations beyond conventional Lie algebra representations. Here, we consider when this function is purely axial, i.e. $\lambda(z)$, and can be expressed in polynomial form as in Eq. \eqref{eq:lambda_pol}.

As a sanity check, we first reproduce the known results for fuzzy spheres with a trivial Poisson multiplier, corresponding to $d_\lambda=0$ and $\lambda_0=1$, in Section \ref{sec:lambda_is_const}. To simplify our numerical search, we set $\hat{Z}_N$ to be diagonal, which is always possible via a unitary transformation\footnote{Under a unitary transformation $\hat{U}$, we simultaneously transform all three matrices as
$$\hat{X}_N \rightarrow \hat{U}^\dagger \hat{X}_N \hat{U} \ , \ 
\hat{Y}_N \rightarrow \hat{U}^\dagger \hat{Y}_N \hat{U} \ , \ \hat{Z}_N \rightarrow \hat{U}^\dagger \hat{Z}_N \hat{U} \ , \ $$
without changing their algebra, i.e. still satisfies Eq. \eqref{eq:implicit_sphere_mat}, Eq. \eqref{eq:gen_Poisson_axial_mat_quant_X}, and Eq. \eqref{eq:gen_Poisson_axial_mat_quant_YZ}.} (thus fixing a gauge freedom). The numerically found quantization parameter $\hbar_N$ agrees with the known analytical result \cite{madore1992fuzzy}, i.e.
\begin{equation}
    \hbar_N=\frac{R}{\lambda_0 \sqrt{N^2-1}} \ .
    \label{eq:usual_fuzzy_sphere}
\end{equation}
We then extend our numerical approach in Section \ref{sec:lambda_is_poly} to nonconstant polynomial Poisson multipliers of degree $d_\lambda \geq 1$, a regime that has received less attention in the literature and for which few analytical results are available.

\begin{figure*}[thbp!]
    \centering
\includegraphics[width=\linewidth]{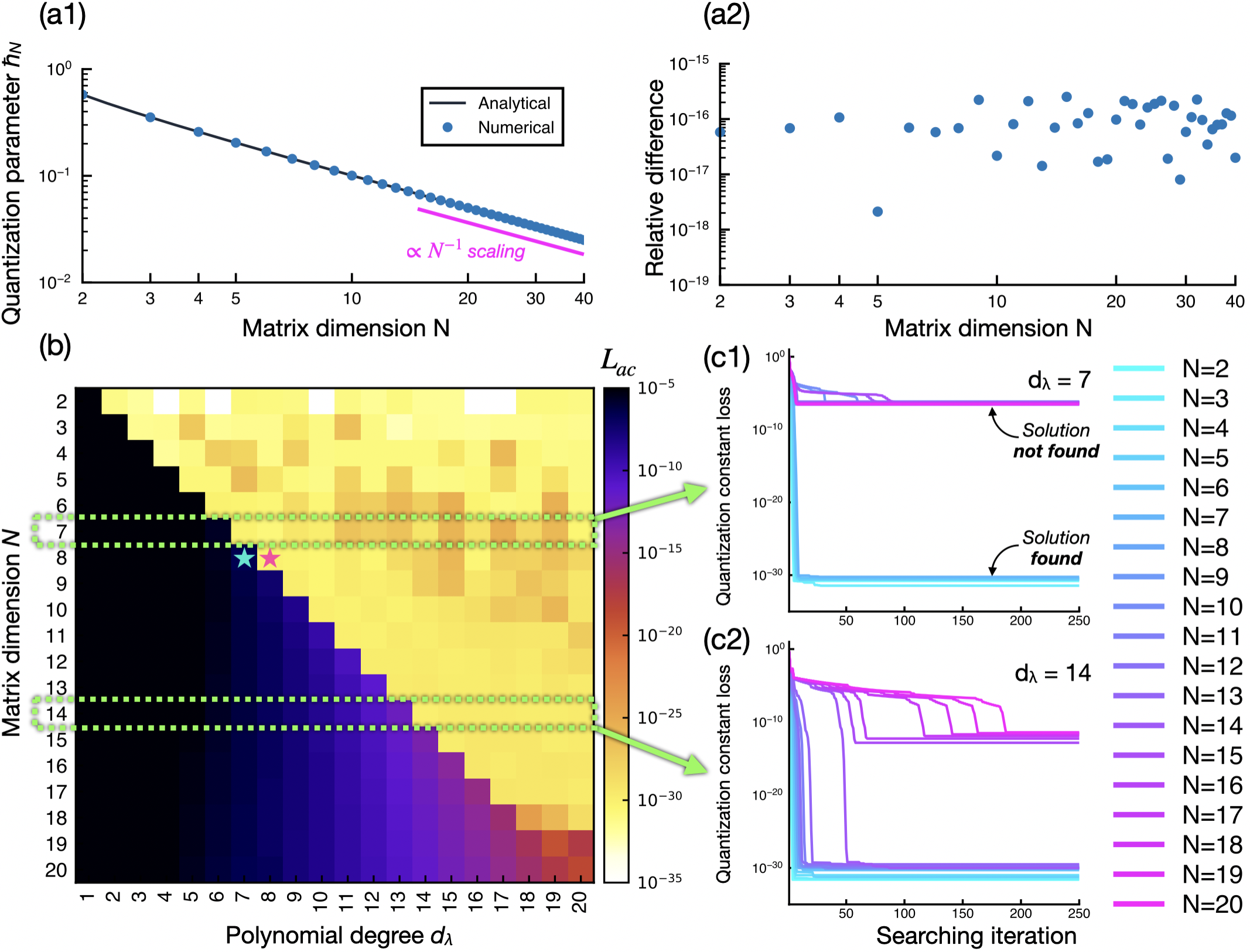}
    \caption{\textbf{Fuzzy sphere searching results.} We set $R=1$. \textbf{(a1)} Numerically obtained quantization parameters $\hbar_N$ compared with the analytical result in Eq. \eqref{eq:usual_fuzzy_sphere} for the constant Poisson multiplier i.e. $\lambda(z)=1$. \textbf{(a2)} Relative difference between the numerical and analytical values. \textbf{(b)} Final algebraic-constraint loss $L_{\text{ac}}$ for nonconstant polynomial Poisson multipliers, across matrix dimensions $N$ and polynomial degrees $d_\lambda$. We examine the starred grids (cyan and magenta) in more detail in Fig. \ref{fig02}b. \textbf{(c1–2)} Total loss histories for $d_\lambda=7$ and $d_\lambda=14$, with colors indicating $N$. The histories show a clear separation between searches that reach small residuals and those that do not.}
\label{fig01}
\end{figure*}

\subsection{Fuzzy Sphere Quantization with Constant Poisson Multiplier, i.e. $d_\lambda = 0$ \label{sec:lambda_is_const}}

We set the sphere's radius to be $R=1$ and the Poisson multiplier to be $\lambda(z)=1$, corresponding to polynomial degree $d_\lambda=0$ with coefficient $\lambda_0 = 1$. Following the standard machine learning approach of loss-based optimization \cite{heaton2018ian}, we formulate the search for quantization representations as a numerical optimization problem. Specifically, we simultaneously optimize the elements of the Hermitian matrix triplet $(\hat{X}_N, \hat{Y}_N,\hat{Z}_N)$ and the quantization parameter $\hbar_N$ to minimize the total loss:
\begin{equation}
    L_{\text{tot}} = c_{\text{ac}} L_{\text{ac}} + c_{\text{irr}}  L_{\text{irr}} \  , 
    \label{eq:L_tot}
\end{equation}
where $L_{\text{ac}}$ and $L_{\text{irr}}$ penalize violations of the following conditions:
\begin{itemize}
    \item \underline{Algebraic constraint:} The loss contribution $L_{\text{ac}}$ enforces the algebraic relations defining the quantization, i.e. Eq. \eqref{eq:implicit_sphere_mat}, Eq. \eqref{eq:gen_Poisson_axial_mat_quant_X}, and Eq. \eqref{eq:gen_Poisson_axial_mat_quant_YZ}. It is defined as follows:
    \begin{equation}
        L_{\text{ac}} = \| \hat{A}_C \|^2 + \| \hat{A}_{XY} \|^2 + \| \hat{A}_{YZ} \|^2 + \| \hat{A}_{ZX} \|^2 \ ,
        \label{eq:L_ac}
    \end{equation}
where $\| ... \|$ is the Hilbert-Schmidt norm \cite{horn1990norms}. The $\hat{A}$-matrices are given by:
$$\hat{A}_C = \hat{X}_N^2 + \hat{Y}_N^2 + \hat{Z}_N^2 - \hat{I}_{N \times N} 
$$
and
\begin{equation}
    \hat{A}_{XY} = \frac{[ \hat{X}_N, \hat{Y}_N]}{i2\hbar_N} - \hat{Z}_N \ , \
        \hat{A}_{YZ} = \frac{[ \hat{Y}_N, \hat{Z}_N]}{i2\hbar_N} - \hat{X}_N \ ,
         \ \hat{A}_{ZX} = \frac{[ \hat{Z}_N, \hat{X}_N]}{i2\hbar_N} - \hat{Y}_N \ .
    \label{eq:A_comm}
\end{equation}
Note that the factor $1/\hbar_N$ for our choices of $\hat{A}_{XY}$, $\hat{A}_{YZ}$, $\hat{A}_{ZX}$ is of critical importance for preventing trivial solutions in which the optimization drives $\hbar_N \rightarrow 0$, causing the commutators between the triplet $(\hat{X}_N,\hat{Y}_N,\hat{Z}_N)$ to vanish and thereby trivially satisfying the commutation constraints.

\item  \underline{Irreducibility:} The loss contribution $L_{\text{irr}}$ should punish proximity of the triplet $(\hat{X}_N,\hat{Y}_N,\hat{Z}_N)$ to
reducibility\footnote{Reducibility means that a common unitary transformation could bring all three matrices into the same nontrivial block-diagonal form.}. To create such quantity, we use the generalized Gell-Mann matrices $\{\hat{T}_a\}^{N^2-1}_{a=1}$, i.e. an orthonormal basis of traceless Hermitian matrices from qudit theory \cite{bertlmann2007bloch}: 
$$\hat{T}_a^\dagger=\hat{T}_a \ , \ 
    \text{Tr}(\hat{T}_a)=0 \ , \ \text{Tr}(\hat{T}_a\hat{T}_b)=\delta_{ab} \ ,
$$
where $\delta_{ab}$ is a Kronecker delta \cite{edwards1987appreciation}. Define the matrix $\hat{\Gamma}$ whose elements are given by:
\begin{equation}
    \Gamma_{ab} = \sum_{\hat{R}_N \in \{ \hat{X}_N,\hat{Y}_N,\hat{Z}_N\} } \text{Tr} \left( [\hat{R}_N,\hat{T}_a]^\dagger [\hat{R}_N,\hat{T}_b]\right) \ ,
    \label{eq:Gamma_irr}
\end{equation}
then the triplet $(\hat{X}_N,\hat{Y}_N,\hat{Z}_N)$ is irreducible if and only if $\hat{\Gamma}$ is positive definite (see Appendix \ref{app:reducibility_test}). We choose
\begin{equation}
    L_{\text{irr}}
    = \text{Tr}(\hat{\Gamma}^{-1}) \ ,
    \label{eq:L_irr}
\end{equation}
which diverges as the smallest eigenvalue approaches $0$, thereby discouraging the optimization from approaching reducible configurations.
\end{itemize}
The weights $c_{\text{ac}}$ and $c_{\text{irr}}$ are hyperparameters that control the relative contributions of the two loss terms.

 Our numerical search proceeds sequentially from matrix dimension $N=2$ to $N=40$. We choose the loss weights to be $c_{\text{ac}}=1$ and $c_{\text{irr}}=10^{-18}$ -- we choose such a small value because the aim is to only gently steer the matrix triplet away from reducible representations, rather than dominate the overall optimization problem\footnote{A similar strategy has been used in structure-preserving neural networks, where a very weak regularization term can prevent learned transformations from collapsing to trivial solutions without substantially interfering with the primary optimization objective \cite{bertalan2019learning}.}. 

At the first matrix dimension, i.e. $N=2$, we initialize the quantization parameter at $\hbar^{\text{(ini)}}_2=1$. We generate two independent complex matrices, drawing all real and imaginary components independently from a standard Gaussian distribution, and symmetrize them to obtain Hermitian matrices $\hat{X}^{\text{(ini)}}_2$ and $\hat{Y}^{\text{(ini)}}_2$. We initialize $\hat{Z}^{\text{(ini)}}_2$ as a diagonal matrix with entries drawn independently from the same Gaussian distribution. The three matrices are then rescaled by a common factor so that
$$\text{Tr} \left[
(\hat{X}^{\text{(ini)}}_2)^2+
(\hat{Y}^{\text{(ini)}}_2)^2 +
(\hat{Z}^{\text{(ini)}}_2)^2
\right] = 1 \ . $$
For each subsequent dimension $N$, we initialize $\hbar^{\text{(ini)}}_N$ using the final value $\hbar^{\text{(fin)}}_{N-1}$ from the preceding optimization. We extend the learned matrices according to
$$\hat{X}_N^{\text{(ini)}}
=\begin{pmatrix}
\hat{X}_{N-1}^{\text{(fin)}}  & \vec{v}_X \\ \vec{v}_X^\dagger & 0
\end{pmatrix} \ \ \text{and} \ \
\hat{Y}_N^{\text{(ini)}}
=\begin{pmatrix}
\hat{Y}_{N-1}^{\text{(fin)}}  &  \vec{v}_Y \\
 \vec{v}_Y^\dagger&0
\end{pmatrix} \ ,
$$
where $\vec{v}_X$ and $\vec{v}_Y$ are independent complex Gaussian vectors, each normalized to Euclidean norm $N^{-1/2}$, thereby reducing the disturbance to the algebraic relations already satisfied by the inherited matrix blocks as the dimension increases. The diagonal matrix is extended as
$$ \hat{Z}_N^{\text{(ini)}}
= \begin{pmatrix}
\hat{Z}_{N-1}^{\text{(fin)}} & \vec{0} \\
\vec{0} & 1
\end{pmatrix} \ .
$$
We use 250 training steps per dimension, including the initial evaluation, with a fixed random seed for reproducibility. We minimize the total loss using a  Levenberg–Marquardt algorithm \cite{levenberg1944method,marquardt1963algorithm}. Training steps are accepted only when they decrease the total loss and pass the irreducibility checks.

We are able reproduce the conventional fuzzy sphere quantization \cite{madore1992fuzzy}, with the numerically obtained quantization parameter $\hbar_N$ agreeing with the known analytical result Eq. \eqref{eq:usual_fuzzy_sphere} to high precision (see Fig. \ref{fig01}a-b). Building on this success, we apply our numerical approach to unexplored territory in Section \ref{sec:lambda_is_poly}, investigating cases in which the Poisson multiplier $\lambda(z)$ is a polynomial of arbitrarily high degree $d_\lambda \geq 1$.

\subsection{Fuzzy Sphere Quantization with Polynomial Poisson Multiplier, i.e. $d_\lambda \geq 1$ \label{sec:lambda_is_poly}}

We further investigate fuzzy sphere quantizations, similar to what has been done Section \ref{sec:lambda_is_const}, but for polynomial Poisson multipliers that are trainable and across various matrix dimension $N$ and polynomial degree $d_\lambda$. We are curious to see what lies beyond conventional fuzzy sphere quantization \cite{madore1992fuzzy}. For simplicity, we fix the coefficient corresponding to the $0$th-degree and the highest-degree to be
$$\lambda_0 = \lambda_{d_\alpha} = 1 $$ 
while setting the rest $\{ \lambda_\alpha\}_{\alpha=1}^{d_\lambda-1}$ to be trainable parameters. 

For this numerical search, we extend to the total loss function in Eq. \eqref{eq:L_tot} by introduce an additional loss contribution $L_{\text{pos}}$, i.e. 
$$ L_{\text{pos}} = \frac1M \sum_{m=1}^{M} \text{ReLu} \big(-\lambda(z_m) \big) \ , $$
where $\text{ReLu}(z)\equiv \max(0,z)$ is the rectified linear unit function \cite{ramachandran2017searching} and $\{z_m\}^M_{m=1}$ are chosen positions in the range $z \in [-R,+R]$, to ensure  the Poisson multiplier remains positive throughout that  range. The total loss function now becomes:
\begin{equation}
    L_{\text{tot}} = c_{\text{ac}} L_{\text{ac}} + c_{\text{irr}}  L_{\text{irr}} + c_{\text{pos}}  L_{\text{pos}}  \  , 
    \label{eq:L_tot_more}
\end{equation}
in which the loss contribution $L_{\text{ac}}$ is given by Eq. \eqref{eq:L_ac} but with the residual matrices in Eq. \eqref{eq:A_comm} replace by:
\begin{equation}
    \hat{A}_{XY} = \frac{[ \hat{X}_N, \hat{Y}_N]}{i2\hbar_N} - \lambda(\hat{Z}_N)\hat{Z}_N \ , \
        \hat{A}_{YZ} = \frac{[ \hat{Y}_N, \hat{Z}_N]}{i2\hbar_N} - S_\lambda [\hat{X}_N ] \ ,
         \ \hat{A}_{ZX} = \frac{[ \hat{Z}_N, \hat{X}_N]}{i2\hbar_N} - S_\lambda[ \hat{Y}_N ] \ ,
         \label{eq:comm_cont_gen}
\end{equation}
and the loss contribution  $L_{\text{irr}}$ stays identical to Eq. \eqref{eq:L_irr}.

For our numerical search, for each matrix dimension from $N=2$ to $N=20$, we perform separate optimizations for polynomial degrees from $d_\lambda=1$ to $d_\lambda=10$. Unlike the optimizations in Section \ref{sec:lambda_is_const}, here the optimizations are initialized independently, without using the final results of preceding runs; the minimization algorithm, on the other hand, remains unchanged. Each optimization runs for $250$ training steps and is initialized using the known solutions for the Hermitian matrix triplet $(\hat{X}_N,\hat{Y}_N,\hat{Z}_N)$ and quantization parameter $\hbar_N$ of the conventional fuzzy sphere \cite{madore1992fuzzy}. All trainable polynomial coefficients are initially set to $0$, i.e. $$\lambda^{\text{(ini)}}(z)=1+z^{d_\lambda} \ . $$
We set the loss weights to $c_{\text{ac}}=1$, $c_{\text{irr}}=10^{-18}$, and $c_{\text{pos}}=1$. The positions $\{z_m\}^M_{m=1}$ are chosen as $M=100$ uniformly spaced points from $-R$ to $+R$.

We then compare the performance of our numerical approach across different irreducible matrix dimensions $N$ and polynomial degrees $d_\lambda$ (see Fig. \ref{fig01}b). Although no universal loss threshold can rigorously establish the existence of an exact solution, the loss histories exhibit a clear separation, often spanning tens of orders of magnitude (see Fig. \ref{fig01}c1–2). This separation allows us to identify pairs $(N,d_\lambda)$ for which our numerical search yields highly accurate solutions, suggesting the possible existence of exact solutions. Given our numerical findings, we observe a sharp boundary in solvability: the loss decreases dramatically for $N\leq d_\lambda$, whereas it remains substantially higher for $N>d_\lambda$. This sharp transition in numerical solvability suggests an underlying no-go theorem, motivating the theoretical analysis that follows, i.e. in Section \ref{sec:theory}. It is important to note that the transition becomes less pronounced at larger matrix dimensions: for $N\gtrsim 19$, $L_{\text{ac}}$ and $L_{\text{irr}}$ become comparable (about $\sim 10^{-18}$), obscuring the separation between successful and unsuccessful searches in the loss histories. This limitation motivates the development of better optimization strategy in future work.

\section{Theoretical Explanation \label{sec:theory}}

We now explain the sharp dimensionality bound observed in Section \ref{sec:lambda_is_poly} that the polynomial degree $d_\lambda$ of the Poisson multiplier imposes a strict upper bound on the irreducible matrix dimension $N$, i.e. Weyl-quantization solutions possible only for $N\leq d_\lambda$. Furthermore, along the way we also discover that the same spectral constraints also yield a lower bound on the quantization parameter $\hbar_N$, which in this Section we establish before deriving the dimensionality bound. Our theoretical analysis  proceeds as follows:
\begin{itemize}
    \item Define $\hat{Z}_N$-spectral direct connectedness from the non-zero elements of $\hat{W}_N \equiv \hat{X}_N + i \hat{Y}_N$ (see Section \ref{sec:spec_dir_conn}).
    \item Show that every direct connection requires $H(z) \equiv \big( z \Lambda(z) \big)'''$ to change sign between the connected eigenvalues (see Section \ref{sec:spec_constraint}).
    \item Prove the dimensionality bound by showing that, under spectral nondegeneracy, consecutive eigenvalues must be connected, so that the degree of $H(z)$, itself bounded by the degree of $\lambda(z)$, bounds the matrix dimension (see Section \ref{sec:bound_nondeg}).
\end{itemize}

\subsection{Spectral Directed Connections \label{sec:spec_dir_conn}}

We denote $\{ v_N^{(j)}\}^{N-1}_{j=0}$ as the set of eigenvalues (all real) and $\{ \lvert j_N\rangle \}^{N-1}_{j=0}$ as the set eigenvectors (all normalized), respectively, of the Hermitian matrix $\hat{Z}_N$. By the spectral theorem \cite{horn2012matrix}, these eigenvectors can be chosen to form an orthonormal basis of the Hilbert space $\mathbb{C}^N$, and hence satisfy the completeness relation:
\begin{equation}
\sum_{j=0}^{N-1}\lvert j_N\rangle\langle j_N\rvert
=\hat I_{N\times N} \ \ \text{and} \ \ \langle j_N \lvert k_N \rangle = 0 \ \forall j \neq k \ .
\label{eq:completeness}
\end{equation}
In Appendix \ref{app:eigval_in}, the Casimir function Eq. \eqref{eq:implicit_sphere_mat} allows us to prove that all eigenvalues $\{ v_N^{(j)}\}^{N-1}_{j=0}$ must be within the interval $[-R,+R]$:
\begin{equation}
    -R \leq v_N^{(0)} \leq v_N^{(1)} \leq ... \leq v_N^{(N-1)} \leq +R \ .
    \label{eq:eigval_ordering}
\end{equation}
as we have labeled them in increasing order. 

Define a non-Hermitian matrix 
\begin{equation}
    \hat{W}_N \equiv \hat{X}_N + i \hat{Y}_N \ ,
    \label{eq:W}
\end{equation}
whose construction is analogous to the angular momentum raising operator \cite{griffiths2018introduction} and is standard in the representation theory underlying the fuzzy sphere \cite{madore1992fuzzy,arnlind2009fuzzy}. Its Hermitian adjoint $\hat{W}_N^\dagger$ plays the corresponding lowering role. These matrices play a crucial role in analyzing the spectral properties of $\hat{Z}_N$. We show in Appendix \ref{app:W_alg} that the following relations hold:
\begin{itemize}
    \item The matrices $\hat{W}_N \hat{W}_N^\dagger$ and $\hat{W}_N^\dagger \hat{W}_N$ can be expressed as
    \begin{equation}
    \begin{split}
        \hat{W}_N \hat{W}_N^\dagger &= R^2 \hat{I}_{N \times N} - \hat{Z}^2_N + 2 \hbar_N \lambda(\hat{Z}_N ) \hat{Z}_N \ , 
        \\
        \hat{W}_N^\dagger \hat{W}_N &= R^2 \hat{I}_{N \times N} - \hat{Z}^2_N - 2 \hbar_N \lambda(\hat{Z}_N ) \hat{Z}_N \ . 
        \label{eq:WW}
    \end{split}
    \end{equation}
    \item Combining the two commutator relations in Eq. \eqref{eq:gen_Poisson_axial_mat_quant_YZ} gives:
\begin{equation}
    [\hat{Z}_N,\hat{W}_N] = 2 \hbar_N S_{\lambda}[\hat{W}_N] \ .
    \label{eq:get_comm_ZW}
\end{equation}
\end{itemize}
We can sandwich both sides of Eq. \eqref{eq:get_comm_ZW} between the bra $\langle j_N\lvert$ and ket $\lvert k_N\rangle$, yielding
\begin{equation}
    \left(v_N^{(j)} - v_N^{(k)} \right) W^{(j,k)}_N = 2\hbar_N \bar{\lambda}(v_N^{(j)},v_N^{(k)})  W_N^{(j,k)} \ ,
    \label{eq:int}
\end{equation}
where 
\begin{equation}
    W_N^{(j,k)} \equiv \langle j_N\lvert \hat{W}_N\lvert k_N\rangle
    \label{eq:W_elements}
\end{equation} 
and $\bar{\lambda}(v_N^{(j)},v_N^{(k)})$ is the average value of $\lambda(z)$ along the $z$-axis between locations $z=v_N^{(j)}$ and $z=v_N^{(k)}$. We provide the derivation for this expression in Appendix \ref{app:get_int}. Given the condition on $\lambda(z)$ Eq.\eqref{eq:lambda_cond} and the positions of the eigenvalues $\{ v_N^{(j)}\}^{N-1}_{j=0}$ Eq. \eqref{eq:eigval_ordering}, we always have $\bar{\lambda}(v_N^{(j)},v_N^{(k)}) > 0$. Following from this, Eq. \eqref{eq:int} implies that 
\begin{equation}
    W^{(j,k)}_N =0 \ \ \forall v_N^{(j)} \leq v_N^{(k)} \ ,
\label{eq:no_back}
\end{equation}
since if it is not the case then the left-hand side (LHS) is a negative multiple of $W^{(j,k)}_N$ while the right-hand side (RHS) is a positive multiple of $W^{(j,k)}_N$ , leading to a contradiction.

For a pair of eigenvalues $v^{(j)}_N $ and $v^{(k)}_N$, we say that $v^{(k)}_N$ is directly connected to $v^{(j)}_N$, i.e. $v^{(k)}_N \dashrightarrow v^{(j)}_N$, if the matrix element $W^{(j,k)}_N$ between them is non-zero, i.e.
$W^{(j,k)}_N \neq 0$. By Eq. \eqref{eq:no_back}, such a connection can only go from a lower-eigenvalue to a higher-eigenvalue.

\subsection{Spectral Constraints from Poisson Multiplier \label{sec:spec_constraint}}

\begin{figure*}[thbp!]
    \centering
\includegraphics[width=\linewidth]{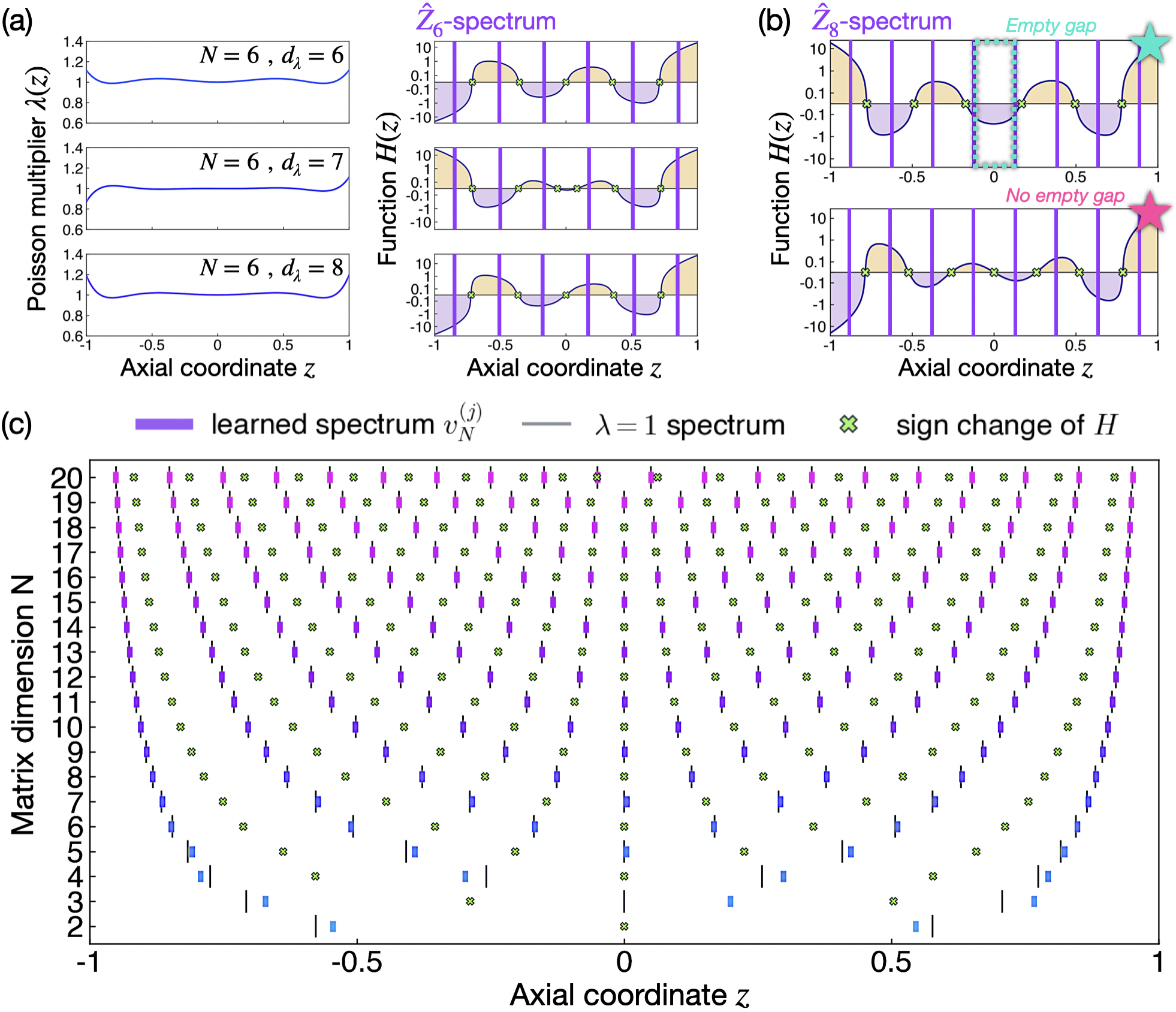}
    \caption{\textbf{Fuzzy sphere axial spectra.} Here we show the axial spectra and the sign-change condition for polynomial fuzzy sphere quantizations. \textbf{(a)} Learned Poisson multipliers $\lambda(z)$ and the corresponding functions $H(z)$, defined in Eq. \eqref{eq:def_H}, for $N=6$ and $d_\lambda\in{6,7,8}$. \textbf{(b)} Comparison of the unsuccessful top-search with $(N,d_\lambda)=(8,7)$ and the successful bottom-search with $(N,d_\lambda)=(8,8)$, corresponding to the starred grids (cyan and magenta, respectively) in Fig. \ref{fig01}b. The highlighted interval in the upper panel contains no sign change of $H(z)$. In \textbf{(a–b)}, vertical purple lines mark axial eigenvalues, green crosses mark sign changes of $H(z)$, and $H(z)$ is plotted on a symmetric logarithmic scale. \textbf{(c)} Learned axial spectra for $N=d_\lambda$, compared with the conventional $\lambda(z)=1$ spectra (gray ticks); green crosses mark sign changes of $H(z)$. Note that the found $(N,d_\lambda)=(20,20)$ solution is likely a failure, as indicated by Fig. \ref{fig01}b, where the sharp transition between successful and unsuccessful searches becomes blurry.}
\label{fig02}
\end{figure*}

From the associative property of matrix multiplication:
\begin{equation}
\begin{split}
(\hat{W}_N \hat{W}^\dagger_N )\hat{W}_N &=  \hat{W}_N (\hat{W}^\dagger_N \hat{W}_N )
\\
\Longrightarrow \ \left( R^2 \hat{I}_{N \times N} - \hat{Z}^2_N + 2 \hbar \lambda(\hat{Z}_N ) \hat{Z}_N \right) \hat{W}_N &= \hat{W}_N \left( R^2 \hat{I}_{N \times N} - \hat{Z}^2_N - 2 \hbar \lambda(\hat{Z}_N ) \hat{Z}_N \right)  \ , 
\end{split}
\label{eq:WWW}
\end{equation}
in which the second-line follows from Eq. \eqref{eq:WW}. Sandwiching both sides between the bra $\langle j_N \lvert$ and ket $\lvert k_N\rangle$, we can then find another relationship beside Eq. \eqref{eq:int} between the Poisson multiplier $\lambda(z)$ with the quantization parameter $\hbar_N$ and the spectral $\{ v_N^{(j)}\}^{N-1}_{j=0}$ of the Hermitian matrix $\hat{Z}_N$ at any matrix dimensionality $N$:
\begin{equation}
   \left( R^2 - (v^{(j)}_N)^2 + 2 \hbar_N v^{(j)}_N \lambda(v^{(j)}_N) \right) W_N^{(j,k)} = \left( R^2 - (v^{(k)}_N)^2 - 2 \hbar_N v^{(k)}_N \lambda(v^{(k)}_N) \right)  W_N^{(j,k)} \ .
   \label{eq:another}
\end{equation}
For any direct connection $v_N^{(k)} \dashrightarrow v_N^{(j)}$, i.e. a non-zero $W^{(j,k)}$, Eq. \eqref{eq:another} gives:
\begin{equation*}
    (v_N^{(j)})^2 - (v_N^{(k)})^2 = 2 \hbar_N \left( v_N^{(j)} \lambda(v_N^{(j)}) + v_N^{(k)} \lambda(v_N^{(k)}) \right) \ .
\end{equation*}
Combining this with Eq. \eqref{eq:int} to eliminates $\hbar_N$, we get a functional equation for $\lambda(z)$:
\begin{equation}
    \left( v_N^{(j)} \lambda(v_N^{(j)}) + v_N^{(k)} \lambda(v_N^{(k)}) \right) - \left(v_N^{(j)} + v_N^{(k)} \right) \bar{\lambda}(v_N^{(j)},v_N^{(k)}) = 0 \ . 
    \label{eq:functional_eq}
\end{equation}
An important insight in our theoretical analysis is that this functional equation can be recasted into an integral constraint involving derivatives of the Poisson multiplier. Define
\begin{equation}
    H(z) \equiv \big( z\Lambda(z) \big)''' = z \lambda''(z) + 3 \lambda'(z) \ , 
    \label{eq:def_H}
\end{equation}
then the LHS of Eq. \eqref{eq:functional_eq} can be rewritten -- by using an integral identity for the trapezoid rule \cite{dragomir2012some} (see Appendix \ref{app:int_eq}) -- as:
$$ \frac{1}{v^{(j)}_N - v^{(k)}_N} \int^{v_N^{(j)}}_{v_N^{(k)}} dz \ (v_N^{(j)} -z)(z-v_N^{(k)})H(z) \ , $$
thus we arrive at an integral constraint between the function $H(z)$ and the spectral of $\hat{Z}_N$:
\begin{equation}
     \int^{v_N^{(j)}}_{v_N^{(k)}} dz \ (v_N^{(j)} -z)(z-v_N^{(k)})H(z) = 0 \ .
    \label{eq:int_eq}
\end{equation} 
Since $(v_N^{(j)}-z)(z-v_N^{(k)})$ is strictly positive on the open interval $z\in(v_N^{(k)},v_N^{(j)})$, this integration constraint implies two possibilities:
\begin{itemize}
    \item \underline{Constant $H(z)$:} Then, $H(z)=0$ everywhere.
    By \eqref{eq:def_H}, for polynomial $\lambda(z)$, this occurs if and only if $\lambda(z)$ is constant, i.e. $\lambda(z)=\lambda_0$. In this case, the conventional fuzzy sphere admits irreducible solutions for every $N\geq2$ \cite{madore1992fuzzy}. The eigenvalues of $\hat{Z}_N$ are equally spaced between
    \begin{equation}
        -R\sqrt{\frac{N-1}{N+1}}
        \ \text{and} \
        +R\sqrt{\frac{N-1}{N+1}} \ .
        \label{eq:usual_fs_spacing}
    \end{equation}

    \item \underline{Varying $H(z)$:}
    Every nonzero matrix element $W_N^{(j,k)}$ requires $H(z)$ to change sign between the corresponding eigenvalues, i.e. $v^{(k)}_N$ and $v^{(j)}_N$. This sign-change condition is illustrated in Fig. \ref{fig02}a, which shows the learned axial spectra and the corresponding functions $H(z)$. Define
    \begin{equation}
        \lambda_{\max} \equiv \max_{z \in [-R,+R]} \lambda(z) 
    \end{equation}
    and let $n_H$ denote the number of zeros for $z \in (-R,R)$ at which $H(z)$ changes sign, then by Eq. \eqref{eq:def_H}: \begin{equation}
    n_H \leq d_\lambda-1 \ .
    \label{eq:nH_bound}
\end{equation}
From this, in Appendix \ref{app:quant_param_bound} we establish a lower-bound for the quantization parameter: 
\begin{equation}
    \hbar_N
    \geq\frac{R}{\lambda_{\max} \sqrt{n_H(n_H+2)}}
    \geq\frac{R}{\lambda_{\max} \sqrt{d_\lambda^2-1}} \ ,
    \label{eq:hbar_oscillation_lower_bound}
\end{equation}
thus for finite-$\hbar_N$ solution of non-constant $H(z)$ to exists we must have $n_H > 0$.
\end{itemize} 

\subsection{Simplification under Spectral Non-Degeneracy \label{sec:bound_nondeg}}

To gain analytical insight into our numerical observation in Section \ref{sec:num}, we introduce an additional assumption that 
\begin{equation}
    \text{the spectral of} \ \ \hat{Z}_N \ \ \text{is non-degenerate} \ ,
    \label{eq:our_simp}
\end{equation}
so that the eigenvalues $\{ v^{(j)}_N\}^{N-1}_{j=0}$ are strictly ordered and Eq. \eqref{eq:eigval_ordering} becomes:
\begin{equation} -R \leq v_N^{(0)} < v_N^{(1)} < ... < v_N^{(N-1)} \leq +R \ . \label{eq:eigval_ordering_new} \end{equation}
As we show in Appendix \ref{app:cons}, following this, it can be proven that
\begin{equation}
    \text{every pair of consecutive eigenvalues must be directly connected} \ .
    \label{eq:cons_conn}
\end{equation}
 The $N$ ordered eigenvalues define $N-1$ disjoint intervals, each containing at least one sign change event of $H(z)$. Therefore:
\begin{equation}
    N-1 \leq n_H \leq d_\lambda-1 \ \ \text{hence} \  \ N \leq n_H + 1 \leq d_\lambda \ .
    \label{eq:comparison}
\end{equation}
Figure \ref{fig02}b illustrates this restriction for $N=8$: the unsuccessful search with $d_\lambda=7$ leaves an interval between consecutive eigenvalues containing no sign change of $H(z)$, whereas the successful search with $d_\lambda=8$ satisfies this condition in every interval. Thus, the degree of the Poisson multiplier imposes an upper bound on the dimension of the irreducible matrix representation of the fuzzy sphere i.e. $N \leq d_\lambda$, which explains what we have observed during our numerical exploration in Section \ref{sec:num}. 

Figure \ref{fig02}c shows the learned axial spectra for $N=d_\lambda$, together with the sign-change locations of $H(z)$ and the conventional $\lambda=1$ spectra for comparison. We observe that, as the degree $d_\lambda$ increases, the learned Poisson multiplier $\lambda(z)$ becomes increasingly flat, approaching a constant profile. Correspondingly, the axial spectra gradually converge toward those of the conventional fuzzy sphere. In contrast, substantial deviations from the conventional spherical spectra remain evident at low $d_\lambda$, clearly indicating the influence of nontrivial Poisson structures.

What if we do not impose nondegeneracy on axial spectrum? Preliminary results suggest that the no-go bound $N\leq d_\lambda$ breaks down for sufficiently large $N$ and $d_\lambda$. A thorough investigation of this regime lies beyond the scope of the present work. We leave establishing lower bounds on the dimensions of such solutions and developing a systematic classification of them for future works\footnote{An explicit construction exists for $(N,d_\lambda)=(35,34)$, although our numerical results suggest that solutions may exist for smaller values.}.

\section{Beyond Polynomial Poisson Multiplier \label{sec:beyond}}

Many directions remain to be explored; the examples we considered here merely scratch the surface of a much deeper subject. What happens if we go beyond Weyl quantization? What if we allow a more general Poisson multiplier $\lambda(x,y,z)$? What if we relax the polynomiality assumption on $\lambda(z)$? In this Section, we investigate the last question by considering a nonpolynomial solution to constant $H(z)=0$ in Eq. \eqref{eq:def_H}, i.e.
\begin{equation}
    \lambda(z) =\lambda_0+\lambda_{-2}z^{-2} \ ,
    \label{eq:lambda_sol}
\end{equation}
which is singular at $z=0$ but nevertheless always satisfies the positivity condition in Eq. \eqref{eq:lambda_cond} for $\lambda_{-2}>0$ (assuming $\lambda_0\geq0$). This family of $\lambda(z)$ is particularly interesting because it always satisfies Eq. \eqref{eq:int_eq}, leaving open the possibility of solutions for arbitrary matrix dimension $N$, without the degree-dependent bound that arises for polynomial $\lambda(z)$ as found in Section \ref{sec:num} and Section \ref{sec:theory}.

For our choice  Eq. \eqref{eq:Weyl_def} of generalizing Weyl's symmetric-ordering for nonpolynomial function, with $\lambda(z)$ given by Eq. \eqref{eq:lambda_sol}:
\begin{equation}
    S_{\lambda}[\hat{A}] = \lambda_0 \hat{A} + \lambda_{-2} \hat{Z}^{-1}_N \hat{A} \hat{Z}^{-1}_N \ ;
    \label{eq:get_Weyl}
\end{equation}
the details of this derivation can be found in Appendix \ref{app:get_Weyl}. In this section, we numerically search for irreducible solutions of the Hermitian matrix triplets $(\hat X_N,\hat Y_N,\hat Z_N )$ in Section \ref{sec:big_num} and then develop a theoretical analysis to explain our observations in Section \ref{sec:big_theo}. Throughout, we assume that $\hat Z_N$ is invertible.

\subsection{Numerical Search \label{sec:big_num}}

We introduce the ratio
$$ \eta \equiv \frac{\lambda_{-2}}{\lambda_0} \ , $$
then set $R=1$ and $\lambda_0 = 1$. Following the procedure described in Section \ref{sec:lambda_is_const}, we search for the solutions of the triplets $(\hat{X}_N,\hat{Y}_N,\hat{Z}_N)$ and their corresponding quantization parameters $\hbar_N$ by minimizing the total loss function $L_{\text{tot}}$, defined in Eq. \eqref{eq:L_tot}. This loss function consists of two contributions:
\begin{itemize}
    \item \underline{Algebraic constraint:} The algebraic constraint loss $L_{\text{ac}}$ follows Eq. \eqref{eq:L_ac}, with the commutation residual retaining the same form as Eq. \eqref{eq:comm_cont_gen} but with the function $\lambda(z)$ and the matrix $S_\lambda[\hat{A}]$ replaced by the expression given in Eq. \eqref{eq:lambda_sol} and Eq. \eqref{eq:get_Weyl}, respectively.
    \item \underline{Irreducibility:} The contribution $L_{\text{irr}}$ stays identical to Eq. \eqref{eq:L_irr}.
\end{itemize}
Here, our numerical search scans through matrix dimensions ranging from $N=2$ to $N=20$, considering five distinct values of the ratio, i.e. $\eta\in \{1, 10^{-1},10^{-2},10^{-3},10^{-4} \}$. For each search, we initialize the optimization using the conventional fuzzy sphere \cite{madore1992fuzzy}, i.e. taking the known analytical solutions for the Hermitian matrices $\hat{X}_N$, $\hat{Y}_N$, $\hat{Z}_N$ and their corresponding quantization parameters $\hbar_N$ Eq. \eqref{eq:usual_fuzzy_sphere} as initial guesses. To avoid singular initial matrices (i.e. $\hat{Z}_N$ is singular for all odd $N$), we also add Gaussian noise with a standard deviation of $5\%$ to each initial matrix element.

\begin{figure*}[thbp!]
    \centering
\includegraphics[width=\linewidth]{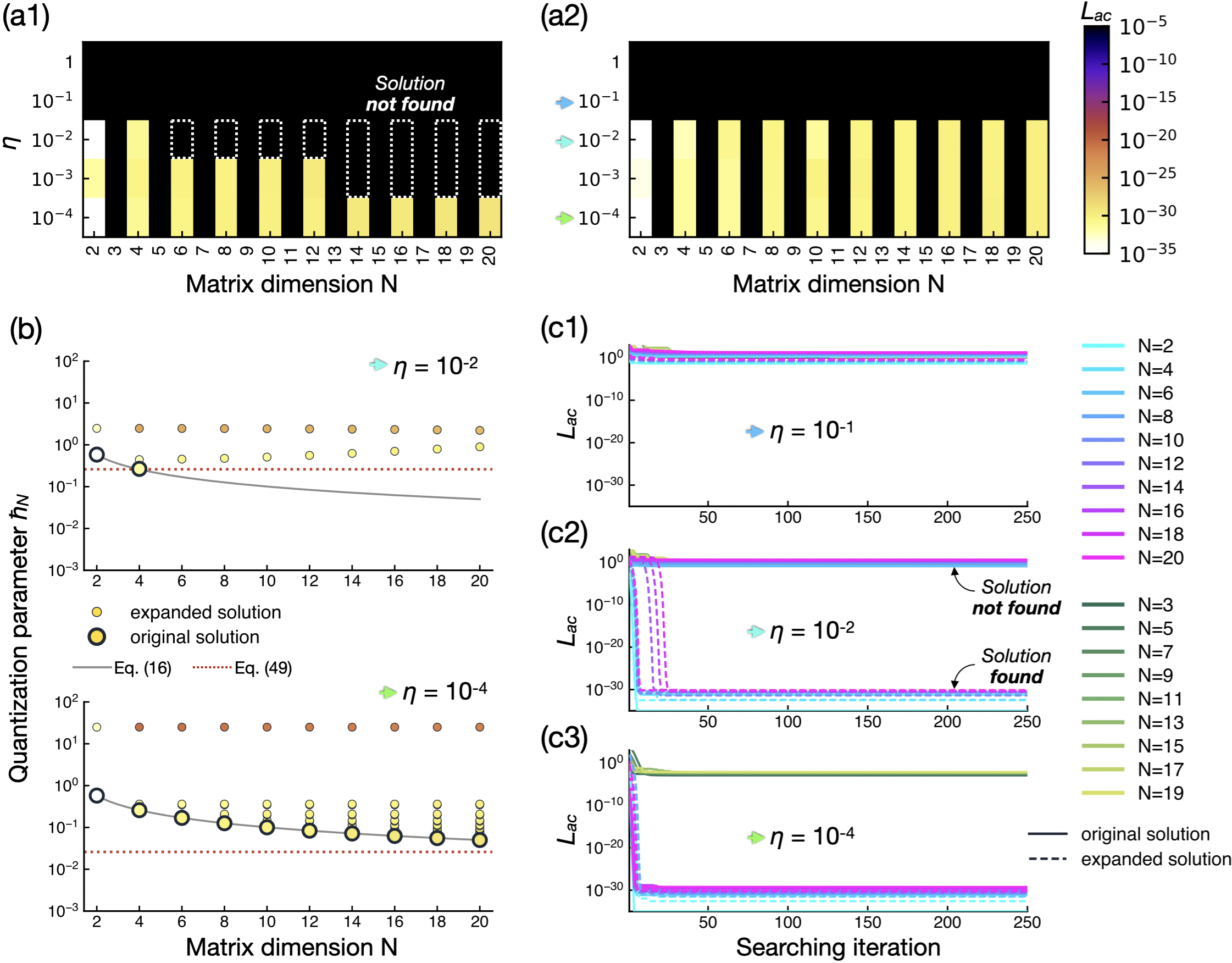}
    \caption{\textbf{Beyond polynomial fuzzy sphere.} Numerical searches for fuzzy sphere quantizations with a nonpolynomial Poisson multiplier; we set $\lambda_0=1$. \textbf{(a1)} Final algebraic-constraint loss $L_{\text{ac}}$ obtained from the initial solution initialization as a function of matrix dimension $N$ and the ratio $\eta$. \textbf{(a2)} Minimum obtained across different initializations, described in appendix K. \textbf{(b)} Numerically obtained quantization parameters $\hbar_N$ from different initializations for $\eta=10^{-2}$ (top) and $\eta=10^{-4}$ (bottom); the color of each dot indicates the corresponding final algebraic-constraint loss, using the same color scale as the colorbar in Fig. \ref{fig03}a. Red dotted lines show the analytical lower bound $\hbar_N$, which has been derived in Eq. \eqref{eq:hbar_bound_below}.  \textbf{(c1-3)} Loss histories for $\eta = 10^{-1},10^{-2},10^{-4}$, respectively. Here we show only the loss trajectories corresponding to the lowest final $L_{\text{ac}}$ values. Solid and dashed curves correspond to the loss for the original solution initialization and the minimum loss over the expanded solution set, respectively, with colors identifying the matrix dimension $N$.}
\label{fig03}
\end{figure*}

We present the final algebraic constraint losses obtained from our numerical searches in Fig. \ref{fig03}a1. Similar to Section \ref{sec:num}, the losses exhibit a clear separation by several tens of orders of magnitude, allowing us to tell apart successful from unsuccessful search. Our results indicate that solutions might be absent for all odd matrix dimensions $N$. Furthermore, the final total loss obtained from the numerical searches remains large for sufficiently large values of $\eta$. These observations motivate a theoretical investigation of potential no-go theorems imposed by the matrix dimension $N$ and the parameter $\eta$, which we pursuit in Appendix \ref{app:big_theory} and report some key findings in Section \ref{sec:big_theo}. 

\subsection{Theoretical Explanation \label{sec:big_theo}}

For the details of our theoretical analysis, see Appendix \ref{app:big_theory}. There, unlike the analysis in Section \ref{sec:bound_nondeg}, we do not require a nondegenerate spectrum of $\hat{Z}_N$. Here, we report only the findings most relevant to our numerical observations in Section \ref{sec:big_num}:

\begin{itemize}
\item \underline{A no-go theorem:} There is no solution in odd matrix dimension $N$, see Appendix \ref{app:big_theory_odd}.
\item \underline{Criterion for finite-dimensional solutions:} The solutions could exist only if 
\begin{equation}
    \frac{\lambda_{-2}}{\lambda_0} \leq \frac{R^2}{7+4\sqrt{3}} \ ,
    \label{eq:eta_range}
\end{equation}
see Appendix \ref{app:big_theory_bound}. This implies that $\lambda_0=0$ admits no finite-dimensional solution.
\end{itemize}
These theoretical findings are consistent with our observations in Section \ref{sec:big_num}. However, there are regions in the $(\eta,N)$-space where solutions are expected to exist (or, at least, where no theoretical obstruction to their existence has been identified), but could not be found numerically (see Fig. \ref{fig03}a1). We conjecture that these solutions may lie far from the conventional fuzzy sphere solution in the search space, making them inaccessible given our original initialization. Finding them may therefore require different initialization strategies.

Indeed, by using theoretically guided initializations, we are able to identify many additional solutions beyond those found in our original searches (see Fig. \ref{fig03}b). We describe the procedure used to identify this expanded set of solutions in Appendix \ref{app:beyondbeyond}. Note that, throughout our numerical searches, we classify a result as a solution if the algebraic constraint loss $L_{ac}$ falls below a threshold of $10^{-18}$. This threshold corresponds to the cases below which successful and unsuccessful searches become numerically indistinguishable in Section \ref{sec:num} (see $N\gtrsim 19$ in Fig. \ref{fig01}b). Nevertheless, for all values of $\eta$, the lowest $L_{\text{ac}}$ values obtained for each $N$ remain well separated from those obtained in cases where theoretically no solution exists (see Fig. \ref{fig03}c1-3).

A possible explanation for why our original searches fail is that the quantization parameter is bounded from below, i.e.
\begin{equation}
    \hbar_{N} \geq \left( \frac{3\sqrt{3}}{2} \right) \frac1{\lambda_0} \sqrt{\frac{\lambda_{-2}}{\lambda_0}} \ ,
    \label{eq:hbar_bound_below}
\end{equation}
as derived in Appendix \ref{app:quant_param}. We can clearly see in Fig. \ref{fig03}b that whenever the quantization parameter associated with the conventional fuzzy sphere falls below this theoretical lower bound, searches using the original initialization yield no solution. Furthermore, the existence of such bound is rather surprising, since it implies that although the matrix dimension can grow without bound, the quantization parameter remains bounded away from zero. Consequently, the fuzzy matrix geometry never becomes commutative, even in the limit $N\to\infty$, and thus admits no classical limit Eq. \eqref{eq:class_lim}.

\section{Discussion \& Conclusion \label{sec:dis_con}}

We have presented a numerical approach to finite-dimensional matrix quantization that searches for irreducible Hermitian matrix representations and compatible Poisson structures under a prescribed matrix ordering. Inspired by machine-learning approaches, we formulate the Casimir constraint, commutation relations, and reducibility condition as algebraic optimization problems, allowing us to explore quantizations that are difficult to construct analytically. Beyond finding individual solutions, systematic numerical searches can uncover unexpected patterns, suggest new conjectures, and point to promising directions for further theoretical investigation. The results of this work demonstrate how such an approach can turn numerical observations into mathematical insights. For Weyl quantized fuzzy spheres with nonconstant polynomial axial Poisson multipliers, our numerical search revealed a sharp relation between polynomial degree and admissible matrix dimension in Section \ref{sec:num}. This observation led us to prove the bound $N\leq d_\lambda$ in Section \ref{sec:theory}, under the assumption of a nondegenerate axial spectrum. Extending the investigation to nonpolynomial multipliers, we further reveal restrictions on admissible representations and establish a positive lower bound on the quantization parameter in Section \ref{sec:beyond}.

Several questions remain open. How do spectral degeneracy (Section \ref{sec:bound_nondeg}) and more general Poisson multipliers (Section \ref{sec:beyond}) affect admissible representations, and which results persist across quantization schemes? Under what conditions does a fuzzy geometry fail to admit a classical limit (Section \ref{sec:big_theo}), despite originating from a well-defined classical geometry? Conversely, one could consider dequantization \cite{steinacker2024quantum}, i.e. starting from a quantum matrix theory and investigating which classical geometries and Poisson structures can emerge in an appropriate classical limit (see Fig. \ref{fig00}). This reverse perspective could reveal whether a given quantum description admits multiple classical interpretations and under what conditions such classical limits exist.

Appendix \ref{app:superman} and Appendix \ref{app:nambu} demonstrate the flexibility of our approach by extending it to more advanced geometric settings, including supermanifolds \cite{grosse1998fuzzy,hasebe2008susy} and generalized Nambu–Poisson structures \cite{ho2016nambu,minic1999m}, respectively. Those fuzzy geometries are relevant to supersymmetric gauge theories \cite{iso2004gauge}, M-theory membrane dynamics \cite{de1999quantum,de1998supermembranes}, and matrix string models of black hole formation \cite{iizuka2013black}. Concrete Hamiltonians or matrix actions, together with appropriate constraints, would allow us to investigate dynamical solutions, including quantum minimal surfaces \cite{arnlind2019quantum,felder2025orthogonal}, and search for new phases or collective phenomena beyond the conventional fuzzy sphere, potentially associated with branes or emergent spacetime. The relation between the Poisson multiplier and the auxiliary Schild density (Section \ref{sec:overview}) further motivates applications to the Schild formulation of string theory \cite{hoppe1996m,yoneya1997schild,ishibashi1997large}, potentially extending the approach from spatial geometries to spacetime embeddings. 

Our long-term goal is to integrate this algebraic approach with methods for inferring classical geometries and Poisson structures from trajectories, building on recent advances in data-driven discovery of classical dynamics, including \cite{bertalan2019learning,cranmer2020lagrangian,phan2025towards,hu2026geometry} and many others. This could enable machine learning methods to infer candidate fundamental quantum theories constrained by classical observations. The combination of numerical construction, conjecture, and analytical explanation demonstrated here provides a practical foundation for extending data-driven discovery into the quantum realm.

\section{Acknowledgement}

This research was supported by internal funding from the Department of Natural Sciences at Scripps and Pitzer Colleges. We thank Kevin Setter and Penelope Messinger for their helpful discussions throughout the project.

\textbf{Data Availability:} The data and code used to generate the figures are available at https://github.com/neutrinopuppy/fuzzy-sphere-quantization.

\textbf{AI Transparency Statement:} OpenAI GPT-6 models and Anthropic Fable models accessed through Codex assisted with initial drafting, language polishing, literature search, and coding for figure generation and manuscript preparation. The ideas, claims, and framing are the authors'. The authors assume responsibility for all content.

\bibliography{main_sphere}
\bibliographystyle{apsrev4-2}

\appendix 

\section{Poisson Structure Overview \label{eq:Poisson_structure}}

\subsection{Poisson Bracket Algebra \label{app:Poi_Alg}}

A Poisson bracket $\{,\}$ is a map from a pair of functions to a single one, satisfying the following algebraic properties for smooth functions $f$, $g$, $h$ and constants $\alpha$, $\beta$:
\begin{itemize}
    \item \underline{Anti-symmetry:} $\{f,g\} = - \{ g,f\}$.
    \item \underline{Linearity:} $\{\alpha f + \beta g, h\} = \alpha \{ f,h\} + \beta\{ g,h \}$.
    \item \underline{Leibniz rule:} $\{f g, h\} = f\{ g,h\} + \{ f,h \} g$.
    \item \underline{Jacobi identity:} $\{f, \{g, h\}\} + \{g, \{h, f\}\} + \{h, \{f, g\}\} = 0$.
\end{itemize}
For polynomial functions of the coordinates $\vec{r}=(r_1, r_2,...,r_n) \in \mathbb{R}^n$, it can be shown that:
\begin{equation}
    \{ f(\vec{r}),g(\vec{r}) \} = \sum^{n}_{\alpha,\beta=1} \frac{\partial}{\partial r_\alpha} f(\vec{r})  \frac{\partial}{\partial r_\beta} g(\vec{r}) \{ r_\alpha, r_\beta \} \ , 
    \label{eq:poly_poisson}
\end{equation}
which demonstrates how the calculus of partial derivatives emerges directly from the algebraic properties. We can generalize this expression to arbitrary smooth functions. Define $\theta^{\alpha \beta}(\vec{r}) \equiv \{ r_\alpha, r_\beta \}$, then follows from \underline{anti-symmetry}, $\theta^{\alpha \beta}(\vec{r})$ is an antisymmetric rank-two contravariant tensor field, transforming as a bivector under coordinate transformations. If $\theta^{\alpha \beta}(\vec{r})$ is invertible, its inverse $\omega_{\alpha\beta}(\vec{r})$ defines an antisymmetric covariant tensor field, i.e.
$$ \omega = \theta^{-1} \ : \ \sum^n_{\gamma=1} \theta^{\alpha \gamma}(\vec{r}) \omega_{\gamma \beta}(\vec{r}) = \delta^\alpha_\beta \ , $$
where $\delta^\alpha_\beta$ is a Kronecker delta \cite{edwards1987appreciation}. The \underline{Jacobi identity} implies that $\omega$ is closed, i.e. $d\omega = 0$, in which $d$ denotes the exterior derivative.

For the sake of completeness, we show the derivation of Eq. \eqref{eq:poly_poisson} directly from the algebraic properties, i.e. applying \underline{anti-symmetry}, \underline{linearity}, and the \underline{Leibniz rule} repeatedly. Here, we present the derivation step-by-step:
\begin{itemize}
    \item Consider the commutation between a single-variable monomial, i.e. $r_\alpha^{k_\alpha}$ with $k_\alpha \in \mathbb{N}$, and a function $g(\vec{r})$. We can apply the \underline{Leibniz rule} repeatedly to obtain:
    \begin{equation}
    \begin{split}
        &\{ r^{k_\alpha},g(\vec{r})\} = \{ r_\alpha^{k_\alpha-1} \cdot r_\alpha,g(\vec{r})\} = r_\alpha^{k_\alpha-1} \{ r_\alpha,g(\vec{r}) \} + \{ r_\alpha^{k_\alpha-1},g(\vec{r})\} r_\alpha
        \\
        & \ \ \ \ =r_\alpha^{k_\alpha-1} \{ r_\alpha,g(\vec{r}) \} + r_\alpha \{ r_\alpha^{k_\alpha-1} ,g(\vec{r})\} = r_\alpha^{k_\alpha-1} \{ r_\alpha,g(\vec{r}) \} + r_\alpha \{ r_\alpha^{k_\alpha-2} \cdot r_\alpha ,g(\vec{r})\} \\
        &\ \ \ \ = r_\alpha^{k_\alpha-1} \{ r_\alpha,g(\vec{r}) \} + r^{k_\alpha-1}_\alpha \{r_\alpha ,g(\vec{r})\} + r_\alpha \{r^{k_\alpha-2}_\alpha ,g(\vec{r}) \} r_\alpha
        \\
        &\ \ \ \ =2 r_\alpha^{k_\alpha-1} \{ r_\alpha,g(\vec{r}) \} + r^2_\alpha \{r_\alpha^{k_\alpha-2} ,g(\vec{r})\} = ...
        \\
        & \ \ \ \ =k_\alpha r_\alpha^{k_\alpha-1} \{ r_\alpha,g(\vec{r}) \} = \frac{\partial}{\partial r_\alpha} (r_\alpha^{k_\alpha}) \{ r_\alpha,g(\vec{r})\} \ .
        \label{eq:1var_mono}
    \end{split}
    \end{equation}
    \item Consider the commutation between a general monomial, i.e. $r_1^{k_1} r_2^{k_2} ... r_n^{k_n}$, and a function $g(\vec{r})$. We can apply Eq. \eqref{eq:1var_mono} and the \underline{Leibniz rule} repeatedly to obtain:
    \begin{equation}
    \begin{split}
        &\{ r_1^{k_1} r_2^{k_2} ... r_n^{k_n},g(\vec{r})\} =  \{ r_1^{k_1} r_2^{k_2} ... r_{n-1}^{k_{n-1}} \cdot r_n^{k_n},g(\vec{r})\}
        \\
        & \ \ \ \ =  r_1^{k_1} r_2^{k_2} ... r_{n-1}^{k_{n-1}} \{  r_n^{k_n},g(\vec{r})\} + \{  r_1^{k_1} r_2^{k_2} ... r_{n-1}^{k_{n-1}} ,g(\vec{r})\} r_n^{k_n} 
        \\
        & \ \ \ \ = r_1^{k_1} r_2^{k_2} ... r_{n-1}^{k_{n-1}} \cdot \frac{\partial}{\partial r_n} (r_n^{k_n}) \{  r_n,g(\vec{r})\} + r_n^{k_n} \{  r_1^{k_1} r_2^{k_2} ...  r_{n-2}^{k_{n-2}} \cdot r_{n-1}^{k_{n-1}} ,g(\vec{r})\}
        \\
        & \ \ \ \ = r_1^{k_1} r_2^{k_2} ... r_{n-1}^{k_{n-1}} \cdot \frac{\partial}{\partial r_n} (r_n^{k_n}) \{  r_n,g(\vec{r})\} 
        \\ &
         \ \ \ \ \ \ \ \ \ \ + r_1^{k_1} r_2^{k_2} ...  r_{n-2}^{k_{n-2}} \cdot r_n^{k_n} \{  r_{n-1}^{k_{n-1}} ,g(\vec{r})\} + r_n^{k_n} \{  r_1^{k_1} r_2^{k_2} ...  r_{n-2}^{k_{n-2}} ,g(\vec{r})\} r_{n-1}^{k_{n-1}}
         \\
        & \ \ \ \ = r_1^{k_1} r_2^{k_2} ... r_{n-1}^{k_{n-1}} \cdot \frac{\partial}{\partial r_n} (r_n^{k_n}) \{  r_n,g(\vec{r})\} 
        \\
        &
         \ \ \ \ \ \ \ \ \ \ + r_1^{k_1} r_2^{k_2} ...  r_{n-2}^{k_{n-2}} \cdot \frac{\partial}{\partial r_{n-1}} (r_{n-1}^{k_{n-1}}) \cdot r_n^{k_n} \{  r_{n-1} ,g(\vec{r})\}
        \\ &
         \ \ \ \ \ \ \ \ \ \ + r_{n-1}^{k_{n-1}} r_n^{k_n} \{  r_1^{k_1} r_2^{k_2} ...  r_{n-2}^{k_{n-2}} ,g(\vec{r})\} = ...
         \\
         & \ \ \ \ = \sum^n_{\alpha=1} r_1^{k_1} r_2^{k_2} ... r_{\alpha-1}^{k_{\alpha-1}} \cdot \frac{\partial}{\partial r_\alpha} (r_\alpha^{k_n}) \cdot r_{\alpha+1}^{k_{\alpha+1}} ... r_n^{k_n} \{r_\alpha,g(\vec{r})\}
        \\ & \ \ \ \ = \sum^n_{\alpha=1} \frac{\partial}{\partial r_\alpha} (r_1^{k_1} r_2^{k_2} ... r_n^{k_n}) \{r_\alpha,g(\vec{r})\} \ .
        \label{eq:nvar_mono}
    \end{split}
    \end{equation}
    \item Consider the commutation between a general polynomial, i.e. a polynomial of degree $d_f \in \mathbb{N}$ that can be written as a sum of monomials
    $$f(\vec{r}) = \sum^{d_f}_{\sum^n_{\alpha=1} k_\alpha \leq d_f} f_{k_1 k_2 ... k_n} r_1^{k_1} r_2^{k_2} ... r_n^{k_n} \ \ \text{with} \ \ f_{k_1 k_2 ... k_n} \in \mathbb{R} \ , $$
    and a function $g(\vec{r})$. We can apply Eq. \eqref{eq:nvar_mono} and \underline{linearity} to obtain:
    \begin{equation}
    \begin{split}
        &\left\{ f(\vec{r}),g(\vec{r})\right\} = \left\{ \sum^{d_f}_{\sum^n_{\alpha=1} k_\alpha \leq d_f} f_{k_1 k_2 ... k_n} r_1^{k_1} r_2^{k_2} ... r_n^{k_n} , g(\vec{r})\right\}
        \\
        & \ \ \ \ =  \sum^{d_f}_{\sum^n_{\alpha=1} k_\alpha \leq d_f} f_{k_1 k_2 ... k_n} \left\{ r_1^{k_1} r_2^{k_2} ... r_n^{k_n} , g(\vec{r})\right\}
        \\
        & \ \ \ \ = \sum^{d_f}_{\sum^n_{\alpha=1} k_\alpha \leq d_f} f_{k_1 k_2 ... k_n} \sum^n_{\alpha=1} \frac{\partial}{\partial r_\alpha} (r_1^{k_1} r_2^{k_2} ... r_n^{k_n}) \{r_\alpha,g(\vec{r})\}
        \\
        & \ \ \ \ =  \sum^n_{\alpha=1} \frac{\partial}{\partial r_\alpha} \left( \sum^{d_f}_{\sum^n_{\alpha=1} k_\alpha \leq d_f} f_{k_1 k_2 ... k_n} r_1^{k_1} r_2^{k_2} ... r_n^{k_n} \right) \{r_\alpha,g(\vec{r})\}
        \\
        & \ \ \ \  =  \sum^n_{\alpha=1} \frac{\partial}{\partial r_\alpha} f(\vec{r}) \{r_\alpha,g(\vec{r})\} \ .
        \label{eq:nvar_poly}
    \end{split}
    \end{equation}
    \item Consider the commutation between two general polynomials $f(\vec{r})$ and $g(\vec{r})$. We can apply Eq. \eqref{eq:nvar_poly} and \underline{anti-symmetry} to obtain:
    \begin{equation*}
    \begin{split}
        &\left\{ f(\vec{r}),g(\vec{r})\right\} =  \sum^n_{\alpha=1} \frac{\partial}{\partial r_\alpha} f(\vec{r}) \{r_\alpha,g(\vec{r})\} = - \sum^n_{\alpha=1} \frac{\partial}{\partial r_\alpha} f(\vec{r}) \{g(\vec{r}),r_\alpha \}
        \\
        & \ \ \ \  = - \sum^n_{\alpha=1} \frac{\partial}{\partial r_\alpha} f(\vec{r}) \sum^n_{\beta=1} \frac{\partial}{\partial r_\beta} g(\vec{r}) \{r_\beta,r_\alpha \} = -\sum^n_{\alpha,\beta=1} \frac{\partial}{\partial r_\alpha} f(\vec{r})  \frac{\partial}{\partial r_\beta} g(\vec{r}) \{r_\beta , r_\alpha \} 
        \\
        & \ \ \ \ = \sum^n_{\alpha,\beta=1} \frac{\partial}{\partial r_\alpha} f(\vec{r})  \frac{\partial}{\partial r_\beta} g(\vec{r}) \{r_\alpha , r_\beta \} \ ,
    \end{split}
    \end{equation*}
    which is identical to Eq. \eqref{eq:poly_poisson}.
\end{itemize}

\subsection{Compatibility with a Casimir Constraint \label{app:Poi_Cas}}

A Casimir constraint is defined by a function $F(\vec r)$ that Poisson commutes with every smooth function $f(\vec r)$, i.e.
$$ \{ F(\vec{r}),g(\vec{r})\} = 0 \ \forall g(\vec{r}) \  , $$
which implies that for every coordinate $r_\alpha$:
\begin{equation}
    \{ F(\vec{r}),r_\alpha \} = 0 \ . 
    \label{eq:Cas_rel}
\end{equation}
Let us consider $n=3$, then the Casimir constraint represents an implicit function describing a two-dimensional surface. We can write Eq. \eqref{eq:Cas_rel} explicitly for all three coordinates $(r_1,r_2,r_3)$:
\begin{equation*}
\begin{split}
    \{ F(\vec{r}),r_1 \} &= \frac{\partial}{\partial r_2} F(\vec{r}) \{r_2,r_1\} + \frac{\partial}{\partial r_3} F(\vec{r}) \{r_3,r_1\} = 0 \ ,
    \\
    \{ F(\vec{r}),r_2 \} &= \frac{\partial}{\partial r_1} F(\vec{r}) \{r_1,r_2\} + \frac{\partial}{\partial r_3} F(\vec{r}) \{r_3,r_2\} = 0 \ ,
    \\
    \{ F(\vec{r}),r_3 \} &= \frac{\partial}{\partial r_1} F(\vec{r}) \{r_1,r_3\} + \frac{\partial}{\partial r_2} F(\vec{r}) \{r_2,r_3\} = 0 \ .
\end{split}
\end{equation*}
These equations can be satisfied if the following three ratios are identical:
$$ \frac{\{r_1,r_2\}}{\displaystyle  \frac{\partial}{\partial r_3} F(\vec{r})} = \frac{\{r_2,r_3\}}{\displaystyle  \frac{\partial}{\partial r_1} F(\vec{r})} = \frac{\{r_3,r_1\}}{\displaystyle  \frac{\partial}{\partial r_2} F(\vec{r})} \ , $$
thus in general there could be a common proportionality factor $\lambda(\vec{r})$ -- called the Poisson multiplier -- such that:
\begin{equation*}
\begin{split}
    \{ r_1,r_2\} &= \lambda(\vec{r}) \frac{\partial}{\partial r_3} F(\vec{r}) \ ,
    \\
    \{ r_2,r_3\} &= \lambda(\vec{r}) \frac{\partial}{\partial r_1} F(\vec{r}) \ ,
    \\
    \{ r_3,r_1\} &= \lambda(\vec{r}) \frac{\partial}{\partial r_2} F(\vec{r}) \ .
\end{split}
\end{equation*}
There is no additional constraint on the $\lambda(\vec r)$ imposed by e.g. the Leibniz rule and the Jacobi identity, therefore the Poisson multiplier can be an arbitrary smooth function. Note that this set of Poisson brackets is identical to Eq. \eqref{eq:gen_Poisson}, with $(r_1,r_2,r_3)=(x,y,z)$.

\section{Irreducibility Criterion \label{app:reducibility_test}}

We prove that a triplet of Hermitian matrices $(\hat{X}_N,\hat{Y}_N,\hat{Z}_N)$ is irreducible if and only if the matrix $\Gamma$ defined in Eq. \eqref{eq:Gamma_irr} is positive definite. Let $\hat{G}=\sum_{a=1}^{N^2-1}g_a\hat{T}_a$ be an arbitrary traceless Hermitian matrix, with real coefficients $\{ g_a\}^{N^2-1}_{a=1}$. Define
$$ C \equiv \sum^{N^2-1}_{a,b=1} g_a g_b \Gamma_{ab} \ , $$
then by construction this sum can be written as:
\begin{equation}
    \begin{split}
        C &=  \sum^{N^2-1}_{a,b=1} g_a g_b \left( \sum_{\hat{R}_N \in \{ \hat{X}_N,\hat{Y}_N,\hat{Z}_N\}} \text{Tr} \left( [\hat{R}_N,\hat{T}_a]^\dagger [\hat{R}_N,\hat{T}_b]\right) \right)
        \\
        &= \sum_{\hat{R}_N \in \{ \hat{X}_N,\hat{Y}_N,\hat{Z}_N\}} \sum^{N^2-1}_{a,b=1} g_a g_b \text{Tr} \left( [\hat{R}_N,\hat{T}_a]^\dagger [\hat{R}_N,\hat{T}_b]\right)
        \\
        &=\sum_{\hat{R}_N \in \{ \hat{X}_N,\hat{Y}_N,\hat{Z}_N\}} \text{Tr} \left( \left(\sum^{N^2-1}_{a=1}[\hat{R}_N,\hat{T}_a] \right)^\dagger \left( \sum^{N^2-1}_{b=1} [\hat{R}_N,\hat{T}_b] \right) \right) 
        \\
        &=\sum_{\hat{R}_N \in \{ \hat{X}_N,\hat{Y}_N,\hat{Z}_N\}} \text{Tr} \left( [\hat{R}_N,\hat{G}]^\dagger [\hat{R}_N,\hat{G}] \right) = \sum_{\hat{R}_N \in \{ \hat{X}_N,\hat{Y}_N,\hat{Z}_N\}} \| [\hat{R}_N,\hat{G}] \|^2
        \\
        & =\|[\hat{G},\hat{X}_N]\|^2
    +\|[\hat{G},\hat{Y}_N]\|^2
    +\|[\hat{G},\hat{Z}_N]\|^2 \ ,
    \end{split}
    \label{eq:Gamma_quadratic_form}
\end{equation}
which implies $C \geq 0$, thus the matrix $\hat{\Gamma}$ is positive semidefinite. Also, $C=0$ happens if and only if $\hat{G}$ commutes with all three matrices. 

We can prove that the triplet $(\hat{X}_N,\hat{Y}_N,\hat{Z}_N)$
is irreducible if and only if $\hat{\Gamma}$ is positive definite:
\begin{itemize}
    \item Suppose the triplet is reducible. Then there exists
    a unitary matrix $\hat{U}$ and an integer $1\leq k<N$ such that:
    $$
        \hat{U}^\dagger \hat{R}_N \hat{U}
        =
        \begin{pmatrix}
            \hat{R}_N^{(1)} & \hat{0} \\
            \hat{0} & \hat{R}_N^{(2)}
        \end{pmatrix} \ ,
        \ 
        \hat{R}_N \in\{\hat{X}_N,\hat{Y}_N,\hat{Z}_N\} \ ,
    $$
    where the two blocks have sizes $k$ and $N-k$ for all
    three matrices. Define
    $$
        \hat{G}
        =
        \hat{U}
        \begin{pmatrix}
            (N-k)\hat{I}_{k\times k} & \hat{0}\\
            \hat{0} & -k\hat{I}_{(N-k) \times (N-k)}
        \end{pmatrix}
        \hat{U}^\dagger \ .
    $$
    This matrix is nonzero and Hermitian, and
    $$\text{Tr}(\hat{G})
        =k(N-k)-(N-k)k=0 \ .
    $$
    Since each block of $\hat{U}^\dagger\hat{G} \hat{U}$ is proportional
    to the identity, $\hat{G}$ commutes with all three matrices, thus $C=0$ and $\hat{\Gamma}$ is not positive definite.

    \item Conversely, suppose $\hat{\Gamma}$ is not positive
    definite. Since it is real symmetric and positive
    semidefinite, it has a zero eigenvalue. Thus, there exist
    real coefficients $\{g_a\}^{N^2-1}_{a=1}$, not all zero, such that $C=0$.
    The corresponding matrix
    $\hat{G}=\sum_a g_a\hat{T}_a$ is nonzero, traceless,
    and Hermitian. Equation~\eqref{eq:Gamma_quadratic_form}
    implies
    $$[\hat{G},\hat{X}_N]
        =[\hat{G},\hat{Y}_N]
        =[\hat{G},\hat{Z}_N]=0 \ . $$
    Diagonalize $\hat{G}$ using a unitary matrix $\hat{U}$:
    $$\hat{U}^\dagger\hat{G} \hat{U}
        = \begin{pmatrix}
        \eta_1 & 0 & ... & 0\\
        0 & \eta_2 & ... & 0\\
        ... & ... & ... & ... \\
        0 & 0 & ... & \eta_N
    \end{pmatrix} \ . $$
    For each $\hat{R}_N \in\{\hat{X}_N,\hat{Y}_N,\hat{Z}_N\}$,
    write $\hat{R}^{\text{(tf)}}_N=\hat{U}^\dagger\hat{R}_N\hat{U}$. The commutation
    relation then gives
    \[
        (\eta_i-\eta_j){R^{\text{(tf)}}_N}_{ij}=0 \ .
    \]
    Therefore, whenever $\eta_i\neq\eta_j$, the corresponding
    entry ${R^{\text{(tf)}}_N}_{ij}$ vanishes in all three matrices.
    Grouping equal eigenvalues of $\hat{G}$ consequently
    puts the triplet into a common block-diagonal form.
    There are at least two blocks: if all $\eta_i$ were equal,
    tracelessness would force $\hat{G}=0$, a contradiction.
    Thus, the triplet is reducible.
\end{itemize}

\section{On the Algebra of $\hat{W}_N$ \label{app:W_alg}}

\subsection{Derivation of Eq. \eqref{eq:WW}}

From the definition of $\hat{W}_N$ in Eq.~\eqref{eq:W}, and since $\hat{X}_N$ and
$\hat{Y}_N$ are Hermitian, we have $\hat{W}_N^\dagger = \hat{X}_N - i\hat{Y}_N$. Consider $\hat{W}_N \hat{W}_N^\dagger$:
\begin{equation*}
    \begin{split}
        \hat{W}_N \hat{W}_N^\dagger
        &= (\hat{X}_N + i\hat{Y}_N)(\hat{X}_N - i\hat{Y}_N) \\
        &= \left( \hat{X}_N^2 + \hat{Y}_N^2 \right) - i\left(\hat{X}_N\hat{Y}_N - \hat{Y}_N\hat{X}_N\right)
        \\
        &= \left( \hat{X}_N^2 + \hat{Y}_N^2 \right) - i [ \hat{X}_N,\hat{Y}_N ] \ .
    \end{split}
\end{equation*}
Using Eq. \eqref{eq:Casimir_to_sandwich} together with the commutator relation Eq. \eqref{eq:gen_Poisson_axial_mat_quant_X}, this equation can be rewritten as:
\begin{equation}
\begin{split}
    \hat{W}_N \hat{W}_N^\dagger &= \left( R^2 \hat{I}_{N \times N} - \hat{Z}^2_N \right) - i \left( i 2\hbar \lambda(\hat{Z}_N) \hat{Z}_N \right) 
    \\
    &= R^2 \hat{I}_{N \times N} - \hat{Z}^2_N + 2 \hbar \lambda(\hat{Z}_N ) \hat{Z}_N \ . 
    \label{eq:WW1}
\end{split}
\end{equation}
Similarly, for $\hat{W}^\dagger_N \hat{W}_N$, we can express it as:
\begin{equation}
    \hat{W}_N^\dagger \hat{W}_N = R^2 \hat{I}_{N \times N} - \hat{Z}^2_N - 2 \hbar \lambda(\hat{Z}_N ) \hat{Z}_N \ . 
    \label{eq:WW2}
\end{equation}

\subsection{Derivation of Eq. \eqref{eq:get_comm_ZW}}

We can rewrite the two commutators in Eq. \eqref{eq:gen_Poisson_axial_mat_quant_YZ} explicitly, using Eq. \eqref{eq:def_S}, as:
\begin{equation*}
\begin{split}
    [ \hat{Z}_N,\hat{X}_N ] = i 2 \hbar_N S_{\lambda}[\hat{Y}_N] &= 2 i \hbar_N \sum^{d_\lambda}_{\alpha=0}  \frac{\lambda_\alpha}{\alpha+1} \sum^\alpha_{\beta=0} \hat{Z}_N^\beta \hat{Y}_N \hat{Z}^{\alpha-\beta}_N \ ,
    \\
    [ \hat{Z}_N, \hat{Y}_N] = -i 2 \hbar_N S_{\lambda}[\hat{X}_N] &= 2 i \hbar_N \sum^{d_\lambda}_{\alpha=0}  \frac{\lambda_\alpha}{\alpha+1} \sum^\alpha_{\beta=0} \hat{Z}_N^\beta \hat{X}_N \hat{Z}^{\alpha-\beta}_N \ .
\end{split}
\end{equation*}
These above expressions are equivalent to:
\begin{equation*}
\begin{split}
    [ \hat{Z}_N,\hat{X}_N ] &= 2 i \hbar_N \sum^{d_\lambda}_{\alpha=0}  \frac{\lambda_\alpha}{\alpha+1} \sum^\alpha_{\beta=0} \hat{Z}_N^\beta i \hat{Y}_N \hat{Z}^{\alpha-\beta}_N \ ,
    \\
    [\hat{Z}_N, i \hat{Y}_N] &= 2 \hbar_N \sum^{d_\lambda}_{\alpha=0}  \frac{\lambda_\alpha}{\alpha+1} \sum^\alpha_{\beta=0} \hat{Z}_N^\beta \hat{X}_N \hat{Z}^{\alpha-\beta}_N \ .
\end{split}
\end{equation*}
Combine the RHSs and the LHSs together, we arrive at:
\begin{equation*}
\begin{split}
    [ \hat{Z}_N,\hat{X}_N ] + [\hat{Z}_N, i \hat{Y}_N]  &= 2 \hbar_N \sum^{d_\lambda}_{\alpha=0}  \frac{\lambda_\alpha}{\alpha+1} \sum^\alpha_{\beta=0} \left( \hat{Z}_N^\beta i \hat{Y}_N \hat{Z}^{\alpha-\beta}_N + \hat{Z}_N^\beta \hat{X}_N \hat{Z}^{\alpha-\beta}_N \right) \ .
    \\
    \Longrightarrow \ [ \hat{Z}_N,\hat{X}_N + i \hat{Y}_N] &= 2 \hbar_N \sum^{d_\lambda}_{\alpha=0}  \frac{\lambda_\alpha}{\alpha+1} \sum^\alpha_{\beta=0} \hat{Z}_N^\beta \left( \hat{X}_N + i \hat{Y}_N \right) \hat{Z}^{\alpha-\beta}_N
    \\
    \Longrightarrow \ [ \hat{Z}_N,\hat{W}_N] &= 2 \hbar_N \sum^{d_\lambda}_{\alpha=0}  \frac{\lambda_\alpha}{\alpha+1} \sum^\alpha_{\beta=0} \hat{Z}_N^\beta \hat{W}_N \hat{Z}^{\alpha-\beta}_N \ ,
\end{split}
\end{equation*}
which is precisely Eq. \eqref{eq:get_comm_ZW}.

\section{Spectrum $\{ v_N^{(j)}\}^{N-1}_{j=0}$ of $\hat{Z}_N$ \label{app:eigval_in}}

Following from the Casimir constraint represented in matrix form Eq. \eqref{eq:implicit_sphere_mat}, we have that
\begin{equation}
    R^2 \hat{I}_{N\times N} - \hat{Z}_N^2 = \hat{X}_N^2 + \hat{Y}_N^2 \ .
    \label{eq:Casimir_to_sandwich}
\end{equation}
Consider sandwiching both sides in between bra $\langle j_N \lvert$ and ket $\lvert j_N \rangle$, then:
\begin{itemize}
    \item For the left-hand side (LHS) of Eq. \eqref{eq:Casimir_to_sandwich}:
        \begin{equation}
            \langle j_N \lvert  R^2 \hat{I}_{N\times N} - \hat{Z}_N^2 \lvert j_N \rangle = R^2 - (v^{(j)}_N)^2
            \label{eq:Cas_LHS}
        \end{equation}
    \item For the right-hand side (RHS) of Eq. \eqref{eq:Casimir_to_sandwich}:
        \begin{equation}
            \langle j_N \lvert  \hat{X}_N^2 + \hat{Y}_N^2 \lvert j_N \rangle = \langle j_N \lvert  \hat{X}^2_N \lvert j_N \rangle + \langle j_N \lvert \hat{Y}^2_N \lvert j_N \rangle = \left\| \hat{X}_N \lvert j_N \rangle \right\|^2 + \left\| \hat{Y}_N \lvert j_N \rangle \right\|^2  \geq 0 \ ,
            \label{eq:Cas_RHS}
        \end{equation}
    where $\|...\|$ is the Hilbert space's norm (always non-negative value).
\end{itemize}
Together, Eq. \eqref{eq:Cas_LHS} and Eq. \eqref{eq:Cas_RHS} imply 
$$R^2 - (v^{(j)}_N)^2 \geq 0 \ \Longrightarrow \ v^{(j)}_N \in [-R,+R] \ . $$
 This argument is general for any eigenvector $\lvert j_N \rangle$ of $\hat{Z}_N$, so it also applies to every corresponding eigenvalue $v_N^{(j)}$, leading to the first-part of Eq. \eqref{eq:eigval_ordering}.

\section{Derivation of Eq. \eqref{eq:int} \label{app:get_int}}

The LHS of Eq. \eqref{eq:get_comm_ZW} after sandwiching:
\begin{equation*}
\begin{split}
    \langle j_N\lvert [\hat{Z}_N,\hat{W}_N] \lvert k_N\rangle &= \langle j_N\lvert \hat{Z}_N \hat{W}_N - \hat{W}_N \hat{Z}_N \lvert k_N\rangle 
    \\
    &= \langle j_N\lvert \hat{Z}_N \hat{W}_N \lvert k_N\rangle - \langle j_N\lvert \hat{W}_N \hat{Z}_N \lvert k_N\rangle
    \\
    &= \langle j_N\lvert v^{(j)}_N \hat{W}_N \lvert k_N\rangle - \langle j_N\lvert \hat{W}_N v^{(k)}_N \lvert k_N\rangle
    \\
    &= (v^{(j)}_N - v^{(k)}_N) \langle j_N\lvert \hat{W}_N \lvert k_N\rangle = (v^{(j)}_N - v^{(k)}_N)  W^{(j,k)}_N \ ,
\end{split}
    \label{eq:sandwich_LHS}
\end{equation*}
which is exactly the same as the LHS of Eq. \eqref{eq:int}.

The RHS of Eq. \eqref{eq:get_comm_ZW} after sandwiching:
\begin{equation}
\begin{split}
    \langle j_N\lvert 2 \hbar_N S_{\lambda}[\hat{W}_N] \lvert k_N\rangle &= 2\hbar_N \sum^{d_\lambda}_{\alpha=0}  \frac{\lambda_\alpha}{\alpha+1} \sum^\alpha_{\beta=0} \langle j_N\lvert \hat{Z}_N^\beta \hat{W}_N \hat{Z}^{\alpha-\beta}_N \lvert k_N\rangle
    \\
    &= 2\hbar_N \sum^{d_\lambda}_{\alpha=0}  \frac{\lambda_\alpha}{\alpha+1} \sum^\alpha_{\beta=0} \langle j_N\lvert (v^{(j)}_N)^\beta \hat{W}_N (v^{(k)}_N)^{\alpha-\beta} \lvert k_N\rangle
    \\
    &= 2\hbar_N \sum^{d_\lambda}_{\alpha=0}  \frac{\lambda_\alpha}{\alpha+1} \sum^\alpha_{\beta=0} (v^{(j)}_N)^\beta  (v^{(k)}_N)^{\alpha-\beta} \langle j_N\lvert \hat{W}_N \lvert k_N\rangle
    \\
    &= 2\hbar_N \sum^{d_\lambda}_{\alpha=0}  \frac{\lambda_\alpha}{\alpha+1} \left[ \frac{({v^{(j)}_N})^{\alpha+1} - (v^{(k)}_N)^{\alpha+1}}{{v^{(j)}_N} - {v^{(k)}_N}}\right] W^{(j,k)}_N \ .
    \label{eq:sandwich_RHS_1}
\end{split}
\end{equation}
We can express the nominator of $[...]$ in an integration form, i.e.
$$ ({v^{(j)}_N})^{\alpha+1} - (v^{(k)}_N)^{\alpha+1} = (\alpha+1) \int^{{v^{(j)}_N}}_{v^{(k)}_N} d\tau \ \tau^\alpha \ , $$
and rewrite Eq. \eqref{eq:sandwich_RHS_1} as:
\begin{equation}
\begin{split}
    &\langle j_N\lvert 2 \hbar_N S_{\lambda}[\hat{W}_N] \lvert k_N\rangle = 2\hbar_N \sum^{d_\lambda}_{\alpha=0}  \frac{\lambda_\alpha}{\alpha+1} \left[ \frac{\displaystyle (\alpha+1) \int^{{v^{(j)}_N}}_{v^{(k)}_N} d\tau \ \tau^\alpha}{{v^{(j)}_N} - {v^{(k)}_N}}\right] W^{(j,k)}_N
    \\
    & \ \ \ \ \ \ \ \ = 2\hbar_N \left[ \frac{\displaystyle \int^{{v^{(j)}_N}}_{v^{(k)}_N} d\tau \ \sum^{d_\lambda}_{\alpha=0} \lambda_\alpha \tau^\alpha }{{v^{(j)}_N} - {v^{(k)}_N}} \right] W_N^{(j,k)} = 2\hbar_N \left[ \frac{\displaystyle \int^{{v^{(j)}_N}}_{v^{(k)}_N} d\tau \ \lambda(\tau) }{{v^{(j)}_N} - {v^{(k)}_N}} \right] W_N^{(j,k)}  \ ,
    \label{eq:sandwich_RHS_2}
\end{split}
\end{equation}
where the last transformation comes directly from the polynomial representation of $\lambda(z)$ as explained in Eq. \eqref{eq:lambda_pol}. The expression inside $[...]$ is precisely the average value of $\lambda(z)$ along the $z$-axis between locations $z=v_N^{(j)}$ and $z=v_N^{(k)}$, which we can denote as $\bar{\lambda}(v_N^{(j)},v_N^{(k)})$, i.e.
\begin{equation}
    \bar{\lambda}(v_N^{(j)},v_N^{(k)}) \equiv \frac{\displaystyle \int^{{v^{(j)}_N}}_{v^{(k)}_N} d\tau \ \lambda(\tau) }{{v^{(j)}_N} - {v^{(k)}_N}}
    \label{eq:lambda_avg}
\end{equation}
And thus, the final expression in Eq. \eqref{eq:sandwich_RHS_2} is therefore identical to Eq. \eqref{eq:int}. It is important to note that, when the two eigenvalues are the same $v_N^{(j)}=v_N^{(k)}$, the expression in $[...]$ and therefore Eq. \eqref{eq:sandwich_RHS_2} seem undefined, but actually we can just go back to the third-line in Eq. \eqref{eq:sandwich_RHS_1} and show that:
\begin{equation}
\begin{split}
    \langle j_N\lvert 2 \hbar_N S_{\lambda}[\hat{W}_N] \lvert k_N\rangle &= 2\hbar_N \sum^{d_\lambda}_{\alpha=0}  \frac{\lambda_\alpha}{\alpha+1} \sum^\alpha_{\beta=0} (v^{(j)}_N)^\beta  (v^{(j)}_N)^{\alpha-\beta} \langle j_N\lvert \hat{W}_N \lvert k_N\rangle 
    \\
    &= 2\hbar_N \sum^{d_\lambda}_{\alpha=0}  \frac{\lambda_\alpha}{\alpha+1} \sum^\alpha_{\beta=0} (v^{(j)}_N)^{\alpha} W^{(j,k)}_N
    \\
    &= 2\hbar_N \sum^{d_\lambda}_{\alpha=0}  \lambda_\alpha (v^{(j)}_N)^{\alpha} W^{(j,k)}_N = 2\hbar_N \lambda(v^{(j)}_N) W^{(j,k)}_N \ .
\end{split}
\end{equation}
This implies the identification $\bar{\lambda}(v_N^{(j)},v_N^{(j)}) = \lambda(v_N^{(j)})$, which is consistent with the averaging operation.

\section{Derivation of Eq. \eqref{eq:int_eq} \label{app:int_eq}}

Consider the following  integral identity for the trapezoid rule \cite{dragomir2012some}:
$$\int_a^b dz \ f(z) = \left( \frac{ f(a)+f(b) }2 \right) (b-a) - \frac12  \int_a^b dz \ (b-z)(z-a)f''(z) \ , $$
which can be rewritten as:
\begin{equation}
    \big( f(a)+f(b) \big) - \frac{2}{b-a} \int_a^b dz \ f(z) =  \frac1{b-a} \int_a^b dz \ (b-z)(z-a)f''(z) \ .
    \label{eq:trap_id}
\end{equation}
Consider $a=v^{(k)}_N$, $b=v^{(j)}_N$, and $f(z) = \big( z\Lambda(z)\big)'$
then:
\begin{itemize}
    \item The \underline{RHS} of Eq. \eqref{eq:trap_id}:
    \begin{equation}
        \frac1{v^{(j)}_N-v^{(k)}_N} \int^{v^{(j)}_N}_{v^{(k)}_N} dz \ (v^{(j)}_N-z)(z-v^{(k)}_N) \big( z \Lambda(z) \big)''' \ .
        \label{eq:RHS_val}
    \end{equation}
    \item The first-term on the \underline{LHS} of Eq. \eqref{eq:trap_id}: Given that
    $$f(z) = \big( z \Lambda(z) \big)' = z\Lambda'(z) + \Lambda(z) = z \lambda(z) + \Lambda(z) \ , $$
    we get:
    \begin{equation} 
    \begin{split}
    f(v^{(j)}_N) + f(v^{(k)}_N) &= \left( v^{(j)}_N \lambda(v^{(j)}_N) + \Lambda(v^{(j)}_N) \right) + \left(   v^{(k)}_N \lambda(v^{(k)}_N) + \Lambda(v^{(k)}_N) \right)
    \\
    &= \left( v^{(j)}_N \lambda(v^{(j)}_N) + v^{(k)}_N \lambda(v^{(k)}_N) \right) + \Lambda(v^{(j)}_N) + \Lambda(v^{(k)}_N)  \ .
    \end{split}
    \label{eq:LHS_t1}
    \end{equation}
    \item The second-term on the \underline{LHS} of Eq. \eqref{eq:trap_id}: The integration can be calculated to be
    $$\int^{v^{(j)}_N}_{v^{(k)}_N} dz \ \big(z \Lambda(z) \big)' = \big(z \Lambda(z) \big) \Bigg|^{z=v^{(j)}_N}_{z=v^{(k)}_N} = v^{(j)}_N \Lambda(v^{(j)}_N) - v^{(k)}_N \Lambda(v^{(k)}_N) \ , $$
    therefore this term can be expressed as:
    \begin{equation}
        \frac{-2v^{(j)}_N}{v^{(j)}_N-v^{(k)}_N} \Lambda(v^{(j)}_N) + \frac{2v^{(k)}_N}{v^{(j)}_N-v^{(k)}_N}  \Lambda(v^{(k)}_N) \ .
        \label{eq:LHS_t2}
    \end{equation}
\end{itemize}
Combine Eq. \eqref{eq:LHS_t1} and Eq. \eqref{eq:LHS_t2}, we obtain the LHS of Eq. \eqref{eq:trap_id} as:
\begin{equation}
    \left( v^{(j)}_N \lambda(v^{(j)}_N) + v^{(k)}_N \lambda(v^{(k)}_N) \right) + \frac{v^{(j)}_N+v^{(k)}_N}{v^{(j)}_N-v^{(k)}_N} \left( \Lambda(v^{(j)}_N) - \Lambda(v^{(k)}_N) \right) \ .
    \label{eq:LHS_next}
\end{equation}
Since the difference between antiderivatives is the same as an integration:
$$\Lambda(v^{(j)}_N) - \Lambda(v^{(k)}_N) = \int^{v^{(j)}_N}_{v^{(k)}_N} dz \ \lambda(z) \ , $$
thus we can use Eq. \eqref{eq:lambda_avg} to express Eq. \eqref{eq:LHS_next} as:
$$ \left( v^{(j)}_N \lambda(v^{(j)}_N) + v^{(k)}_N \lambda(v^{(k)}_N) \right) + \left( v^{(j)}_N+v^{(k)}_N \right) \bar{\lambda}(v^{(j)}_N,v^{(k)}_N) \ . $$
This should be equal to the RHS of Eq. \eqref{eq:trap_id}, i.e. Eq. \eqref{eq:RHS_val}. The finding here allows us to rewrite Eq. \eqref{eq:functional_eq} as Eq. \eqref{eq:int_eq}.

\section{Quantization Constraints from Poisson Multiplier \label{app:quant_param_bound}}

\subsection{Incoming and Outgoing Terminal Eigenvalues \label{app:lh_evals}}

Consider an eigenvector $\lvert j_N \rangle$, then we can calculate the Hilbert space's norm of $\hat{W}_N^\dagger \lvert j_N \rangle$ and $\hat{W}_N \lvert j_N \rangle$ as follows:
\begin{equation}
    \begin{split}
    \left\| \hat{W}^\dagger_N \lvert j_N \rangle\right\|^2 &= \langle j_N \lvert \hat{W}_N  \hat{W}^\dagger_N \lvert j_N \rangle 
    \\
    &= \langle j_N \lvert R^2 \hat{I}_{N \times N} - \hat{Z}^2_N + 2 \hbar \lambda(\hat{Z}_N ) \hat{Z}_N \lvert j_N \rangle 
    \\
    &= R^2 -(v^{(j)}_N)^2 + 2 \hbar_N v^{(j)}_N \lambda(v^{(j)}_N) \ ,
    \\
    \left\| \hat{W}_N \lvert j_N \rangle\right\|^2 &= \langle j_N \lvert \hat{W}^\dagger_N  \hat{W}_N \lvert j_N \rangle 
    \\
    &= \langle j_N \lvert R^2 \hat{I}_{N \times N} - \hat{Z}^2_N - 2 \hbar \lambda(\hat{Z}_N ) \hat{Z}_N \lvert j_N \rangle 
    \\
    &= R^2 -(v^{(j)}_N)^2 - 2 \hbar_N v^{(j)}_N \lambda(v^{(j)}_N) \ ,
    \label{eq:norm_Wstate}
    \end{split}
\end{equation}
where we have used Eq. \eqref{eq:WW1} and Eq. \eqref{eq:WW2} to rewrite the operators $\hat{W}_N  \hat{W}_N^\dagger$ and $\hat{W}_N^\dagger  \hat{W}_N$. 

We can use the completeness relation Eq. \eqref{eq:completeness} to express $\hat{W}^\dagger_N \lvert j_N \rangle$ as:
\begin{equation}
    \hat{W}_N^\dagger \lvert j_N \rangle = \hat{I}_{N \times N} \hat{W}^\dagger_N \lvert j_N \rangle = \sum^{N-1}_{\ell=0} \lvert \ell_N \rangle \langle \ell_N \lvert \hat{W}^\dagger_N \lvert j_N \rangle = \sum^{N-1}_{\ell=0} {W_N^{(j,\ell)}}^* \lvert \ell_N \rangle
    \label{eq:Wstate_in_basis}
\end{equation}
Define incoming terminal eigenvalues as those that have no incoming direct connection to it, then if $v^{(j^{\text{in}})}_N$ is an incoming terminal eigenvalue, $W_N^{(j^{\text{in}},\ell)}=0$ for all index $\ell$ and 
$$ \hat{W}_N^\dagger \lvert j^{\text{in}}_N \rangle = 0 \ . $$
Due to this, the Hilbert space's norm of $\hat{W}^\dagger_N \lvert j^{\text{in}}_N \rangle$ should vanishes, therefore
\begin{equation}
\begin{split}
    R^2 -(v^{(j^{\text{in}})}_N)^2 + 2 \hbar_N v^{(j^{\text{in}})}_N \lambda(v^{(j^{\text{in}})}_N) = 0
\end{split}
\label{eq:Wnorm_ground}
\end{equation}
as a consequence of Eq. \eqref{eq:norm_Wstate}. Since the lowest-eigenvalue $v^{(0)}_N$ is always an incoming terminal eigenvalue, following from \eqref{eq:no_back}, it must also satisfies Eq. \eqref{eq:Wnorm_ground}. Similarly, we can define outgoing terminal eigenvalues via
$$ \hat{W}_N \lvert j^{\text{out}}_N \rangle = 0 \ , $$
its corresponding eigenvalue $v^{(j^{\text{out}})}_N$ must satisfies
\begin{equation}
     R^2 -(v^{(j^{\text{out}})}_N)^2 - 2 \hbar_N v^{(j^{\text{out}})}_N \lambda(v^{(j^{\text{out}})}_N) = 0 \ ,
\label{eq:Wnorm_top}
\end{equation}
and the highest-eigenvalue $v^{(N-1)}_N$ -- which is always be an outgoing terminal eigenvalue -- should satisfies Eq. \eqref{eq:Wnorm_top}.

\subsection{Derivation of Eq. \eqref{eq:hbar_oscillation_lower_bound}}

The behavior of $H(z)$ constraints the quantization parameter $\hbar_N$. Consider a direct connected path that starts at eigenvalue $v^{(\ell_0)}_N$ and ends at eigenvalue $v^{(\ell_K)}_N$, i.e.
\begin{equation}
    v_N^{(\ell_0)}
    \dashrightarrow v_N^{(\ell_1)}
    \dashrightarrow\cdots
    \dashrightarrow v_N^{(\ell_K)} \ \ \text{where} \ \ K \leq n_H
    \label{eq:K_bound}
\end{equation}
and such path must exists -- otherwise $\hat{W}_N=0$, which by Eq. \eqref{eq:W} implies vanishing $\hat{X}_N=\hat{Y}_N=0$. As defined from Appendix \ref{app:lh_evals}, $v_N^{(\ell_0)}$ is an incoming terminal eigenvalue and $v_N^{(\ell_K)}$ is an outgoing terminal eigenvalue, therefore they must satisfy:
\begin{equation*}
\begin{split}
    R^2 -(v^{(\ell_0)}_N)^2 &= - 2 \hbar_N v^{(\ell_0)}_N \lambda(v^{(\ell_0)}_N) \ , 
    \\
    R^2 - (v^{(\ell_K)}_N)^2 &= 2 \hbar_N v^{(\ell_K)}_N \lambda(v^{(\ell_K)}_N) \ ,
\end{split}
\label{eq:hbar_endpoint_equations}
\end{equation*}
then from Eq. \eqref{eq:lambda_cond} and Eq. \eqref{eq:eigval_ordering} we get $v^{(0)}_N <0< v^{(N-1)}_N$.
These equations can be rearranged and recast into the following inequalities:
\begin{equation}
\begin{split}
    R^2
    =(v^{(\ell_0)}_N)^2 + 2 \hbar_N |v^{(\ell_0)}_N| \lambda(v^{(\ell_0)}_N)
      &\leq (v^{(\ell_0)}_N)^2 + 2 \hbar_N |v^{(\ell_0)}_N| \lambda_{\max} \ ,
      \\
    R^2
    = (v^{(\ell_K)}_N)^2 + 2 \hbar_N v^{(\ell_K)}_N \lambda(v^{(\ell_K)}_N)
      &\leq (v^{(\ell_K)}_N)^2 + 2 \hbar_N v^{(\ell_K)}_N \lambda_{\max} \ .
\end{split}
\end{equation}
Solving these yields a lower bound:
\begin{equation}
\begin{split}
    |v^{(\ell_0)}_N| , v^{(\ell_K)}_N
    &\geq\sqrt{R^2+\hbar_N^2 \lambda_{\max}^2}-\hbar_N \lambda_{\max} 
    \\
    \Longrightarrow \ v^{(\ell_K)}_N-v^{(\ell_0)}_N &\geq 2\left( \sqrt{R^2+\hbar_N^2 \lambda_{\max}^2}-\hbar_N \lambda_{\max} \right) \ .
\end{split} 
\label{eq:hbar_endpoint_lower_bound}
\end{equation}
For each consecutive pair of eigenvalues  $v^{(\ell_{k})}_N \dashrightarrow v^{(\ell_{k+1})}_N$, we can use Eq. \eqref{eq:int} to arrive at the upper bound:
\begin{equation}
\begin{split}
 v^{(\ell_{k+1})}_N-v^{(\ell_k)}_N
    &=2\hbar_N\bar\lambda(v^{(\ell_{k+1})}_N,v^{(\ell_k)}_N)
    \leq 2\hbar_N \lambda_{\max}
    \\
    \Longrightarrow \ v^{(\ell_K)}_N-v^{(\ell_0)}_N &\leq 2 K \hbar_N \lambda_{\max} \leq 2 n_H \hbar_N \lambda_{\max} \leq 2 (d_\lambda - 1) \hbar_N \lambda_{\max} \ ,
\end{split}
\label{eq:hbar_increment_bound}
\end{equation}
in which we take Eq. \eqref{eq:nH_bound} and Eq. \eqref{eq:K_bound} into consideration for the second-line. Combining both Eq. \eqref{eq:hbar_endpoint_lower_bound} and Eq. \eqref{eq:hbar_increment_bound}, we can eliminate $v^{(\ell_K)}_N-v^{(\ell_0)}_N$ from these inequalities and yield
\begin{equation*}    2\left(\sqrt{R^2+\hbar_N^2 \lambda_{\max}^2}-\hbar_N \lambda_{\max} \right) \leq 2 n_H \hbar_N \lambda_{\max} \leq 2 (d_\lambda-1) \hbar_N \lambda_{\max} \ .
\end{equation*}
This establishes a lower-bound for the quantization parameter: 
\begin{equation*}
    \hbar_N
    \geq\frac{R}{\lambda_{\max} \sqrt{n_H(n_H+2)}}
    \geq\frac{R}{\lambda_{\max} \sqrt{d_\lambda^2-1}} \ ,
    \label{eq:hbar_oscillation_lower_bound}
\end{equation*}
thus for finite-$\hbar_N$ solution of non-constant $H(z)$ to exists we must have $n_H > 0$.

\section{Spectral Connectivity of Consecutive Eigenvalues \label{app:cons}}

Sandwiching $\hat{W}_N\hat{W}_N^\dagger$ between bra $\langle j_N \lvert$ and ket $\lvert k_N\rangle$, since $\hat{W}_N\hat{W}_N^\dagger$ is a function of $\hat{Z}_N$ as found in Eq. \eqref{eq:WW1} and  $|j_N\rangle$, $|k_N\rangle$ are eigenvectors of $\hat{Z}_N$, it follows that
\begin{equation}
    \langle j_N \lvert \hat{W}_N\hat{W}_N^\dagger \lvert k_N\rangle = 0 \ \forall j \neq k \ .
    \label{eq:go_to_0}
\end{equation}
We can use the completeness relation Eq. \eqref{eq:completeness} to insert an identity matrix between $\hat{W}_N$ and $\hat{W}_N^\dagger$ in Eq. \eqref{eq:go_to_0} to arrive at:
\begin{equation}
\begin{split}
    \langle j_N \lvert \hat{W}_N\hat{W}_N^\dagger \lvert k_N\rangle &= \langle j_N \lvert \hat{W}_N \hat{I}_{N \times N} \hat{W}_N^\dagger \lvert k_N\rangle = \langle j_N \lvert \hat{W}_N \sum^{N-1}_{\ell = 0} \lvert \ell_N \rangle \langle \ell_N \lvert \hat{W}_N^\dagger \lvert k_N\rangle
    \\
    &= \sum^{N-1}_{\ell = 0} \langle j_N \lvert \hat{W}_N \lvert \ell_N \rangle \langle \ell_N \lvert \hat{W}_N^\dagger \lvert k_N\rangle = \sum^{N-1}_{\ell = 0} W^{(j,\ell)}_N {W^{(k,\ell)}_N}^* = 0 \ \forall j \neq k \ .
\end{split}
\label{eq:WW_decomp}
\end{equation}

\underline{Suppose} there exists an index $m\neq j$, $m\neq k$, and $m\in{0,1,\ldots,N-1}$ such that both $W_N^{(j,m)}$ and $W_N^{(k,m)}$ are non-zero. Then $W_N^{(j,m)}{W_N^{(k,m)}}^*\neq 0$. However, since Eq. \eqref{eq:WW_decomp} requires the sum of these products to vanish for $j\neq k$, this non-zero contribution must be canceled by at least one other non-zero term. Therefore, there must exist an index $\tilde{m}\neq j$, $\tilde{m}\neq k$, and $\tilde m\neq m$ such that $W_N^{(j,\tilde m)}\neq 0$ and $W_N^{(k,\tilde m)}\neq 0$. From Eq. \eqref{eq:int}, for the matrix element $W^{(j,m)}\neq 0$:
\begin{equation}
    v_N^{(j)} - v_N^{(m)} = 2\hbar_N \bar{\lambda}(v^{(j)}_N,v^{(m)}_N)  \ .
    \label{eq:get_v}
\end{equation}
We can use the integral representation of the mean value $ \bar{\lambda}(v^{(j)}_N,v^{(m)}_N)$, i.e. 
$$ \bar{\lambda}(v^{(j)}_N,v^{(m)}_N) = \frac{\displaystyle \int^{v^{(j)}_N}_{v^{(m)}_N} dz \ \lambda(z)}{v^{(j)}_N - v^{(m)}_N} \ , $$
to rewrite Eq. \eqref{eq:get_v} as:
$$ \left( v^{(j)}_N - v^{(m)}_N \right)^2 = 2\hbar_N \int^{v^{(j)}_N}_{v^{(m)}_N} dz \ \lambda(z) \ . $$
Similarly, we can do the same for the matrix elements $W^{(k,m)}$, $W^{(j,\tilde{m})}$, and $W^{(k,\tilde{m})}$. Then, without those expressions, we can show that:
\begin{equation*}
    \begin{split}
        \left( v^{(j)}_N - v^{(m)}_N \right)^2 - \left( v^{(k)}_N - v^{(m)}_N \right)^2 &= 2\hbar_N \int^{v^{(j)}_N}_{v^{(m)}_N} dz \ \lambda(z) - 2\hbar_N \int^{v^{(k)}_N}_{v^{(m)}_N} dz \ \lambda(z) 
        \\
        &= 2\hbar_N \int^{v^{(j)}_N}_{v^{(k)}_N} dz \ \lambda(z) \ ,
        \\
        \left( v^{(j)}_N - v^{(\tilde{m})}_N \right)^2 - \left( v^{(k)}_N - v^{(\tilde{m})}_N \right)^2 &= 2\hbar_N \int^{v^{(j)}_N}_{v^{(\tilde{m})}_N} dz \ \lambda(z) - 2\hbar_N \int^{v^{(k)}_N}_{v^{(\tilde{m})}_N} dz \ \lambda(z) 
        \\
        &= 2\hbar_N \int^{v^{(j)}_N}_{v^{(k)}_N} dz \ \lambda(z) \ ,
    \end{split}
\end{equation*}
thus
$$ \left( v^{(j)}_N - v^{(m)}_N \right)^2 - \left( v^{(k)}_N - v^{(m)}_N \right)^2 = \left( v^{(j)}_N - v^{(\tilde{m})}_N \right)^2 - \left( v^{(k)}_N - v^{(\tilde{m})}_N \right)^2 \ . $$
Subtracting the LHS from the RHS gives:
$$ 2 \left( v^{(j)}_N - v^{(k)}_N \right) \left( v^{(m)}_N - v^{(\tilde{m})}_N \right) = 0 \ . $$
But this cannot be true if the spectral of $\hat{Z}_N$ is non-degenerate, as assumed by Eq. \eqref{eq:our_simp}. Therefore, our \underline{supposition} is wrong, thus 
\begin{equation}
    \text{for each index} \ \ \ell \ \ \text{there is at most one index} \ \ k \neq \ell \ \ \text{such that} \ \  W^{(k,\ell)}_N \neq 0 \ .
    \label{eq:k_to_l}
\end{equation}
Similarly, we can also arrive at
\begin{equation}
    \text{for each index} \ \ \ell \ \ \text{there is at most one index} \ \ k \neq \ell \ \ \text{such that} \ \  W^{(\ell,k)}_N \neq 0 \ .
    \label{eq:l_to_k}
\end{equation}
In terms of direct connection, this means for every eigenvalues $v^{(\ell)}_N$ (which are all distinct) there should be at most one outgoing direct connection from it (i.e. $v^{(\ell)}_N \dashrightarrow ... $) and at most one incoming direct connection to it (i.e. $... \dashrightarrow v^{(\ell)}_N$).

The irreducibility of the matrix triplet $(\hat{X}_N,\hat{Y}_N,\hat{Z}_N)$ implies that the only matrices $\hat{G}$ commuting with all of them: 
\begin{equation}
    [\hat{G},\hat{X}_N] =  [\hat{G},\hat{Y}_N] = [\hat{G},\hat{Z}_N] = 0
    \label{eq:reducible}
\end{equation}
are scalar multiples of the identity (i.e. $\hat{G} \propto \hat{I}_{N \times N}$). \underline{\underline{Suppose}} there exists an eigenvalue $v^{(j)}_N$ that is not the highest-eigenvalue (i.e. $j \neq N-1$) and have no outgoing direct connection from it. Following its incoming direct connections backward whenever possible must eventually terminate, since every direct connection points toward a higher eigenvalue and the spectrum is finite, yielding the following path:
\begin{equation*}
    v_N^{(\ell_0)}
    \dashrightarrow v_N^{(\ell_1)}
    \dashrightarrow\cdots
    \dashrightarrow v_N^{(\ell_K)} \ \ \text{where} \ \
    \ell_0<\ell_1< ... <\ell_K=j \ .
\end{equation*}
Then, it is possible to construct the following non-identity Hermitian matrix 
$$ \hat{G} \equiv \sum^{K}_{\mu=0} \lvert {\ell_\mu}_N \rangle \langle {\ell_\mu}_N \lvert $$
that satisfies Eq. \eqref{eq:reducible}:
\begin{itemize}
    \item $[\hat{G},\hat{Z}_N]=0$: This is equivalent to $\hat{G}\hat{Z}_N=\hat{Z}_N \hat{G}$, which can be shown by:
    \begin{equation*}
    \begin{split}
        \hat{G}\hat{Z}_N &= \sum^{K}_{\mu=0} \lvert {\ell_\mu}_N \rangle \langle {\ell_\mu}_N \lvert \hat{Z}_N = \sum^{K}_{\mu=0} \lvert {\ell_\mu}_N \rangle \langle {\ell_\mu}_N \lvert v^{(\ell_\mu)}_N 
        \\
        &= \sum^{K}_{\mu=0}  v^{(\ell_\mu)}_N \lvert {\ell_\mu}_N \rangle \langle {\ell_\mu}_N \lvert = \sum^{K}_{\mu=0}  \hat{Z}_N \lvert {\ell_\mu}_N \rangle \langle {\ell_\mu}_N \lvert = \hat{Z}_N \hat{G}_N \ .
    \end{split}
    \end{equation*}
    Here we have used that $\hat{Z}_N$ is a Hermitian matrix, therefore its eigenvalues must be purely real.
    \item $[\hat{G},\hat{X}_N]=[\hat{G},\hat{Y}_N]=0$: To prove this, we first need to show that $\hat{G}\hat{W}_N=\hat{W}_N \hat{G}$. From the uniqueness of the direct connection path, i.e. Eq. \eqref{eq:k_to_l} and Eq. \eqref{eq:l_to_k}, we have:
    $$ \hat{W}_N \lvert {\ell_\mu}_N \rangle = \begin{cases}
        W_N^{(\ell_{\mu+1},\ell_\mu)} \lvert {\ell_{\mu+1}}_N \rangle & \ \text{for} \ 0 \leq \mu < K
        \\
        0 & \ \text{for} \ \mu=K 
    \end{cases} \ , $$
    and 
    \begin{equation*}
    \begin{split}
        & \hat{W}_N^\dagger \lvert {\ell_\mu}_N \rangle  = \begin{cases}
        0 & \ \text{for} \ \mu=0
        \\
        \left( W_N^{(\ell_{\mu},\ell_{\mu-1})} \right)^* \lvert {\ell_{\mu-1}}_N \rangle & \ \text{for} \ 0 < \mu \leq K 
    \end{cases}
        \\[1em]
        \Longrightarrow \ & \langle {\ell_\mu}_N \lvert \hat{W}_N = \begin{cases}
        0 & \ \text{for} \ \mu=0
        \\
        \langle {\ell_{\mu-1}}_N \lvert  W_N^{(\ell_{\mu},\ell_{\mu-1})}  & \ \text{for} \ 0 < \mu \leq K 
    \end{cases} \ .
    \end{split}
    \end{equation*}
    Using these to represent $\hat{G}\hat{W}_N$ and $\hat{W}_N \hat{G}$, we obtain identical expressions, i.e.
    \begin{equation*}
    \begin{split}
        \hat{G} \hat{W}_N &= \sum^K_{\mu=0} \lvert {\ell_\mu}_N\rangle \langle {\ell_\mu}_N \lvert \hat{W}_N = \sum^{K}_{\mu=1}  \lvert {\ell_\mu}_N\rangle \langle {\ell_{\mu-1}}_N \lvert W_N^{(\ell_\mu,\ell_{\mu-1})}
        \\
        &=\sum^{K}_{\mu=1}  W_N^{(\ell_\mu,\ell_{\mu-1})} \lvert {\ell_\mu}_N\rangle \langle {\ell_{\mu-1}}_N \lvert
    \end{split}
    \end{equation*}
    and
    \begin{equation*}
    \begin{split}
        \hat{W}_N \hat{G}  &= \sum^K_{\mu=0}  \hat{W}_N \lvert {\ell_\mu}_N\rangle \langle {\ell_\mu}_N \lvert = \sum^{K-1}_{\mu=0}  W_N^{(\ell_{\mu+1},\ell_\mu)} \lvert {\ell_{\mu+1}}_N\rangle \langle {\ell_\mu}_N \lvert
        \\
        &=\sum^{K}_{\mu=1}  W_N^{(\ell_\mu,\ell_{\mu-1})} \lvert {\ell_\mu}_N\rangle \langle {\ell_{\mu-1}}_N \lvert \ ,
    \end{split}
    \end{equation*}
    thus they are equal. Given that $\hat{G}\hat{W}_N=\hat{W}_N \hat{G}$, we take the Hermitian adjoint on both sides, and since $\hat{G}^\dagger =\hat{G}$, we also get $\hat{G}\hat{W}^\dagger_N=\hat{W}_N^\dagger \hat{G}$. From Eq. \eqref{eq:W}, we have
    $$ \hat{X}_N =\frac{\displaystyle \hat{W}_N + \hat{W}_N^\dagger}2 \ \ \text{and} \ \ \hat{Y}_N =\frac{\displaystyle \hat{W}_N-\hat{W}_N^\dagger}{2i} \ , $$
    and since $\hat{G}$ commutes with both $\hat{W}_N$ and $\hat{W}_N^\dagger$, it also commutes with $\hat{X}_N$ and $\hat{Y}_N$.
\end{itemize}
Therefore, our \underline{\underline{supposition}} contradicts irreducibility. Similarly, an eigenvalue $v^{(j)}_N$ that is not the lowest-eigenvalue (i.e. $j \neq N-1$) and have no incoming direct connection from it should not exists. Together, these results imply that the direct connections form a single path passing through all eigenstates, with each connection linking consecutive eigenvalues:
$$v^{(0)}_N \dashrightarrow v^{(1)}_N \dashrightarrow ... \dashrightarrow v^{(N-1)}_N \ . $$

\section{Derivation of Eq. \eqref{eq:get_Weyl} \label{app:get_Weyl}}

Consider $\lambda(z)=z^{-2}$, then we can calculate the antiderivative $\Lambda(z)$ from solving $\Lambda'(z)=\lambda(z)$, which gives $\Lambda(z) =  z^{-1}$ (up to an additional constant). Let us define an inverse matrix:
\begin{equation}
    \hat{M}(t) = \left( \hat{Z}_N + t \hat{A} \right)^{-1} \ ,
    \label{eq:M_def}
\end{equation}
then from the definition of matrix inversion:
$$ \left( \hat{Z}_N + t \hat{A} \right) \hat{M}(t) = \hat{I}_{N \times N} \ . $$
Differentiate this expression, i.e. apply $d/dt$ on it, gives:
$$\hat{A} \hat{M}(t) + \left( \hat{Z}_N + t \hat{A} \right) \hat{M}'(t) = 0 \ . $$
At $t=0$, since $\hat{M}(0) = \hat{Z}_N^{-1}$ from Eq. \eqref{eq:M_def}, we obtain:
$$ \hat{A} \hat{M}(0) + \left( \hat{Z}_N + t \hat{A} \right) \hat{M}'(0) =  \hat{A} \hat{Z}_N^{-1} + \hat{Z}_N \hat{M}'(0) = 0 \ . $$
Multiply both sides from the left by $\hat{Z}_N^{-1}$, we arrive at:
\begin{equation}
    \hat{Z}_N^{-1} \hat{A} \hat{Z}_N^{-1} + \hat{M}'(0) = 0 \ \Longrightarrow \  \hat{M}'(0) = - \hat{Z}_N^{-1} \hat{A} \hat{Z}_N^{-1} \ .
    \label{eq:derM_def}
\end{equation}
Finally, from the extension formula Eq. \eqref{eq:Weyl_def} of Weyl's symmetric ordering:
\begin{equation*}
\begin{split}
    S_{z^{-2}}[\hat{A}] 
    &= \frac{d}{dt} \Lambda(\hat{Z}_N + t \hat{A})\Big|_{t=0} = \frac{d}{dt} (\hat{Z}_N + t \hat{A})^{-1} \Big|_{t=0} = -\hat{M}'(0) = \hat{Z}_N^{-1} \hat{A} \hat{Z}_N^{-1} \ ,
\end{split}
\end{equation*}
where in the last transformation we have used Eq. \eqref{eq:derM_def}. Note that this expression is much simpler than the naive choice i.e. treating $\hat{Z}_N^{-2}$ just as $\hat{Z}^2_N$:
$$ S_{z^2}[\hat{A}] = \frac{\hat{Z}_N^2 \hat{A} + \hat{Z}_N \hat{A} \hat{Z}_N + \hat{A} \hat{Z}_N^2}3 \ \Longrightarrow \ S_{z^{-2}}[\hat{A}] \overset{?}{=} \frac{\hat{Z}_N^{-2} \hat{A} + \hat{Z}^{-1}_N \hat{A} \hat{Z}^{-1}_N + \hat{A} \hat{Z}^{-2}_N}3 \ . $$
Here we use $\overset{?}{=}$ to specify the naive choice.

\section{Theory for Fuzzy Sphere with Non-Polynomial Poisson Multiplier \label{app:big_theory}}

We begin by collecting several useful algebraic relations:
\begin{itemize}
    \item The antiderivative $\Lambda(z)$ of the $\lambda(z)$ in Eq. \eqref{eq:lambda_sol} can be calculated -- upon setting the arbitrary additive constant to $0$ -- as:
    \begin{equation}
        \Lambda'(z) \equiv \lambda(z) = \lambda_0 + \lambda_{-2} z^{-2} \ \Longrightarrow \ \Lambda(z) = \lambda_0 z - \lambda_{-2} z^{-1} \ .
        \label{eq:Lambda_nonpol}
    \end{equation}
    \item Invertibility of $\hat{Z}_N$ implies that all its eigenvalues $\{ v^{(j)}_N\}^{N-1}_0$ are non-zero, i.e.
    \begin{equation}
        v^{(j)}_N \neq 0 \ \forall j \ .
        \label{eq:spec_nonzero}
    \end{equation}
    \item Define
    \begin{equation}
        \Theta_N (z) \equiv R^2 - z^2 + 2\hbar_N \left( \lambda_0 z + \lambda_{-2} z^{-1} \right) \ ,
        \label{eq:theta}
    \end{equation}
    then 
    Eq. \eqref{eq:WW} can be expressed explicitly as:
    \begin{equation}
        \hat{W}_N \hat{W}_N^\dagger = \Theta_N(\hat{Z}_N) \ \ \text{and} \ \
        \hat{W}_N^\dagger \hat{W}_N = \Theta_N(-\hat{Z}_N) \ .
    \label{eq:WW_nonpol}
    \end{equation}
    Consequently, the orthonormal eigenvector basis $\{\lvert j_N\rangle\}_{j=0}^{N-1}$ of $\hat{Z}_N$ is also an orthonormal eigenvector basis for both $\hat{W}_N\hat{W}_N^\dagger$ and $\hat{W}_N^\dagger\hat{W}_N$, with respective spectra
$$
\left\{\Theta_N(v_N^{(j)})\right\}_{j=0}^{N-1} \ \ \text{and} \ \
\left\{\Theta_N(-v_N^{(j)})\right\}_{j=0}^{N-1} \ .
$$
\item Using Eq. \eqref{eq:get_Weyl}, Eq. \eqref{eq:gen_Poisson_axial_mat_quant_X} can be expressed as:
\begin{equation}
    [\hat{Z}_N,\hat{W}_N] = 2\hbar_N \left( \lambda_0 \hat{W}_N + \lambda_{-2} \hat{Z}_N^{-1} \hat{W}_N \hat{Z}_N^{-1} \right) \ .
    \label{eq:ZW_comm_nonpol}
\end{equation} 
\end{itemize}

\subsection{Allowed Matrix Dimensionality \label{app:big_theory_odd}}

We can further show that the spectra of $\hat{W}_N\hat{W}_N^\dagger$ and $\hat{W}_N^\dagger\hat{W}_N$ are non-negative, identical, and can be mapped one-to-one:
\begin{itemize}
    \item Sandwiching $\hat{W}_N\hat{W}_N^\dagger$ and $\hat{W}_N^\dagger\hat{W}_N$ between bra $\langle j_N \lvert$ and ket $\lvert j_N \rangle$, we obtain expressions for the squared Hilbert–Schmidt norms, which are necessarily nonnegative:
    $$
        \langle j_N \lvert \hat{W}_N\hat{W}_N^\dagger \lvert j_N \rangle = \| \hat{W}_N^\dagger \lvert j_N \rangle \|^2 \geq 0 \ , \ \langle j_N \lvert \hat{W}_N^\dagger \hat{W}_N \lvert j_N \rangle = \| \hat{W}_N \lvert j_N \rangle \|^2 \geq 0 \ .
    $$
    On the other hand, using Eq. \eqref{eq:WW_nonpol}, these expressions are equal to $\Theta_N(v_N^{(j)})$ and $\Theta_N(-v_N^{(j)})$, respectively. Thus, $\Theta_N(v_N^{(j)})$ and $\Theta_N(-v_N^{(j)})$ must be non-negative, hence the spectra of $\hat{W}_N\hat{W}_N^\dagger$ and $\hat{W}_N^\dagger\hat{W}_N$ are non-negative.
    \item For any positive eigenvalue, e.g. $\Theta_N(v_N^{(j)}) > 0$:
    $$ \hat{W}_N\hat{W}_N^\dagger \lvert j_N \rangle = \Theta_N(v_N^{(j)}) \lvert j_N \rangle \ , $$
    then $\hat{W}_N^\dagger \lvert j_N \rangle \neq 0$ (else the LHS equal to $0$, which contradicts with the RHS). Apply both sides with $\hat{W}_N^\dagger$, we get: 
    $$ \hat{W}_N^\dagger \hat{W}_N\hat{W}_N^\dagger \lvert j_N \rangle = \hat{W}_N^\dagger \Theta_N(v_N^{(j)}) \lvert j_N \rangle \ \Longrightarrow \ \hat{W}_N^\dagger \hat{W}_N \left( \hat{W}_N^\dagger \lvert j_N \rangle \right) = \Theta_N(v_N^{(j)}) \left( \hat{W}_N^\dagger \lvert j_N \rangle \right) \ . $$
    Thus, $\Theta_N(v_N^{(j)})$ should also be an eigenvalue of $\hat{W}_N^\dagger \hat{W}_N$, with the corresponding non-vanishing eigenvector $\hat{W}_N^\dagger \lvert j_N \rangle$. The map $\lvert j_N \rangle \rightarrow \hat{W}_N^\dagger \lvert j_N \rangle$ is invertible, i.e. we can uniquely recover $\lvert j_N \rangle$ from $\hat{W}_N^\dagger \lvert j_N \rangle$ as:
    $$ \frac{\hat{W}_N \left( \hat{W}_N^\dagger \lvert j_N \rangle \right)}{\Theta_N(v_N^{(j)})} = \lvert j_N \rangle \ , $$
    therefore we can map all positive eigenvalues of $\hat{W}_N\hat{W}_N^\dagger$ and $\hat{W}_N^\dagger\hat{W}_N$ one-to-one.
    \item Since the positive eigenvalues of $\hat{W}_N\hat{W}_N^\dagger$ and $\hat{W}_N^\dagger\hat{W}_N$ can be paired one-to-one, their spectra must also contain the same number of zero eigenvalues. Thus, their zero eigenvalues can likewise be paired one-to-one, establishing that a one-to-one map between identical eigenvalues through out their entire spectra always exists.
\end{itemize}

Now let us look at the spectrum $\{ v_N^{(j)} \}^{N-1}_{j=0}$ of $\hat{Z}_N$ and remove all pairs of eigenvalues with opposite signs. \underline{Assume} that we are left with a non-empty set:
\begin{equation}
    \{ v_N^{(\ell_k)} \}^{K}_{k=0} \ \ \text{where} \ v_N^{(\ell_k)} + v_N^{(\ell_{\tilde{k}})}\neq 0 \ \forall k \neq \tilde{k} \ .
    \label{eq:no_pair_left}
\end{equation}
Note that this removal procedure preserves the property that every element in $\{ \Theta(v_N^{(\ell_k)}) \}^{K}_{k=0}$ remains equal to at least one element in $\{ \Theta(-v_N^{(\ell_k)}) \}^{K}_{k=0}$. Order this set of remained eigenvalues so that
\begin{equation}
    \Lambda(v^{(\ell_0)}) \leq \Lambda(v^{(\ell_1)}) \leq ... \leq \Lambda(v^{(\ell_K)}) \ ,
    \label{eq:order_of_the_phoenix}
\end{equation}
then we select an element $\Theta(-v_N^{(\ell_{\tilde{0}})}) \in \{ \Theta(-v_N^{(\ell_k)}) \}^{K}_{k=0}$ such that:
$$ \Theta(v_N^{(\ell_0)}) = \Theta(-v_N^{(\ell_{\tilde{0}})}) \ ; $$
using Eq. \eqref{eq:theta} gives:
\begin{equation}
    \begin{split}
        0 &= \Theta(v_N^{(\ell_0)}) -  \Theta(-v_N^{(\ell_{\tilde{0}})}) \\
        &= -(v_N^{(\ell_0)})^2 + (v_N^{(\ell_{\tilde{0}})})^2 + 2\hbar_N \left[ \lambda_0 \left( v_N^{(\ell_0)} + v_N^{(\ell_{\tilde{0}})} \right) + \lambda_{-2} \left( \frac1{v_N^{(\ell_0)}} + \frac1{v_N^{(\ell_{\tilde{0}})}} \right) \right]
        \\
        &= \left( v_N^{(\ell_0)} + v_N^{(\ell_{\tilde{0}})} \right) \left[ -\left( v_N^{(\ell_0)} -v_N^{(\ell_{\tilde{0}})} \right) + 2\hbar_N \left( \lambda_0 + \frac{\lambda_{-2}}{v_N^{(\ell_0)} v_N^{(\ell_{\tilde{0}})}} \right) \right] \ .
    \end{split}
\end{equation}
Following from Eq. \eqref{eq:no_pair_left} that $v_N^{(\ell_0)} + v_N^{(\ell_{\tilde{0}})} \neq 0$, this implies
\begin{equation}
    0 =  -\left( v_N^{(\ell_0)} -v_N^{(\ell_{\tilde{0}})} \right) + 2\hbar_N \left( \lambda_0 + \frac{\lambda_{-2}}{v_N^{(\ell_0)} v_N^{(\ell_{\tilde{0}})}} \right) \ \Longrightarrow \ \frac{v_N^{(\ell_0)} -v_N^{(\ell_{\tilde{0}})}}{2\hbar_N} = \lambda_0 + \frac{\lambda_{-2}}{v_N^{(\ell_0)} v_N^{(\ell_{\tilde{0}})}}  \ . 
    \label{eq:equality}
\end{equation} 
If $v_N^{(\ell_0)} = v_N^{(\ell_{\tilde{0}})}$, then:
\begin{itemize}
    \item The LHS of Eq. \eqref{eq:equality}: Vanish, i.e. exactly equal to $0$,
    \item The RHS of Eq. \eqref{eq:equality}: Using Eq. \eqref{eq:lambda_sol}, we can rewrite it as
    $$\lambda_0 + \frac{\lambda_{-2}}{(v_N^{(\ell_0)})^2} = \lambda(v_N^{(\ell_0)} ) > 0 \ , $$
\end{itemize}
leading to a contradiction. Thus:
\begin{equation}
    v_N^{(\ell_0)} \neq v_N^{(\ell_{\tilde{0}})}
    \label{eq:diff_v} \ .
\end{equation}

Multiplying both sides of Eq. \eqref{eq:equality} with $v_N^{(\ell_0)} -v_N^{(\ell_{\tilde{0}})}$ gives:
\begin{equation}
    \begin{split}
        \frac{\left(v_N^{(\ell_0)} -v_N^{(\ell_{\tilde{0}})} \right)^2}{2\hbar_N} &= \left( v_N^{(\ell_0)} -v_N^{(\ell_{\tilde{0}})} \right) \left( \lambda_0 + \frac{\lambda_{-2}}{v_N^{(\ell_0)} v_N^{(\ell_{\tilde{0}})}} \right) 
        \\
        &= \lambda_0 \left(v_N^{(\ell_0)} -v_N^{(\ell_{\tilde{0}})} \right) - \lambda_{-2} \left(\frac1{v_N^{(\ell_0)}} - \frac1{v_N^{(\ell_{\tilde{0}})}} \right)
        \\
        &= \Lambda(v_N^{(\ell_0)}) - \Lambda(v_N^{(\ell_{\tilde{0}})}) \ ,
    \end{split}
    \label{eq:contradictory}
\end{equation}
where we have used Eq. \eqref{eq:Lambda_nonpol}. Here, we can show a contradiction:
\begin{itemize}
    \item The LHS of Eq. \eqref{eq:contradictory}: Given Eq. \eqref{eq:diff_v}, the expression here must be positive,
    \item The RHS of Eq. \eqref{eq:contradictory}: Following the ordering in Eq. \eqref{eq:order_of_the_phoenix}, the expression here must be non-positive,
\end{itemize}
therefore the \underline{assumption} is wrong, and the set in Eq. \eqref{eq:no_pair_left} must be an empty set. Given that the removal procedure always goes in pairs, therefore the number of eigenvalues -- which is equal to the matrix dimension $N$ -- can only be an even number, i.e. $N \in 2\mathbb{N}_{>0}$.

\subsection{Criterion for Solutions to Exist \label{app:big_theory_bound}}

In Appendix \ref{app:big_theory_odd} we have shown that the spectrum of $\hat{Z}_N$ is non-zero and always come in pairs of opposite sign. Thus, there must be negative and positive eigenvalues. Select the smallest positive eigenvalue, e.g. say $0 < v^{(j_+)}_N \leq R$. Following the first-line of Eq. \eqref{eq:norm_Wstate}:
\begin{equation}
    \| \hat{W}^\dagger_N \lvert {j_+}_N \rangle \|^2 = \left( R^2 - (v^{(j_+)}_N )^2 \right) + 2 \hbar_N v^{(j_+)}_N \lambda(v^{(j_+)}_N) \ .
    \label{eq:Wnorm_now}
\end{equation}
In this equation, the RHS cannot be $0$, thus $\| \hat{W}^\dagger_N \lvert {j_+}_N \rangle \|^2 > 0$, and we can then follow Eq. \eqref{eq:Wstate_in_basis} to arrive at:
$$ \hat{W}^\dagger_N \lvert {j_+}_N \rangle = \sum^{N-1}_{j=0} {W_N^{(j_+,j)}}^* \lvert j_N \rangle > 0 \ . $$
In other words, there exists at least one direct connection going to $v^{(j_+)}_N$, i.e. $... \dashrightarrow v^{(j_+)}_N $.

Consider one such direct connection, e.g. say $v^{(j_-)}_N \dashrightarrow v^{(j_+)}_N$. Sandwiching both sides of Eq. \eqref{eq:ZW_comm_nonpol} between bra $\langle {j_+}_N \lvert$ and ket $\lvert {j_-}_N \rangle$:
\begin{equation}
    \begin{split}
    \langle {j_+}_N \lvert [\hat{Z}_N,\hat{W}_N] \lvert {j_-}_N \rangle = 2\hbar_N \left( \lambda_0 \langle {j_+}_N \lvert \hat{W}_N \lvert {j_-}_N \rangle + \lambda_{-2} \langle {j_+}_N \lvert \hat{Z}_N^{-1} \hat{W}_N \hat{Z}_N^{-1} \lvert {j_-}_N \rangle \right) &
    \\
    \Longrightarrow \ \left( v^{(j_+)}_N - v^{(j_-)}_N  \right) W^{(j_+,j_-)}_N = 2\hbar_N \left( \lambda_0 + \frac{\lambda_{-2}}{v^{(j_+)}_N v^{(j_-)}_N} \right) W^{(j_+,j_-)}_N &
    \\
    \Longrightarrow \ v^{(j_+)}_N - v^{(j_-)}_N = 2\hbar_N \left( \lambda_0 + \frac{\lambda_{-2}}{v^{(j_+)}_N v^{(j_-)}_N} \right) & \ .
    \end{split}
    \label{eq:matrix_rel}
\end{equation} 
If $v^{(j_-)}_N>0$, then the RHS should be positive, and LHS therefore implies $v^{(j_+)}_N>v^{(j_-)}_N$. This contradicts with $v^{(j_+)}_N$ being the smallest positive eigenvalues of $\hat{Z}_N$, hence 
$$v^{(j_-)}_N<0 \ \ \text{and} \ \ v^{(j_-)}_N = -|v^{(j_-)}_N| \ . $$
Define
$$\eta = \frac{\lambda_{-2}}{\lambda_0} \ \ \text{and} \ \rho = v^{(j_+)}_N |v^{(j_-)}_N| \ , $$
then Eq. \eqref{eq:matrix_rel} can be rewritten as:
\begin{equation}
    \left( \frac{\rho}{\rho - \eta} \right) \left(v^{(j_+)}_N + |v^{(j_-)}_N| \right) = 2 \hbar_N \lambda_0 \ ,
    \label{eq:equality_to_use}
\end{equation}
which implies $\rho > \eta$.

To proceed, we need to constraints $v^{(j_+)}_N$ and $|v^{(j_-)}_N|$ further; one of such direction is through the non-negativity of following the Hilbert-Schmidt norms. Start by using Eq. \eqref{eq:norm_Wstate} to obtain the following inequalities:
\begin{equation*}
\begin{split}
    \| \hat{W}_N \lvert {j_+}_N \rangle \|^2 &=  R^2 - (v^{(j_+)}_N )^2 - 2 \hbar_N v^{(j_+)}_N \lambda(v^{(j_+)}_N) \geq 0 \ ,
    \\
    \| \hat{W}^\dagger_N \lvert {j_-}_N \rangle \|^2 &=  R^2 - |v^{(j_-)}_N |^2 - 2 \hbar_N |v^{(j_-)}_N| \lambda(|v^{(j_-)}_N|) \geq 0 \ .
\end{split}    
\end{equation*}
Rearranging these inequalities and then writing $\lambda(z)$ explicitly gives:
\begin{equation*}
\begin{split}
    R^2 &\geq (v^{(j_+)}_N )^2 + 2 \hbar_N  \left( \lambda_0 v^{(j_+)}_N + \frac{\lambda_{-2}}{v^{(j_+)}_N} \right) \ ,
    \\
     R^2 &\geq |v^{(j_-)}_N |^2 + 2 \hbar_N \left( \lambda_0 |v^{(j_-)}_N| + \frac{\lambda_{-2}}{|v^{(j_-)}_N|} \right) \ .
\end{split}    
\end{equation*}
Adding the LHSs and RHSs of these inequalities yields
\begin{equation*}
\begin{split}
    2R^2 &\geq (v^{(j_+)}_N)^2+ |v^{(j_-)}_N |^2 + 2\hbar_N \lambda_0 \left[ v^{(j_+)}_N + |v^{(j_-)}_N | + \eta \left( \frac1{v^{(j_+)}_N } + \frac1{|v^{(j_-)}_N |} \right) \right]
    \\
    &\geq (v^{(j_+)}_N)^2+ |v^{(j_-)}_N |^2 + 2\hbar_N \lambda_0 \left( v^{(j_+)}_N + |v^{(j_-)}_N | \right) \left( \frac{\rho+\eta}{\rho} \right)
\end{split}    
\end{equation*}
We can use Eq. \eqref{eq:equality_to_use} to replace $ 2 \hbar_N \lambda_0$ and obtain:
$$
    2R^2 \geq (v^{(j_+)}_N)^2+ |v^{(j_-)}_N |^2 + \left( v^{(j_+)}_N + |v^{(j_-)}_N | \right)^2 \left( \frac{\rho+\eta}{\rho - \eta} \right) \ . $$
Note that $(v^{(j_+)}_N)^2+ |v^{(j_-)}_N |^2 = \left( v^{(j_+)}_N + |v^{(j_-)}_N | \right)^2 - 2\rho$, therefore this inequality becomes:
$$
    2R^2 \geq \left( v^{(j_+)}_N + |v^{(j_-)}_N | \right)^2 - 2\rho+ \left( v^{(j_+)}_N + |v^{(j_-)}_N | \right)^2 \left( \frac{\rho+\eta}{\rho - \eta} \right) = \left( v^{(j_+)}_N + |v^{(j_-)}_N | \right)^2 \left( \frac{2\rho}{\rho-\eta} \right) -2\rho \ . $$
We use the inequality of arithmetic and geometric means (also known as the AM-GM inequality) \cite{bullen2013handbook} to get 
\begin{equation}
    \left( v^{(j_+)}_N + |v^{(j_-)}_N | \right)^2 \geq 4\rho \ ,
    \label{eq:AM_GM}
\end{equation}
which allow us to simplify the inequality even further:
$$
    2R^2 \geq 4\rho \left( \frac{2\rho}{\rho-\eta} \right) -2\rho \ \Longrightarrow \ R^2 \geq \frac{4\rho^2}{\rho-\eta} - \rho \ . $$
Since 
$$ \min_{\rho > \eta} \left( \frac{4\rho^2}{\rho-\eta} - \rho \right) = \left( 7 + 4\sqrt{3} \right) \eta \ , $$
we arrive at the upper bound for $\eta$:
$$ R^2 \geq \left( 7 + 4\sqrt{3} \right) \eta \ \Longrightarrow \ \eta \leq \frac{R^2}{7 + 4\sqrt{3}} \ . $$

\subsection{No Classical Limit \label{app:quant_param}}

We can show that the quantization parameter $\hbar_N$ is bound from below, by using the AM-GM inequality Eq. \eqref{eq:AM_GM}, i.e. 
$$ v^{(j_+)}_N + |v^{(j_-)}_N | \geq 2\sqrt{\rho} \ , $$
in Eq. \eqref{eq:equality_to_use} to obtain:
$$\left( \frac{\rho}{\rho - \eta} \right) \left( 2\sqrt{\rho} \right) \leq 2 \hbar_N \lambda_0 \ \Longrightarrow \ \hbar_N \geq \frac1{\lambda_0} \left( \frac{\rho \sqrt{\rho}}{\rho-\eta} \right) \ . $$
Since 
$$ \min_{\rho > \eta} \left( \frac{\rho \sqrt{\rho}}{\rho-\eta} \right) = \left( \frac{3\sqrt{3}}2 \right) \sqrt{\eta} \ , $$
we arrive at the lower bound for $\hbar_N$ at any matrix dimension $N$:
$$ \hbar_N \geq \frac1{\lambda_0} \left( \frac{3\sqrt{3}}2 \right) \sqrt{\eta} = \left( \frac{3\sqrt{3}}2 \right) \frac1{\lambda_0} \sqrt{\frac{\lambda_{-2}}{\lambda_0}} \ . $$

\section{Additional Solutions for Beyond Polynomial Poisson Multiplier \label{app:beyondbeyond}}

To explore solutions that deviate substantially from the conventional fuzzy sphere, we must modify the initializations of the Hermitian matrix triplets $(\hat{X}_N, \hat{Y}_N,\hat{Z}_N)$. Guided by the theoretical analysis in Appendix \ref{app:big_theory}, we construct our initializations based on the predicted algebraic structure, rather than relying on arbitrary perturbations of the conventional fuzzy-sphere matrices.

Using Eq. \eqref{eq:Lambda_nonpol}, the average value of $\bar{\lambda}(z)$ in Eq. \eqref{eq:int} can be calculated to give:
\begin{equation}
  \bar{\lambda}\big(v_N^{(j)} v_N^{(k)}\big)=\frac{\Lambda(v_N^{(j)})-\Lambda(v_N^{(k)})}{v_N^{(j)}-v_N^{(k)}}=\lambda_0+\frac{\lambda_{-2}}{v_N^{(j)}v_N^{(k)}} \ ,
  \label{eq:lbar}
\end{equation}
which is negative whenever the two eigenvalues have opposite signs and satisfy:
\begin{equation}
    |v_N^{(j)}v_N^{(k)}|<\eta \ .
  \label{eq:connection_except}
\end{equation}
Thus Eq. \eqref{eq:no_back} no longer holds for every pair of eigenvalues; it is possible that a higher eigenvalue may have a direct connection to a lower one, provided that the connection crosses through $0$ (i.e. between opposite sign eigenvalues) and obeys the inequality Eq. \eqref{eq:connection_except}.

Let us order these eigenvalues. We know that the non-zero elements of $\hat{W}_n$ link the $N$ eigenvalues into a single connected sequence
\begin{equation}
  p_0\dashrightarrow p_1\dashrightarrow\dots\dashrightarrow p_{N-1} \,,
  \label{eq:sequence}
\end{equation}
in which $\{p_k\}_{k=0}^{N-1}$ is the spectrum of $\hat{Z}_N$ in the order visited by $\hat{W}_N$. The conventional fuzzy sphere is the strictly increasing sequence ($p_i = v^{(i)}_N$). Importantly, the optimization algorithm we use to learn the matrices representing the fuzzy sphere keeps the connected sequence of its initialization, so the initialization determines which solution can be found. Thus, using the theory laid out by Appendix \ref{app:big_theory}, we can therefore initialize other solutions that also solve the fuzzy sphere.

For the Poisson multiplier $\lambda(z) = \lambda_0 + \lambda_{-2}z^{-2}$, all solutions with a single connected sequence can be listed exactly. By
Eq. \eqref{eq:WW_nonpol}, the first element of the sequence has no incoming connection and the last has no outgoing one, i.e.
$\hat{W}_N^\dagger\ket{p_0}=0$ and $\hat{W}_N\ket{p_{N-1}}=0$, while every pair of consecutive elements satisfies Eq.~\eqref{eq:int} with
Eq.~\eqref{eq:lbar}:
\begin{equation}
  \Theta_N(p_0)=0 \,,\qquad p_{k+1}-p_k=2\hbar_N\left(\lambda_0+\frac{\lambda_{-2}}{p_k\,p_{k+1}}\right) ,\qquad
  \Theta_N(-p_{N-1})=0 \,.
  \label{eq:chain}
\end{equation}
For a given $\hbar_N$, the first condition is a cubic equation for $p_0$ and each connection a quadratic equation for $p_{k+1}$; the
last condition then fixes $\hbar_N$. With the pairing $p_{N-1-k}=-p_k$ of Appendix \ref{app:big_theory}, the connected sequences can be categorized by the
number of times they cross $z=0$, i.e. $1,3,\dots,N-1$\footnote{Note that the total number of crossings must be odd because we have an even number of dimensions $N,$ and so there are $N-1$ connections between eigenvalues, where our first eigenvalue is below $0$ and our last is above.}; once is the traditional solution. For $\eta\to0$ each of these $N/2$ sequences admits two solutions, so that
there are $N$ solutions at every even $N$.

Each solution of Eq.~\eqref{eq:chain} is a sequence $p_0,\dots,p_{N-1}$ together with its $\hbar_N$. Unlike the optimizations in Section~III\,A, here the optimizations are initialized independently, without using the final results of preceding runs: for each even matrix dimension $N$ and each solution of Eq.~\eqref{eq:chain}, we initialize $\hbar_N^{(ini)}=\hbar_N$, $\hat{Z}_N^{(ini)}$ as the diagonal matrix of the sequence in ascending order, and connect the $N$ eigenvalues in the order of the sequence,
\begin{equation}
  \begin{split}
  \hat{X}_N^{(ini)}&=\frac{\hat{W}_N^{(ini)}+\hat{W}_N^{(ini)\dagger}}{2} \quad\text{and}\quad
  \hat{Y}_N^{(ini)}=\frac{\hat{W}_N^{(ini)}-\hat{W}_N^{(ini)\dagger}}{2i} \\
  \text{with}\quad
  \hat{W}_N^{(ini)}&=\sum_{k=0}^{N-2} w_k\,\ket{p_{k+1}}\bra{p_k}
  \quad\text{and}\quad
  w_k=\sqrt{\Theta_N\big(p_{k+1}\big)} \,,
  \end{split}
  \label{eq:XYext}
\end{equation}
where $\ket{p}$ is the eigenvector of $\hat{Z}_N^{(ini)}$ with eigenvalue $p$ and $\Theta_N$ is Eq.~(J3) evaluated at $\hbar_N$, so that the weights $w_k$ are those that Eq.~\eqref{eq:WW_nonpol} assigns to the connections. The minimization algorithm, on the other hand, remains unchanged: we use the loss weights $c_{ac}=1$ and $c_{irr}=10^{-18}$, keep $\hat{Z}_N$ diagonal, and run 250 training steps per optimization, including the initial evaluation.

We are able to find every solution of Eq.~\eqref{eq:chain}: for even matrix dimensions $N=2,\dots,24$, the 156 solutions at $\eta=10^{-4}$ and the 26 solutions at $\eta=10^{-2}$ are each reached by the training (see Fig.~\ref{fig03}).

\section{Fuzzy Supersphere \label{app:superman}}

\subsection{Poisson Structures on Supermanifold}

A supermanifold has bosonic coordinates $r_j$ (i.e. even $\mathbb{Z}_2$-parity) and fermionic coordinates
$\theta_\alpha$ (i.e. odd $\mathbb{Z}_2$-parity), understood as even and odd elements of a Grassmann algebra, respectively. The fermionic coordinates satisfy $\theta_\alpha\theta_\beta=-\theta_\beta\theta_\alpha$, whereas the bosonic coordinates commute with all coordinates. Note that bosonic does not imply purely real-valued: the coordinates $r_j$ may contain even nilpotent contributions\footnote{An element is called nilpotent if some positive integer power of it vanishes.}, e.g. terms proportional to $\theta_1\theta_2$ in addition to their ordinary real parts.  Define \(|...|\) as the $\mathbb{Z}_2$-parity of a function, taking the value $0$ for bosonic functions and $1$ for fermionic functions, then on a supermanifold, a Poisson bracket satisfies \underline{linearity} and the following properties, which generalizes those mentioned in Appendix \eqref{app:Poi_Alg}:
\begin{itemize}
    \item \underline{Graded anti-symmetry:} $\{f,g\}=-(-1)^{|f||g|}\{g,f\}$,
    \item \underline{Graded Leibniz rule:} 
    $\{f,gh\}=\{f,g\}h+(-1)^{|f||g|}g\{f,h\}$,
    \item \underline{Graded Jacobi identity:} $$(-1)^{|f||h|}\{f,\{g,h\}\} + (-1)^{|g||f|}\{g,\{h,f\}\} + (-1)^{|h||g|}\{h,\{f,g\}\}= 0 \ , $$
\end{itemize}
where $f$, $g$, and $h$ are homogeneous functions, meaning that each has a definite $\mathbb{Z}_2$-parity (either purely bosonic or fermionic).

Consider a classical supersphere \cite{grosse1998fuzzy} that has 2 bosonic dimensions and 2 fermionic dimensions, embedded in a superspace with 3 bosonic coordinates and 2 fermionic coordinates, i.e. $\mathbb{R}^{3|2}$. This supersurface is described by the following Casimir constraints:
\begin{equation}
    F(\vec{r},\{ \theta_\mu\}^2_{\mu=1}) = \sum^3_{j=1} r_j^2 + \left( \theta_1  \theta_2 - \theta_2 \theta_1 \right) - R^2 = 0  \ ,
    \label{eq:SUSY_Cas}
\end{equation}
where $\vec{r}= (r_1,r_2,r_3)$. This constraint is an equality in a Grassmann algebra, requiring both its ordinary and nilpotent parts to vanish. Ignoring all nilpotent contributions recovers the ordinary sphere, i.e. $$ \sum_{j=1}^{3} \left( r_j\big|_0 \right)^2=R^2 \ , $$ 
where $r_j \big|_0$ denotes the ordinary part of $r_j$ \cite{bertolini2012super}.

Similar to Appendix \ref{app:Poi_Cas}, requiring the Casimir constraint to satisfy $\{F,r_j\}=\{F,\theta_\mu\}=0$ yields the most general Poisson brackets between supercoordinates to be of the form:
\begin{equation}
\begin{split}
\{r_j,r_k\}&=2\lambda(\vec r) \sum^3_{\ell=1} \epsilon_{jk\ell }r _\ell
                 +B_{ij}(\vec r)\theta_1\theta_2 \ ,\\
\{r_j,\theta_\mu\}&=\sum^2_{\nu=1} {C_j}_{\mu \nu}(\vec r) \theta_\nu \ ,\\
\{\theta_\mu,\theta_\nu\}
&=D_{\mu \nu}(\vec r)
                 +E_{\mu \nu}(\vec r)\theta_1 \theta_2 \ ,
\end{split}
\label{eq:super_general}
\end{equation}
where $\epsilon_{jk\ell}$ is a three-dimensional Levi-Civita symbol \cite{levi1977absolute} and we have the freedom to choose:
\begin{itemize}
    \item A smooth function $\lambda(\vec r)\ne0$.
    \item A smooth symmetric and invertible matrix $\hat{D}$, i.e.
    \begin{equation}
        \hat{D}(\vec r)=
    \begin{pmatrix}
        D_{11}(\vec r)&D_{12}(\vec r)\\
        D_{21}(\vec r)&D_{22}(\vec r)
    \end{pmatrix} \ \ \text{with} \
    \ D_{12}(\vec{r})=D_{21}(\vec{r})
    \label{eq:def_D}
    \end{equation}
    and
    $$\det\big(\hat{D}(\vec{r})\big) \equiv D_{11}(\vec r) D_{22}(\vec r) - D_{12}(\vec r) D_{21}(\vec r) \neq 0 \ . $$
    \item A smooth vector field $\vec u(\vec r)$
    tangent to the spheres, i.e.
    $\sum^3_{j=1} r_j u_j =0$.
\end{itemize}
Geometrically, this freedom corresponds to the symplectic
structure on the ordinary sphere, a nondegenerate symmetric
metric in the odd directions, and a compatible connection,
as in Rothstein's description of supersymplectic
supermanifolds \cite{rothstein2005structure}. 

Other functions in Eq. \eqref{eq:super_general}, i.e. $B_{jk}(\vec{r})$, ${C_j}_{\mu\nu}(\vec{r})$, and $E_{\mu\nu}(\vec{r})$, can be determined from the chosen freedom. For the sake of completeness, let us show them in details. Define:
$$\vec{w}(\vec{r}) = \frac{\vec{r}}{\vec{r}^2} + \vec{u}(\vec{r}) \ , \ P_{jk} (\vec{r}) = 2 \lambda(\vec{r}) \sum^3_{\ell = 1} \epsilon_{jk\ell} r_\ell \ , $$
then:
\begin{itemize}
    \item The $C$-functions: $${C_j}_{\mu\nu}(\vec{r}) = \left[ \frac12  \left( \sum^3_{k=1} P_{jk}(\vec{r}) \frac{\partial}{\partial r_k} \hat{D} (\vec{r}) \right) \hat{D}^{-1} + w_j(\vec{r})  \hat{D}(\vec{r}) \hat{\varepsilon} \right]_{\mu\nu} \ , $$
    where $\hat{D}^{-1}(\vec{r})$ is the inverse matrix of $\hat{D}(\vec{r})$ in Eq. \eqref{eq:def_D} and $\hat{\varepsilon}$ is the two-dimensional Levi-Civita matrix \cite{levi1977absolute} i.e.:
    $$\hat{\varepsilon} = \begin{pmatrix}
        \varepsilon_{11} & \varepsilon_{12} \\ \varepsilon_{21} & \varepsilon_{22}
    \end{pmatrix} = \begin{pmatrix}
        1 & 0 \\ -1 & 0
    \end{pmatrix} \ . $$
    \item Define a connection-curvature matrix $\hat{\mathcal{R}}$, whose elements are given by:
    \begin{equation*}
         \begin{split}{\mathcal{R}_{jk}}_{\mu\nu}(\vec{r}) = \sum^3_{\ell=1} \left( P_{j\ell}(\vec{r}) \frac{\partial}{\partial r_\ell} {C_k}_{\mu\nu} (\vec{r}) - P_{k\ell}(\vec{r}) \frac{\partial}{\partial r_\ell} {C_j}_{\mu\nu} (\vec{r}) - \frac{\partial}{\partial r_\ell} P_{jk}(\vec{r}){C_\ell}_{\mu\nu} (\vec{r}) \right) &
         \\
          - \sum^2_{\kappa=1} \left( {C_j}_{\mu\kappa}(\vec{r}) {C_k}_{\kappa \nu}(\vec{r}) - {C_k}_{\mu\kappa}(\vec{r}) {C_j}_{\kappa \nu}(\vec{r}) \right) & \ .
         \end{split}
    \end{equation*}
    The $B$-functions can be calculated from:
    $$B_{jk}(\vec{r}) = -\frac12 \sum^2_{\mu,\nu=1} \big( \hat{D}(\vec{r}) \hat{\varepsilon} \big)^{-1}_{\mu\nu} {\mathcal{R}_{jk}}_{\nu\mu}(\vec{r}) \ , $$
    where $\big( \hat{D}(\vec{r}) \hat{\varepsilon} \big)^{-1}$ is the inverse matrix of $\hat{D}(\vec{r}) \hat{\varepsilon}$.
    \item The $E$-functions:
    \begin{equation*}
        \begin{split}
            E_{\mu\nu}(\vec{r}) &= \sum^3_{j=1} w_j(\vec{r}) \frac{\partial}{\partial r_j} D_{\mu\nu}(\vec{r}) 
            \\
            & + \frac1{4 \det(\hat{D}(\vec{r}))} \sum^3_{j,k=1} P_{jk}(\vec{r}) \left[ \left( \frac{\partial}{\partial r_j} \hat{D} (\vec{r}) \right) \hat{\varepsilon} \left( \frac{\partial}{\partial r_k} \hat{D} (\vec{r}) \right)  \right]_{\mu\nu} \ . 
        \end{split}
    \end{equation*}
\end{itemize}

\subsection{Weyl Quantization of Supersphere}

To quantize the supersphere, we replace the bosonic coordinates
$r_j$ by block-diagonal matrices and the fermionic coordinates
$\theta_\mu$ by block-off-diagonal matrices, i.e.
\begin{equation}
r_j \rightarrow
\hat R_j=
\begin{pmatrix}
\hat{R}_j^{(p \times p)}&0\\
0& \hat{R}_j^{(q \times q)}
\end{pmatrix} \ \ \text{and} \ \ 
\theta_\mu \rightarrow 
\hat\Theta_\mu=
\begin{pmatrix}
0&\hat{\Theta}^{(p \times q)}_\mu\\
\hat{\Theta}^{(q \times p)}_\mu &0
\end{pmatrix} \ ,
\label{eq:super_matrix_blocks}
\end{equation}
in which $\hat{R}_j^{(p \times p)}$ and $\hat{R}_j^{(q \times q)}$ are matrices of
sizes $p \times p$ and $q \times q$, while $\hat{\Theta}^{(p \times q)}_\mu$ and $\hat{\Theta}^{(q \times p)}_\mu$
have sizes $p \times q$ and $q \times p$, respectively. Their matrix elements are ordinary complex numbers and
the block structure encodes their even or odd $\mathbb{Z}_2$-parity.
All matrices (i.e. the bosonic $\{ \hat{R}_j\}^3_{j=1}$ and the fermionic $\{ \hat{\Theta}_{\mu}\}^2_{\mu=1}$) therefore have size $N\times N$, with
$N=p+q$:
\begin{itemize}
    \item Under Weyl quantization, the Casimir constraint Eq. \eqref{eq:SUSY_Cas} in matrix form (no ordering ambiguity here so any matrix quantization scheme would give the same result) is given by:
\begin{equation}
    \mathcal{W} \left[F(\{ r_j\}^3_{j=1},\{ \theta_{\mu}\}^2_{\mu=1})\right] = \sum^3_{j=1} \hat{R}_j^2 + \left( \hat{\Theta}_1 \hat{\Theta}_2 - \hat \Theta_2 \hat \Theta_1 \right) - R^2 \hat{I}_{(p+q)\times(p+q)} = 0  \ .
    \label{eq:SUSY_Cas_mat}
\end{equation}
Define the superadjoint operation $^{\ddagger}$:
$$ \begin{pmatrix}
\hat{A}_j^{(p \times p)}& \hat{A}_j^{(p \times q)} \\
\hat{A}_j^{(q \times p)} & \hat{A}_j^{(q \times q)}
\end{pmatrix}^{\ddagger} = \begin{pmatrix}
(\hat{A}_j^{(p \times p)})^\dagger &  (\hat{A}_j^{(q \times p)})^\dagger
\\ -(\hat{A}_j^{(p \times q)})^\dagger & (\hat{A}_j^{(q \times q)})^\dagger
\end{pmatrix}   \ , $$
then:
\begin{itemize}
    \item For the bosonic matrices $\{ \hat{R}_j\}^3_{j=1}$, similar to Eq. \eqref{eq:Herm_XYZ}, we impose Hermiticity condition:
    $$\hat{R}_j = \hat{R}_j^{\ddagger} \ \Longrightarrow \ \hat{R}_j^{(p\times p)} = (\hat{R}_j^{(p\times p)})^\dagger \ , \ \hat{R}_j^{(q\times q)} = (\hat{R}_j^{(q\times q)})^\dagger \ . $$
    \item For the fermionic matrices $\{ \hat{\Theta}_{\mu}\}^2_{\mu=1}$, we impose
$$\hat\Theta_1^\ddagger=\hat\Theta_2 \ , \   \hat\Theta_2^\ddagger=-\hat\Theta_1 \ \Longrightarrow \ \hat \Theta_2^{(p \times q)} =(\hat \Theta_1^{(q \times p)})^\dagger \ , \ \hat \Theta_2^{(q \times p)} = -(\hat \Theta_1^{(p \times q)})^\dagger \ . $$
This is the analogue of the bosonic Hermiticity condition, which ensures that the fermionic contribution
$\hat\Theta_1\hat\Theta_2-\hat\Theta_2\hat\Theta_1$
to the Casimir constraint in matrix form Eq. \eqref{eq:SUSY_Cas_mat}, is Hermitian.
\end{itemize} 
\item Weyl quantization of the Poisson brackets between supercoordinates Eq. \eqref{eq:super_general} gives:
\begin{equation}
\begin{split}
[\hat R_j,\hat R_k]
&=i\hbar_{(p+q)} \,\mathcal W
 \left[
 2\lambda(\vec r)\sum_{\ell=1}^3\epsilon_{jk\ell}r_\ell
 +B_{jk}(\vec r)\theta_1\theta_2
 \right],\\
[\hat R_j,\hat\Theta_\mu]
&=i\hbar_{(p+q)} \mathcal W
 \left[
 \sum_{\nu=1}^2{C_j}_{\mu\nu}(\vec r)\theta_\nu
 \right],\\
[ \hat\Theta_\mu,\hat\Theta_\nu]_+
&=i\hbar_{(p+q)} \mathcal W
 \left[
 D_{\mu\nu}(\vec r)
 +E_{\mu\nu}(\vec r)\theta_1\theta_2
 \right] \ ,
\end{split}
\label{eq:super_matrix_general}
\end{equation}
where $[...,...]_+$ is the anti-commutator, i.e.
$$[\hat{U}_1,\hat{U}_2]_+ = \hat{U}_1 \hat{U}_2 + \hat{U}_2 \hat{U}_1 \ \forall \hat{U}_1, \hat{U}_2 \ , $$
and Weyl ordering of supermanifold coordinates is defined by replacing classical coordinates with their corresponding matrices and averaging over all permutations of the factors, including a factor of $-1$ for each interchange of two odd $\mathbb{Z}_2$-parity factors, e.g.
$$\mathcal W [\theta_1\theta_2]
=\frac12
 \left(\hat\Theta_1\hat\Theta_2-\hat\Theta_2\hat\Theta_1\right) \ .$$
\end{itemize}

\subsection{A Benchmark Test for Numerical Fuzzy Supersphere}

\begin{figure*}[hbp!]
    \centering
\includegraphics[width=\linewidth]{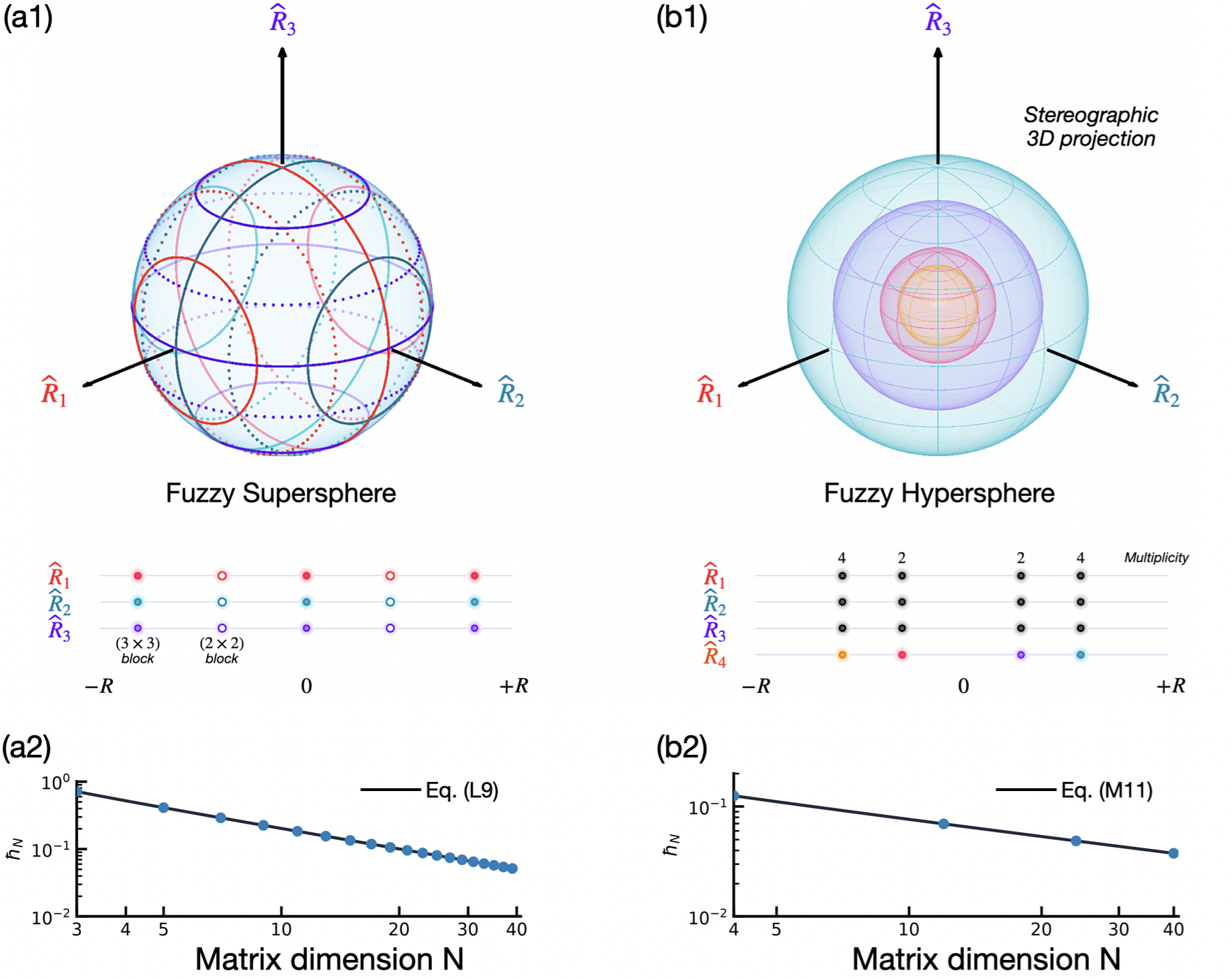}
    \caption{\textbf{Quantization of more general classical geometries into fuzzy matrix geometries.} Here we show the fuzzy supersphere (i.e. a $N=5$ matrix quantization of a classical supermanifold) in \textbf{(a1)} and the fuzzy hypersphere (i.e. a $N=12$ matrix quantization of a classical hypersurface equipped with a Nambu structure) in \textbf{(b1)}. Numerical searches can reproduce known analytical results, in \textbf{(a2)} and in \textbf{(b2)} respectively.}
\label{figS01}
\end{figure*}

For a benchmark, consider the
$\mathfrak{osp}(1|2)$  fuzzy supersphere \cite{hasebe2008susy} (see Fig. \ref{figS01}a1):
$$ \lambda(\vec{r}) =1 \ , \ \hat{D} (\vec{r}) =-i \hat\varepsilon\sum_{j=1}^3 \hat{\sigma}_j r_j = \begin{pmatrix}
    r_2-ir_1 & i r_3 \\ i r_3 & r_2 + i r_1
\end{pmatrix}  \ , \ u_j (\vec{r}) = 0 \ , $$
where $\hat{\sigma}_j$ are the Pauli matrices \cite{griffiths2018introduction}, then:
\begin{equation}
B_{jk}(\vec{r}) =E_{\mu\nu}(\vec{r}) =0 \ , \
{C_j}_{\mu\nu}(
\vec{r}
) =-i ( \hat{\sigma}_j^T )_{\mu\nu} \ .
\end{equation}
The matrix dimensions $(p,q)$ satisfy the relation $p=q+1$, which means the total dimension $N$ is always an odd number, i.e. 
$$N = 2q+1 \in 2\mathbb{N}_{>0}+1 \ . $$
Eq. \eqref{eq:super_matrix_general} reduces to
\begin{equation}
\begin{split}
[\hat R_j,\hat R_k]
&=2i\hbar_N\sum_{\ell=1}^3
  \epsilon_{jk\ell}\hat R_\ell \ , 
  \\
[\hat R_j,\hat\Theta_\mu]
&=\hbar_N\sum_{\nu=1}^2
  (\hat{\sigma}_j^T)_{\mu\nu} \hat\Theta_\nu \ ,\\
[\hat\Theta_\mu,\hat\Theta_\nu]_+
&=\hbar_N\sum_{j=1}^3
  (\hat\varepsilon \hat{\sigma}_j)_{\mu\nu}\hat R_j \ .
\end{split}
\label{eq:super_matrix_benchmark}
\end{equation}
For irreducible representations, the quantization parameter is given by:
\begin{equation}
\hbar_N
=\frac{2R}{\sqrt{N^2-1}} \ .
\label{eq:super_hbar_benchmark}
\end{equation}
For simplicity, we set $R=1$.

Our numerical approach is similar to Section \ref{sec:num}, with the total loss consisting of the algebraic constraints -- Eq. \eqref{eq:SUSY_Cas_mat} and Eq. \eqref{eq:super_matrix_benchmark} -- and the reducibility punishment -- for the quintet $(\{\hat{R}_j\}^3_{j=1},\{ \hat \Theta_\mu \}^2_{\mu=1})$. The trainable parameters
are:
\begin{itemize}
    \item the three Hermitian block pairs
$\{ (\hat R_j^{((q+1) \times (q+1))},\hat R_j^{(q \times q)}) \}^3_{j=1}$,
    \item the complex matrices $\{ \hat \Theta_1^{((q+1)\times q)} , \hat \Theta_1^{(q\times (q+1))} \}$, and
    \item the quantization parameter $\hbar_N>0$.
\end{itemize} 
The results from our numerical search agree with the known analytical values Eq. \eqref{eq:super_hbar_benchmark} (see Fig. \ref{figS01}a2).

\section{Fuzzy Hypersphere \label{app:nambu}}

\subsection{Generalized Nambu-Poisson Structures}

An $m$-ary Nambu-Poisson bracket, with $m\geq2$, assigns
a function $\{f_1,f_2,...,f_m\}$ to $m$ functions $f_1$, $f_2$, ..., $f_m$; special cases include:
\begin{itemize}
\item For $m=2$, the bracket defines Poisson structures.
\item For $m=3$, the bracket defines Nambu structures.
\end{itemize}
In general, this bracket is linear in each argument (i.e. satisfies \underline{linearity}), changes sign when any two
function arguments are exchanged (i.e. satisfies \underline{anti-symmetry}), and follows the \underline{Leibniz rule}:
$$
\{f_1,f_2,...,f_{m-1},gh\}
=
\{f_1,f_2,...,f_{m-1},g\}h
+g\{f_1,f_2,...,f_{m-1},h\} \ .
$$
Furthermore, it also satisfies the \underline{fundamental identity}
\cite{takhtajan1994foundation,ho2016nambu}:
\begin{equation}
    \begin{split}
        &\{f_1,f_2,...,f_{m-1},\{g_1,g_2,...,g_m\}\}
\\
& \ \ \ \ \ \ \ \ = \sum_{r=1}^{m}
\{g_1,g_2, ...,g_{r-1}\{f_1,f_2,...,f_{m-1},g_r\},g_{r+1},...,g_m\} \ ,
    \end{split}
    \label{eq:nambu_FI}
\end{equation}
which reduces to the \underline{Jacobi identity} for $m=2$. Given the coordinates $(r_1,r_2,...,r_d)$, the bracket takes the form
\begin{equation}
\{f_1,f_2,...,f_m\}
=
\sum_{j_1,j_2,...,j_m=1}^{d}
\frac{\partial}{\partial r_{j_1}} f_1 \frac{\partial}{\partial r_{j_2}} f_2
...
\frac{\partial}{\partial r_{j_m}} f_m \ \{r_{j_1},r_{j_2},...,r_{j_m}\} \ ,
\end{equation}
where $
\{r_{j_1},r_{j_2},...,r_{j_m}\}
$ are the components of the antisymmetric Nambu-Poisson tensor. These components must satisfy the constraints imposed by
Eq. \eqref{eq:nambu_FI}.

For a regular hypersurface defined by $F(\vec{r})=0$ in $\mathbb R^{m+1}$,
the most general $m$-ary Nambu-Poisson bracket having $F(\vec{r})$ as a Casimir constraint:
\begin{equation}
\{r_{j_1},r_{j_2},...,r_{j_m}\}
=
\lambda(\vec r)
\sum_{k=1}^{m+1}
\epsilon_{j_1 j_2 ... j_m k}
\frac{\partial}{\partial r_k} F(\vec{r}) \ ,
\label{eq:nambu_general}
\end{equation}
where $\epsilon_{j_1 j_2 ... j_m k}$ is the $(m+1)$-dimensional Levi-Civita symbol \cite{levi1977absolute} and $\lambda(\vec{r})$ can be any smooth function, i.e. the \underline{fundamental identity} for $m$-ary Nambu-Poisson bracket in $(m+1)$-dimensional space adds no restriction on $\lambda(\vec{r})$.

Consider an $m$-dimensional hypersphere embedded in
$\mathbb R^{m+1}$, with the Casimir constraint:
\begin{equation}
F(\vec r)=\sum_{j=1}^{m+1}r_j^2-R^2=0 \ .
\label{eq:Cas_h3s}
\end{equation}
Consequently, Eq. \eqref{eq:nambu_general} becomes:
\begin{equation}
\{r_{j_1},r_{j_2},...,r_{j_m}\}
=
2 \lambda(\vec r)
\sum_{k=1}^{m+1}
\epsilon_{j_1 j_2 ... j_m k}
r_k \ .
\label{eq:nambu_general_hs}
\end{equation}

\subsection{Weyl Quantization of Hypersphere}

Let us focus on the $m=3$ hypersphere, on which we have the Nambu structures, i.e. Eq. \eqref{eq:nambu_general_hs} becomes:
\begin{equation}
\{r_{j_1},r_{j_2},r_{j_3}\}
=
2 \lambda(\vec{r})
\sum_{k=1}^{4}
\epsilon_{j_1 j_2 j_3 k}
r_k \ ,
\label{eq:nambu_general_h3s}
\end{equation}
To quantize the hypersphere, we need to map four coordinates $\{r_j \}^4_{j=1}$ to four Hermitian matrices $\{\hat{R}_j \}^4_{j=1}$ of size $N \times N$. This geometry emerges in M2-M5-brane systems from M-theory \cite{basu2005m2}, where multiple M2-branes end on an M5-brane and their expanding transverse cross-sections are described by fuzzy hyperspheres.

 For $m=2$, quantization promotes the Poisson bracket to matrix commutation, which also satisfies the Poisson algebra properties (\underline{linearity}, \underline{anti-symmetry}, \underline{Leibniz rule}, and \underline{Jacobi identity}). For $m=3$, however, the promotion of the Nambu bracket is less straightforward, e.g. the totally anti-symmetrized product of three matrices generally fails to satisfy the defining properties of a Nambu-Poisson algebra, specifically the \underline{Leibniz rule} and the \underline{fundamental identity} \cite{ho2016nambu}. A matrix prescription must therefore be specified separately, with a choice of which classical properties of Nambu algebra it could preserve\footnote{Preserving the full Nambu algebra requires departing from the ordinary matrix product. Zariski quantization achieves this on an enlarged algebra with a deformed commutative product \cite{dito1997deformation}. It thus differs from the finite-dimensional Hermitian matrix framework considered here.}.

For an even matrix dimension $N$, i.e. $N \in 2\mathbb{N}_{>0}$, a conventional promotion choice would be \cite{basu2005m2,guralnik2001polarization}:
\begin{equation}
    \{ \hat{U}_1,\hat{U}_2,\hat{U}_3\} \rightarrow  [\hat{U}_1,\hat{U}_2,\hat{U}_3,\hat K]_4 \ \ \text{where} \ \ \hat K=\begin{pmatrix}\hat I_{N/2}&0\\0&-\hat I_{N/2}\end{pmatrix} \ .
    \label{eq:comm_pro_h3s}
\end{equation}
The matrix operation $[...,...,...,...]_4$ denotes the fully
anti-symmetrized product of four matrices, i.e.   $$ [\hat{U}_1,\hat{U}_2,\hat{U}_3,\hat{U}_4]_4
=\frac{1}{4!}\sum_{\pi\in S_4}
\operatorname{sgn}(\pi)
\hat{U}_{\pi(1)}\hat{U}_{\pi(2)}\hat{U}_{\pi(3)} \hat{U}_{\pi(4)} \ , $$
where $S_4$ denotes the set of all $4!=24$ permutations
of $(1,2,3,4)$. Each permutation $\pi$ specifies an ordering
of the four matrices, and
\[
\operatorname{sgn}(\pi)=
\begin{cases}
+1, & \text{if the ordering is obtained by an even number of swaps},\\
-1, & \text{if the ordering is obtained by an odd number of swaps}.
\end{cases}
\]
Here each swap exchanges two entries, starting from
$(1,2,3,4)$, e.g.
$\hat U_1\hat U_2\hat U_3\hat U_4$ has a positive sign,
while $\hat U_2\hat U_1\hat U_3\hat U_4$ has a negative sign. This promotion choice preserves \underline{linearity} and
\underline{anti-symmetry}. Also, if the multiplier $\lambda(\vec{r})$ is a constant in Eq. \eqref{eq:nambu_general_h3s}, the
\underline{fundamental identity} holds for all matrix quantization scheme when every argument
is a linear combination of $\{ \hat{R}_j \}^4_{j=1}$. However, in general the \underline{Leibniz rule} fails.

Under Weyl quantization:
\begin{itemize}
    \item The Casimir constraint Eq. \eqref{eq:Cas_h3s} in matrix form (no ordering ambiguity here so any matrix quantization scheme would give the same result):
    \begin{equation}
    \mathcal{W} \left[F(\vec{r})\right] = \sum^4_{j=1} \hat{R}_j^2 - R^2 \hat{I}_{N \times N} = 0  \ .
    \label{eq:Cas_h3s_mat}
\end{equation}
\item The Nambu brackets Eq. \eqref{eq:nambu_general_h3s} in matrix form:
\begin{equation}
[\hat{R}_{j_1},\hat{R}_{j_2},\hat{R}_{j_3},\hat{K}]_4
=
2 \hbar_N
\sum_{k=1}^{4}
\epsilon_{j_1 j_2 j_3 k}
\mathcal{W} [\lambda(\vec{r}) r_k] \ .
\label{eq:nambu_spec_h3s_mat}
\end{equation}
\end{itemize}

\subsection{A Benchmark Test for Numerical Fuzzy Hypersphere}

For a benchmark, consider fuzzy hypersphere  \cite{basu2005m2,guralnik2001polarization} corresponding to constant
$\lambda(\vec{r})$, i.e. $\lambda(\vec{r})=1$ (see Fig. \ref{figS01}b). Then, Eq. \eqref{eq:nambu_spec_h3s_mat} becomes
\begin{equation}
    [\hat{R}_{j_1},\hat{R}_{j_2},\hat{R}_{j_3},\hat{K}]_4
=
2 \hbar_N
\sum_{k=1}^{4}
\epsilon_{j_1 j_2 j_3 k}
\hat{R}_k
\label{eq:nambu_spec_h3s_1}
\end{equation}
and every matrix quantization scheme is the same (no ordering ambiguity anywhere). For irreducible representations, the allowable matrix dimension $N$ is given by
$$N = \frac{(n+1)(n+3)}2 \ , $$
where $n$ is an odd number i.e. $n \in 2\mathbb{N}_{>0}+1$, and the corresponding quantization parameter can be calculated with
\begin{equation}
\hbar_N
=\frac{(n+2)R^2}{3(n+1)(n+3)} \ .
\label{eq:hyper_hbar_benchmark}
\end{equation}
For simplicity, we set $R=1$.

Our numerical approach is similar to Section \ref{sec:num}, with the total loss consisting of the algebraic constraints -- Eq. \eqref{eq:Cas_h3s_mat} and Eq. \eqref{eq:nambu_spec_h3s_1} -- and the reducibility punishment. The trainable parameters
are:
\begin{itemize}
    \item the four Hermitian matrices
$\{ \hat R_j \}^4_{j=1}$, and
    \item the quantization parameter $\hbar_N>0$.
\end{itemize} 
The results from our numerical search agree with the known analytical values Eq. \eqref{eq:hyper_hbar_benchmark} (see Fig. \ref{figS01}b2).

\end{document}